\documentclass[%
 aip,
 amsmath,amssymb,
 reprint,%
]{revtex4-1}

\usepackage{graphicx}% Include figure files
\usepackage{dcolumn}% Align table columns on decimal point
\usepackage{bm}% bold math
\usepackage[utf8]{inputenc}
\usepackage[T1]{fontenc}
\usepackage{mathptmx}
\usepackage{etoolbox}
\usepackage[squaren]{SIunits}
\makeatletter
\def\@email#1#2{%
 \endgroup
 \patchcmd{\titleblock@produce}
  {\frontmatter@RRAPformat}
  {\frontmatter@RRAPformat{\produce@RRAP{*#1\href{mailto:#2}{#2}}}\frontmatter@RRAPformat}
  {}{}
}%
\makeatother
\begin{document}

\preprint{AIP/123-QED}

\title[Sample title]{Levitons in correlated nano-scale systems}
% Force line breaks with \\
\author{F. Ronetti}
% \altaffiliation[Also at ]{Physics Department, XYZ University.}%Lines break automatically or can be forced with \\
\author{B. Bertin-Johannet}
\author{A. Popoff}
\altaffiliation[Also at ]{Coll\`ege de Tipaerui, BP 4557 - 98713 Papeete, Tahiti, French Polynesia}
\author{J. Rech}
\author{T. Jonckheere}
\author{B. Grémaud}
\author{L. Raymond\\}
\author{T. Martin}
\email{flavio.ronetti@univ-amu.fr}
\affiliation{ 
Aix Marseille Univ, Universit\'e de Toulon, CNRS, CPT, IPhU, AMUTECH, Marseille, France%\\This line break forced with \textbackslash\textbackslash
}%

\date{\today}% It is always \today, today,
             %  but any date may be explicitly specified

\begin{abstract}
In this short review (written to celebrate David Campbell's 80th birthday), we provide a theoretical description of quantum transport in nanoscale systems in the presence of single-electron excitations generated by Lorentzian voltage drives, termed \textit{Levitons}. These excitations allow to realize the analog of quantum optics experiments using electrons instead of photons. Importantly, electrons in condensed matter systems are strongly affected by the presence of different types of non-trivial correlations, with no counterpart in the domain of photonic quantum optics. After providing a short introduction about Levitons in non-interacting systems, we focus on how they operate in the presence of two types of strong electronic correlations in nanoscale systems, such as those arising in the fractional quantum Hall effect or in superconducting systems. Specifically, we consider Levitons in a quantum Hall bar of the fractional quantum Hall effect, pinched by a quantum point contact, where anyons with fractional charge and statistics tunnel between opposite edges. In this case, a Leviton-Leviton interaction can be induced by the strongly correlated background. Concerning the effect of superconducting correlations on Levitons, we show that, in a normal metal system coupled to BCS superconductors, half integer Levitons minimize the excess noise in the Andreev regime. Interestingly, energy-entangled electron states can be realized on-demand in this type of hybrid setups by exploiting crossed Andreev reflection. The results exposed in this review have potential applications in the context of quantum information and computation with single-electron flying qubits.
\end{abstract}

%Parallel to the birth of mesoscopic physics, in the 1980's pioneering experiments on second order interference were performed in quantum optics as single photon sources became available: it was possible by photon down conversion on a nonlinear crystal to generate near identical photon pairs pulses and to exploit such pairs of photons in Hong-Ou-Mandel (HOM) collisions at a half silvered mirror showing second order interference effects as the wave packets of the two photons overlap. 
%%%%%%%%%%%%%%%%%%%%%%%%%%%%%%%%%%%%%%%%%%%%%%%%%%%%%%%%%%%%%%%%%%%%%%%%%%%%%%%%%%%%%%%%%

\maketitle

\begin{quotation}
A quarter of a century ago,\cite{lee1993,levitov1996} Lorentzian voltage pulses were introduced in the context of electron quantum transport as a time domain version of the Anderson catastrophe\cite{anderson1967}: unless the time integral of the injected bias voltage corresponds to an integer number of flux quanta, the number of created electron-hole pairs diverges. With the development of electronic quantum optics in nanophysics/mesoscopic physics, these Leviton excitations have been used to implement scenarios which mimic the Hanbury Brown and Twiss partitioning\cite{brown1956} and Hong-Ou-Mandel collision\cite{hong1987} experiments for single photon sources. Here, we examine to what degree these protocols can be extended to situations where electron correlations are operating. First, we consider Lorentzian voltage pulses for Laughlin fractions of the fractional quantum Hall effect and show that they give rise to minimal excess noise when multiples of electron charge are injected as opposed to fractional charges. Second, we show that the strongly correlated background induce an effective interaction between Levitons propagating in the edge states. Finally, we show that Levitons in the presence of superconducting correlations can realize a Cooper pair beam splitter operating in the time domain, thus generating an on-demand source of delocalized energy entangled electron states.    
\end{quotation}

\section{\label{introduction}Introduction}

In modern condensed matter physics, the fabrication of nanoscale systems at low temperature allows to access the wave-like nature of electrons which manifests in coherent phenomena such as interferences, similar to those associated with photons.
This gave rise to a field of research called mesoscopic physics. It was pioneered roughly about half a century ago by the ground-breaking discovery interference effects in disordered metals soon to be followed by that of the integer quantum Hall effect.\cite{klitzing1980} Initially, transport experiments were carried out at a constant electrical bias and many peculiar phenomena were observed, such as the quantization of the Hall conductance. Later on, the interest for a deeper exploration of these systems pushed the mesoscopic physics community  to explore dynamical aspects of quantum transport by superposing to the DC drive an AC voltage. 

The idea of pushing this investigation down to the single-electron level gave rise during the last two decades to Electron Quantum Optics (EQO). The wave-like nature of electrons traveling in one-dimensional edge states of quantum Hall systems bears strong analogies with the propagation of photons in wave-guides. Using equivalents of beam-splitters and optical fibers, the electronic equivalents of optical setups can be implemented in a solid state system and used to investigate mesoscopic transport at the single-electron limit. These optical-like experiments provide a powerful tool to improve the understanding of electron propagation in quantum conductors. Inspired by the controlled manipulations of the quantum state of light, the recent development of single electron emitters has opened the way to the controlled preparation, manipulation and characterization of single to few electronic excitations that propagate in EQO setups. 
However, these experiments go beyond the simple transposition of optics concepts in electronics as several major differences occur between electrons and photons. Photons are neutral particles that interact weakly and obey Bose-Einstein statistics, while electrons are fermions, they bear a charge and thus interact strongly with their environment, and they are accompanied by a Fermi sea. 

In the context of EQO, a remarkable effort has been put forth by the condensed matter community to implement on-demand sources of electronic wave-packets in mesoscopic systems~\cite{dubois2013b, grenier2013,misiorny2018,glattli2016b,bauerle2018}. Here, we focus on a type of on-demand source of electrons which is based on the application of a time-dependent voltage to a quantum conductor \cite{glattli2016b,glattli2016a,ferraro2018,misiorny2018,moskalets2016}. 
However, an arbitrary AC voltage generally excites unwanted neutral electron-hole pairs, thus spoiling at its heart the idea of a single-electron source.
This issue was overcome thanks to the theoretical prediction by Levitov and co-workers that a superposition of quantized Lorentzian-shaped pulses, carrying an integer number of particles per period, is able to inject single-electron excitations devoid of any additional electron-hole pairs, then termed \textit{Levitons} \cite{levitov1996,ivanov1997,keeling2006}. Indeed, this kind of single-electron source is simple to realize and operate, since it relies on usual electronic components, and potentially provides a high level of miniaturization and scalability. For their fascinating properties \cite{moskalets2015}, Levitons have been proposed as flying qubits \cite{glattli2018} and as a source of entanglement \cite{dasenbrook2015,dasenbrook2016b,dasenbrook2016} with appealing applications for quantum information processing. Moreover, quantum tomography protocols able to reconstruct their single-electron wave-functions have been proposed \cite{grenier2011,ferraro2013,ferraro2014} and experimentally realized \cite{jullien2014}. 

\begin{figure}
    \centering
    \includegraphics[width=\linewidth]{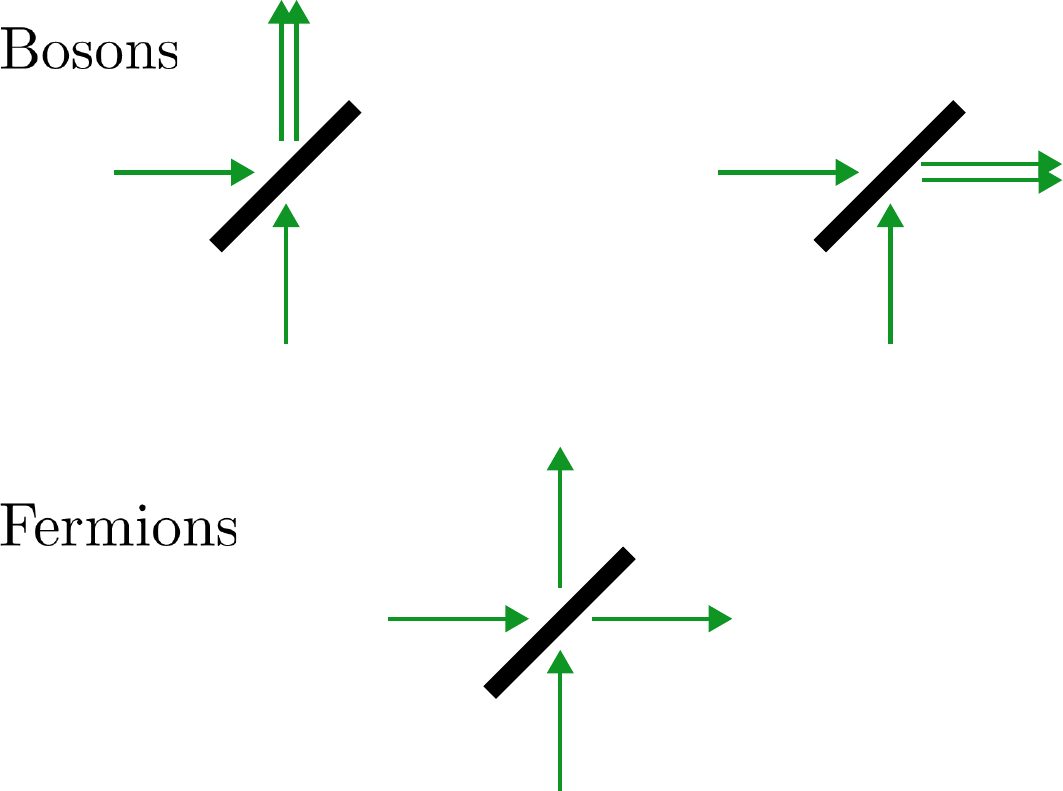}
    \caption{HOM collisions between pairs of bosons and pairs of electrons at the location of a semi-transparent mirror or its equivalent, a QPC, for electrons. Bosons show a bunching effect and the two particle end up at the same output, while electrons exit at opposite outputs}
    \label{fig:hom}
\end{figure}

These single-electron sources allow the on-demand injection of individual excitations into mesoscopic devices mimicking the conventional photonic quantum optics, with quantum Hall edge channels\cite{klitzing1980} behaving as waveguides. For instance, the role of the half-silvered mirror of conventional optics, can be played by a quantum point contact (QPC), where electrons are reflected or transmitted with a tunable probability. In this sense, one seminal example is the Hanbury Brown and Twiss (HBT) interferometer \cite{brown1956}, where a stream of electronic wave-packets is excited along ballistic channels and is partitioned at the location of a QPC \cite{bocquillon2012}. Another fundamental achievement of EQO has been the implementation of the Hong-Ou-Mandel (HOM) interferometer,\cite{hong1987} (see Fig. \ref{fig:hom}) where electrons are incident on the opposite side of a QPC with a tunable delay \cite{bocquillon2013,glattli2016b,dubois2013b}. By performing this kind of collision experiments, it is possible to gather information about the forms of the impinging electronic wave-packets and to measure their degree of indistinguishability \cite{jonckheere2012,ferraro2018,ferraro2015}. %The shot noise signal, generated due to the granular nature of electrons \cite{Martin_Houches,Moskalets17}, was employed to probe the single-electron nature of Levitons in a non-interacting two-dimensional electron gas \cite{glattli2016a,dubois13-nature}. 

%For instance, when two indistinguishable and coherent electronic states collide simultaneously (zero time delay) at the QPC charge current fluctuations are known to vanish at zero temperature, thus showing the so called Pauli dip \cite{Ferraro14b,glattli2016b,dubois13-nature}. This dip can be interpreted in terms of anti-bunching effects related to the Fermi statistics of electrons \cite{Wahl14,Ferraro14,Freulon15,Marguerite16,Cabart18}.

As opposed to photons propagating in vacuum, electrons in solid-state systems are affected by Coulomb interaction and its interplay with other degrees of freedom in the material. As a result, the research field of EQO in correlated nanoscale devices present many fascinating and appealing features which have no counterparts in standard quantum optics. Moreover, a full understanding of the propagation of single-electron excitations in correlated systems, such as the fractional quantum Hall edge states and hybrid-superconducting systems, is extremely relevant for the potential implementation of quantum information and computation schemes. Indeed, in very recent years, a huge interest has been devoted to the concept of electron flying qubits, which can be an efficient and fully scalable solution towards solid-state quantum computation. 
To this end, a better control on the realization of electron flying qubits and their interactions would have to be achieved.
This requires a deeper investigation of Levitons in correlated systems. In this review, we will introduce the powerful theoretical framework for the description of transport properties of Levitons in the FQH regime and in superconducting systems, or photo-assisted shot-noise (PASN) formalism, and we will present the main results that have been obtained so far in this field.

The scope of the paper is as follows. Levitons and Electronic quantum optics of the integer quantum Hall effect are introduced pedagogically in Sec. \ref{integer}. The photo-assisted shot noise formalism for the weak back-scattering regime of the FQHE is then considered in Sec. \ref{chap_charge}, and Levitons are discussed in this context in Sec. \ref{levitons_fqhe}. Sec. \ref{entangled} focuses on a time dependent version of the Cooper pair beam spitter. Perspectives and extensions are described in Sec. \ref{conclusion}.

%%%%%%%%%%%%%%%%%%%%%%%%%%%%%%%%%%%%%%%%%%%%%%%%%%%%%%%%%%%%%%%%%%%%%%%%%%%%%%%%%%%%%%%%%%%%
 \section{Levitons in non-interacting systems\label{noninteracting}}
\label{integer}
 
Levitons have been first introduced in systems where correlations between electrons play no role. As a starting point for our main discussions, we begin by focusing on the emission and propagation of Levitons in non-interacting systems. In this way, we can introduce the general theoretical framework for the description of fundamental concepts of EQO. In order to set the stage for our later discussions, we present our calculations in the context of the Integer Quantum Hall effect (IQHE)\cite{klitzing1980} at $\nu = 1$, where a correct theoretical interpretation can be provided even by neglecting electronic correlations. As a matter of fact, the results involving Levitons presented in this case can be generalized to any non-interacting system where quantum channels can be realized, e.g., a two-dimensional electronic gas in the presence of a QPC.

As a first step, we demonstrate that Lorentizan pulses carrying an integer number of electronic charges $e$ behave as a controlled source of single-electron excitations devoid of hole-like particles. 

\subsection{Edge state driven by a time-dependent voltage\label{Levitons}}

Let us consider one edge of a quantum Hall bar at filling factor $\nu=1$, connected to a reservoir driven by a generic time-dependent voltage $V(t)$. For the following discussion, it does not matter on which chirality (right or left moving fermions) we focus on. A single edge state of a quantum Hall bar at $\nu=1$ is described by the edge Hamiltonian (we put $\hbar=1$ throughout the paper)
\begin{equation}
H_{0}=\int_{-\infty}^{+\infty}\mathrm{d}x\, :\Psi^{\dagger}(x)(- i u \partial_x-\mu)\Psi(x):\,,
\label{eq:ham_ferm2}
\end{equation}
where $\mu=u k_F$ is the chemical potential and
\begin{equation}
\label{eq:psi_ferm2}
\Psi(x)=\frac{1}{\sqrt{2\pi u}}\int_{-\infty}^{+\infty}\mathrm{d}\, \epsilon e^{ i\epsilon \frac{x}{u}}a(\epsilon).
\end{equation}

The fermionic operators $a(\epsilon)$ satisfy the following average values over the equilibrium configuration at temperature $\theta$
\begin{align}
\label{eq:av_value_a1}
\langle a^{\dagger}(\epsilon)a(\epsilon')\rangle=\delta(\epsilon-\epsilon') f(\epsilon),\\
\label{eq:av_value_a2}
\langle a(\epsilon)a^{\dagger}(\epsilon')\rangle=\delta(\epsilon-\epsilon') \left(1-f(\epsilon)\right),
\end{align}
where $f(\epsilon)=\left[1+e^{\frac{\epsilon-\mu}{\theta}}\right]^{-1}$ is the Fermi distribution function at temperature $\theta$.

The time-dependent current flowing along this edge state is given by $J(t) = e^2/(2\pi) V(t)$. The total charge $\mathcal{C}$ injected by the drive into the edge states can be written as
\begin{equation}
\mathcal{C}=\int_{-\infty}^{+\infty}\mathrm{d}t\, J(t)=\frac{e^2}{2\pi}\int_{-\infty}^{+\infty}\mathrm{d}t\, V(t).\label{eq:total charge}
\end{equation} 

When time-dependent voltage $V(t)$ is applied on a conducting channel, the electrons that exit the contact and enter the conducting channels have acquired a time-dependent phase $e^{i \chi(t)}$, with
\begin{equation}
\chi(t)=e \int_{-\infty}^{t-\frac{x}{u}}\mathrm{d}t'\, V(t'),
\label{chi_def}\end{equation}
where we assumed a right-moving chirality for simplicity. It is useful to introduce the Fourier transform of this voltage phase, which reads
\begin{equation}
\label{eq:p_epsilon}
p(\epsilon)=\int_{-\infty}^{+\infty} \mathrm{d}t\, e^{i \epsilon t}e^{i \chi (t)}.
\end{equation}
The fermion field operator in Eq. \eqref{eq:psi_ferm2} can be recast in terms of this Fourier transform as
\begin{align}
\Psi(x,t)=\frac{1}{\sqrt{2\pi v}}\int_{-\infty}^{+\infty}d \epsilon e^{-i\epsilon \left(t-\frac{x}{u}\right)}\tilde{a}(\epsilon),
\end{align}
where we defined
\begin{equation}
\tilde{a}(\epsilon)=\int_{-\infty}^{+\infty}\mathrm{d}\epsilon_1 p(\epsilon_1)a(\epsilon-\epsilon_1),\label{eq:tilde_a}
\end{equation}
which is expressed as a superposition of fermionic operators $a(\epsilon)$ weighted by the coefficients $p(\epsilon)$. The latter can physically be understood as the amplitude of probability of absorption and emission of a photon with energy $\epsilon$ from the voltage drive. In the presence of a time-dependent drive, electrons are excited above the chemical potential when photons are absorbed and holes are generated below the chemical potential when photons are emitted, with respect to the equilibrium situation. In the following, we employ this description of fermion fields in the presence of an external voltage  to find out the drive shape required to emit on-demand single-electron excitations.

\subsection{Lorentzian drive as an on-demand single-electron source}

While the average value of the emitted charge can be controlled by properly choosing the parameters of the voltage [see Eq.~\eqref{eq:total charge}], additional neutral particles can be excited in the edge state, such as electron-hole pairs. The number of holes can be calculated in terms of fermionic operators $\tilde{a}(\epsilon)$ as
\begin{equation}
N_h=\int_{-\infty}^{\mu}\mathrm{d}\epsilon\, \left\langle\tilde{a}(\epsilon)\tilde{a}^{\dagger}(\epsilon)\right\rangle\,,
\end{equation}
where the average is calculated at equilibrium. The integration is performed exclusively over energies \textit{below} the chemical potential. In the zero temperature limit, one finds
\begin{align}
N_h=\int_{-\infty}^{0}\mathrm{d}\epsilon\,\int_{-\infty}^{\epsilon}d \epsilon_1|p(\epsilon_1)|^2\,.\label{eq:n_h_fin}
\end{align}
As a result, we linked the number of holes to the function $p(\epsilon)$ and, as a consequence, to the shape of the time-dependent voltage drive. 
A sufficient condition for having no electron-hole pairs on average is thus given by $p(\epsilon)=0$ for $\epsilon < 0$. This imposes a constraint on the structure of
 $e^{i \chi (t)}$: the integral in Eq.~\eqref{eq:p_epsilon} vanishes when $e^{i \chi (t)}$ has no pole in the lower half plane, but at least one pole in the upper half plane, since it has to be non-zero somewhere. The simplest choice of function that fulfills this requirement is
\begin{equation}
\label{eq:phase_lor}
e^{i \chi (t)}=\frac{t+ i W}{t- i W},
\end{equation}
where $W$ is a positive real number. By using the definition Eq. (\ref{chi_def}), one can find the corresponding voltage, which is a Lorentzian voltage pulse
\begin{equation}
V_{lor}(t)=\frac{1}{(-e)}\frac{2W}{t^2+W^2},\label{eq:vlor}
\end{equation} 
where W can now also be viewed as the half width at half height. One can show that a quantized Lorentzian drive with a single peak creates a quantum state of the form~\cite{keeling2006,grenier2013}
\begin{equation}
\left|\Psi\right\rangle = \int \mathrm{d}x\, ~\varphi_1^*(x)\Psi^{\dagger}(x)\left|F\right\rangle,
\end{equation}
where $\Psi^{\dagger}(x)$ creates an electron at the position $x$, $\left|F\right\rangle$ is the ground state of the system and
\begin{align}
	\label{eq:wave-leviton}
	\varphi_1(x-ut) = \sqrt{\frac{W u}{\pi}}\frac{1}{x-ut+i u W},
\end{align}
is the wave-function of a single Leviton propagating in a chiral edge state. 
a
For an implementation which can be compared to experiments one has to consider periodic voltage pulses. The expression for a train of quantized Lorentzian pulses with period $\mathcal{T}$ is
\begin{equation}\label{eq:lorentzian_periodic}
V_{lor}(t)=\sum_{k=-\infty}^{\infty}\frac{1}{(-e)}\frac{2W}{(t-k\mathcal{T})^2+W^2}.
\end{equation}
In this case, the wave-function becomes
\begin{equation}
\label{eq:wave-levitond_period}
\varphi_1(x-ut) = \sqrt{\frac{u \sinh\left(\frac{2\pi W}{\mathcal{T}}\right)}{2}} \left\{\sin\left[\frac{\pi \left(t-\frac{x}{u}-iW\right)}{\mathcal{T}}\right]\right\}^{-1}.
\end{equation}

\subsection{Current noise generated by Levitons in a QPC geometry\label{subsec:noise}}
\begin{figure}
\centering
\includegraphics[width=\linewidth]{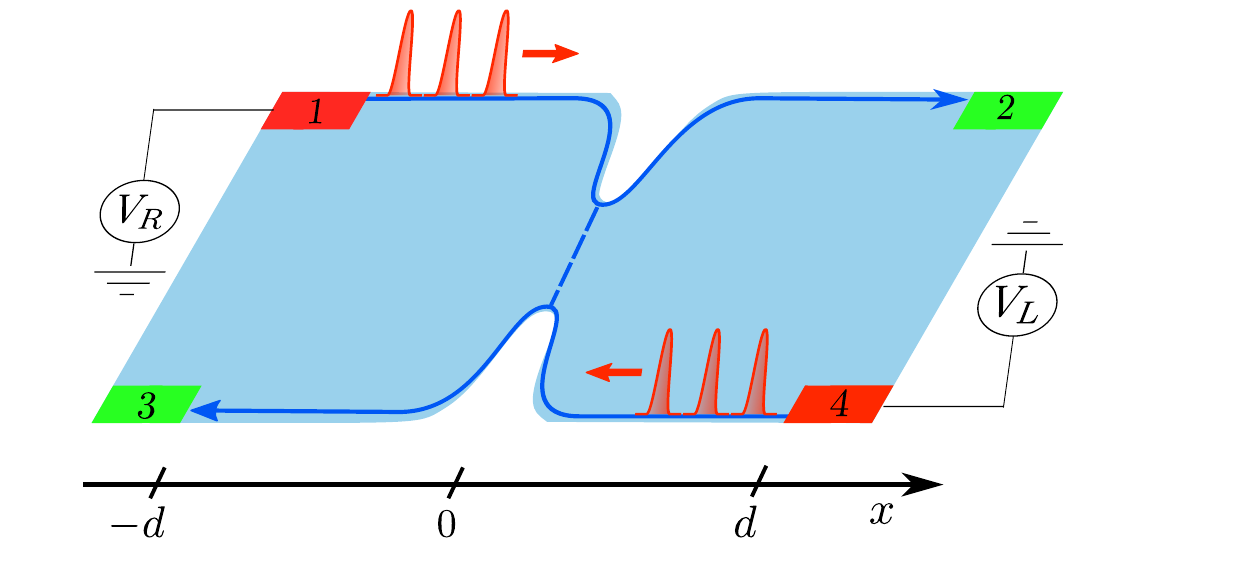}
\caption{Four-terminal setup for EQO experiments. Contact 1 and 4 are used as input terminals, while contact 2 and 3 are the output terminals where current and noise are measured.}
\label{fig:interf_setup}
\end{figure}

Here, we consider a quantum Hall bar at filling factor $\nu=1$ in a four-terminal QPC geometry, presented in Fig. \ref{fig:interf_setup}. Two periodic voltages $V_R(t)$ and $V_L(t)$, with period $\mathcal{T}=\frac{2\pi}{\Omega}$, are applied to reservoirs $1$ and $4$, respectively. In this setup, reservoir $1$ and $4$ (resp. $2$ and $3$) plays the role of sources (resp. detectors) for right-movers and left-movers. The derivation of current noise will be carried out for generic periodic voltages $V_{R/L}$. At the end of the calculations, we will focus on specific configurations for the external drive and we will specify the form for $V_{R/L}$.

Tunneling at the QPC is treated within the \textit{Scattering Matrix Theory} \cite{blanter2000,martin2005,nazarov2009,lesovik2011}. This kind of approach provides a phenomenological description of tunneling processes in the presence of a QPC, without resorting to a microscopic model, in terms of a scattering matrix mixing incoming and outgoing fermion states. We remark that this theory is valid only for free electrons and, when dealing with interacting systems, it can no longer be applied. Fermion fields incoming into the edge states from the two reservoirs are given by
\begin{align}
\Psi_{in,R/L}(x,t)\equiv&\Psi_{R/L}(x,t)=e^{i \chi_{R/L}(t\mp\frac{x}{u})}\psi_{R/L}(x,t),\label{eq:instates}
\end{align}
where $\chi_{R/L}(t)=e\int_{-\infty}^{t}\mathrm{d}t\, V_{R/L}(t)$ and $\psi_{R/L}(x,t)$ are fermion fields in the absence of the voltage drive exiting from terminals $1$ and $4$, respectively. Electronic fields outgoing from the QPC are termed $\Psi_{out,R/L}(x,t)$, whether they enter into reservoirs $2$ or $3$.
Thus, one has
\begin{equation}
    \left(\begin{matrix}
        \Psi_{out,L}\\\Psi_{out,R}
    \end{matrix}\right)=\left(\begin{matrix}
        \sqrt{T} \hspace{4mm} i \sqrt{R}\\  -i \sqrt{R}\hspace{4mm}\sqrt{T}
    \end{matrix}\right)\left(\begin{matrix}
    \Psi_{in,L}\\\Psi_{in,R}
    \end{matrix}\right)\,,\label{eq:scattering_matrix_ele}
\end{equation}
where $T$ is the transmission probability for incoming electrons and $R$ the reflection probability.
The zero-frequency current noise is defined as ($\alpha$ and $\beta $ can assume the value $2$ or $3$)
\begin{equation}\label{eq:noise_eqo}
    \mathcal{S}_{\alpha \beta}=\int_{-\frac{\mathcal{T}}{2}}^{\frac{\mathcal{T}}{2}}\mathrm{d}t\, \int_{-\infty}^{+\infty} \mathrm{d}t'\, \langle \delta J_{\alpha}(x_{\alpha},t)\delta J_{\beta}(x_{\beta},t')\rangle\,,
\end{equation}
where $\delta J_{\alpha}(x_{\alpha},t)\equiv J_{\alpha}(x_{\alpha},t) - \langle J_{\alpha}(x_{\alpha},t)\rangle $ is the variation of the chiral current operator incoming into reservoir $\alpha$ at position $x_{\alpha}$. The current operator can be expressed as a balance of fermionic operators entering or exiting reservoirs as
\begin{align}
\label{current operator}
J_{2}(x,t)=-e v \Psi^{\dagger}_{out,R}(x,t)\Psi_{out,R}(x,t),\\
J_{3}(x,t)=-e v =-e v \Psi^{\dagger}_{out,L}(x,t)\Psi_{out,L}(x,t).
\end{align}

In the following, we focus on the cross-correlator of reservoirs $2$ and $3$, which we indicate by $\mathcal{S}$. Below, we present the results for the current noise in two relevant configurations of EQO, namely Hanburry Brown and Twiss (HBT) and the Hong-Ou-Madel experiment (HOM).

%%%%%%%%%%%%%%%%%%%%%%%%%%%%%%%%%%%%%%%%%%%%%%%%%%%%%%%%%%%%%%%%%%%%%%%%

\subsubsection{Hanbury Brown and Twiss setup}

In the HBT setup, a single voltage drive is turned on
\begin{align}
V_R(t)=V(t), \hspace{4mm} V_L(t)=0,
\label{HBT_conf}
\end{align}
where $V(t)=V_{dc}+V_{ac}(t)$, with $V_{ac}$ a generic function with period $\mathcal{T}$ and satisfying $\int_{-\frac{\mathcal{T}}{2}}^{\frac{\mathcal{T}}{2}}\frac{\mathrm{d}t}{\mathcal{T}} V_{ac}(t)=0$.
We consider a periodic voltage source as it is relevant for the experimental case where data acquisition is achieved over long times. 

In order to conveniently deal with periodic voltage phases, we introduce the following Fourier decomposition \cite{dubois2013}
\begin{equation}
e^{i\chi_{ac}(t)}=\sum_{l=-\infty}^{+\infty}p_{l}e^{-i l \Omega t},
\end{equation}
where $\chi_{ac}(t)=e \int_{0}^{t}\mathrm{d}t'\, V_{ac}(t')$ is a function with period $\mathcal{T}$. Here, we have introduced the Fourier coefficients
\begin{align}
p_{l}&=\int_{-\frac{\mathcal{T}}{2}}^{\frac{\mathcal{T}}{2}} \mathrm{d}t\, e^{i \chi_{ac} (t)}e^{i l \Omega t}=\int_{-\frac{\mathcal{T}}{2}}^{\frac{\mathcal{T}}{2}} \mathrm{d}t\, e^{i \chi (t)}e^{i (l+q) \Omega t}, \label{eq:p_period}
\end{align}
where we defined $q$, the number of particle injected by $V(t)$ in a period as
\begin{equation}
q=-\frac{e}{h}\int_{-\frac{\mathcal{T}}{2}}^{\frac{\mathcal{T}}{2}} \mathrm{d}t\, V(t).
\end{equation}
Eq. \eqref{eq:p_period} is the probability amplitude for particles to absorb or emit an energy $l  \Omega$. The phase $e^{i\chi_{ac}(t)}$ and the coefficients $p_l$ are the analog in the periodic case of the quantity $e^{i\chi(t)}$ and $p(\epsilon)$ appearing in Eq.~\eqref{eq:p_epsilon}. This discretization of energy shifts can be interpreted in terms of emission or absorption of finite quanta of energy corresponding to photons of the electromagnetic field (typically microwaves) generated by $V_{ac}$. These energy transfers due to an AC drive are called \textit{photo-assisted processes} and the $p_l$ in Eq.~\eqref{eq:p_period} are called \textit{photo-assisted coefficients} \cite{dubois2013,glattli2016a}.

In terms of the photo-assisted coefficients, the zero-frequency noise at finite temperature $\theta$ due to a QPC with transmission $T$ in the HBT configuration is 
\begin{equation}
    \label{eq:noise_pasn_temp}
    \begin{aligned}
        \mathcal{S}^{HBT}=-\frac{e^2}{2\pi}RT\sum_{l}|p_{l}|^2 (l+q)\Omega &\coth\left[\frac{(q+l)\Omega}{2 \theta}\right].
    \end{aligned}
\end{equation}
This quantity is called photo-assisted shot-noise and carries information about the properties of the driving voltage due to the presence of the coefficients $p_l$. In the absence of the voltage drive, one has $p_l = \delta_{l,0}$ and Eq.~\eqref{eq:noise_pasn_temp} gives the equilibrium thermal noise
\begin{equation}
\mathcal{S}_{th} = -RT \frac{e^2}{\pi} \frac{\theta}{\Omega}.
\end{equation}

At zero temperature, the noise in Eq. \eqref{eq:noise_pasn_temp} becomes
\begin{equation}
\label{eq:noise_pasn_zero}
    \mathcal{S}^{HBT}=-S_{0}\sum_{l}|p_{l}|^2 |l+q|,
\end{equation}
where we introduced $S_0=\frac{e^2}{\mathcal{T}}RT$.

The noise in the HBT geometry is typically used to verify the single-electron nature of Levitons~\cite{dubois2013b}. 
This can be understood by computing the number of electrons and holes generated by the drive.
Following~\cite{keeling2006}, let us count the number of electrons generated above the Fermi level (that we set to $\mu=0$) or holes below it on a single right-moving edge channel. At $\theta=0$, these quantities are defined as
\begin{equation}
\label{eq:num_eh}
\begin{aligned}
    N_{e}&=\int_{-\infty}^{+\infty}\frac{d \epsilon}{2\pi} \Theta(\epsilon) \left \langle \tilde{a}^{\dagger}_{in,R}(\epsilon)\tilde{a}_{in,R}(\epsilon)\right \rangle,\\
    N_h&=\int_{-\infty}^{+\infty}\frac{d \epsilon}{2\pi} \Theta(-\epsilon) \left \langle \tilde{a}_{in,R}(\epsilon)\tilde{a}^{\dagger}_{in,R}(\epsilon)\right \rangle,
\end{aligned}
\end{equation}
where $\Theta(E)$ is the Heavyside distribution and we introduced the operator:
\begin{equation}
    \tilde{a}_{R}(\epsilon)=\frac{1}{\sqrt{2\pi v}}\int_{-\infty}^{+\infty}\mathrm{d}t\,~ e^{i \epsilon\left(t-\frac{x}{v}\right)} \Psi_{in,R}(x,t),
\end{equation}
Substituting the definition of operators $\tilde{a}_R(\epsilon)$ into Eq.~\eqref{eq:num_eh}, one finds
\begin{equation}
    \label{eq:NehInt}
    \begin{aligned}
        N_{e}&=\frac{1}{\left(4\pi v\right)^2}\sum_{l>-q}\left|p_l\right|^2 \left|l+q\right|\Omega\\
        N_{h}&=\frac{1}{\left(4\pi v\right)^2}\sum_{l<-q}\left|p_l\right|^2 \left|l+q\right|\Omega\,.
    \end{aligned}
\end{equation}
Therefore, the number of electrons plus the number of holes created by the drive is proportional to the total noise~\eqref{eq:noise_pasn_zero}.
As will be clear below, it is interesting to define the excess noise:
\begin{equation}
    \Delta \mathcal{S}=\mathcal{S}^{HBT}-\mathcal{S}_{dc},
    \label{eq:exc_IQH}
\end{equation}
where $\mathcal{S}_{dc}$ is the noise due solely to $V_{dc}$ (equivalent to setting $p_l=\delta_{l,0}$).
Indeed one can show, from Eq.~\eqref{eq:noise_pasn_zero}, that it reads
\begin{equation}
    \label{eq:noise_excIQHzero}
    \Delta\mathcal{S}=-S_{0}\sum_{l<-q}|p_{l}|^2 |l+q|\,,
\end{equation}
i.e. it is proportional to the number of holes created by the drive.
As a result, at zero temperature, $\Delta \mathcal{S}$ vanishes when the drive generates electron-like excitations only.

\begin{figure}
\centering
\includegraphics[width=\linewidth]{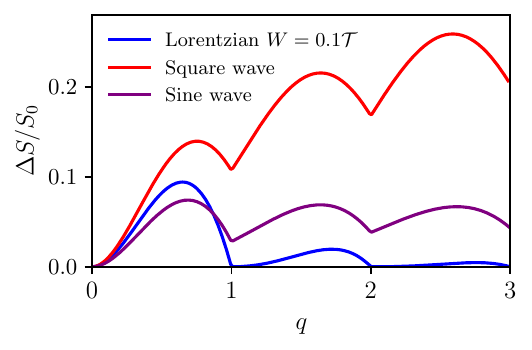}
\caption{Theoretical expectation for the excess-noise $\Delta \mathcal{S}$ as a function of $q$ for a cosine drive (purple line), a square drive (red line) and a Lorentzian drive (blue line), in units of $S_0$, at zero temperature. The width of Lorentzian pulses is $W=0.1 \mathcal{T}$.}
\label{fig:exc_zero}
\end{figure}

Below, we give the theoretical results for the excess noise for some experimentally relevant voltage drives. In particular, we choose a sinusoidal (abbr. sin) drive, a square (abbr. sqr) drive and a Lorentzian (abbr. lor) drive, given respectively by
\begin{align}
V_{sin}(t)&=V_{dc}(1-\cos \left(\Omega t\right)),\label{eq:sine_drive}\\
V_{sqr}(t)&=2V_{dc}\sum_{k=-\infty}^{\infty}\Theta\left(t-k\mathcal{T}\right)\Theta\left(\frac{\mathcal{T}}{2}-t+k \mathcal{T}\right),\label{eq/square_drive}\\
V_{lor}(t)&=\frac{V_{dc}}{\pi}\sum_{k=-\infty}^{+\infty}\frac{W}{W^2+\left(t-k\mathcal{T}\right)^2}.\label{eq:lorentzian_drive}
\end{align}
The corresponding form for the $p_{l}$ coefficients can be calculated and the excess noise associated to each particular drive is reported in Fig. \ref{fig:exc_zero} for the case of zero temperature~\cite{rech2017}.
A common feature to all voltages is the appearance of minima in correspondence of integer values of $q=-\frac{e V_{dc}}{\Omega}$, showing that an integer amount of charge always minimize the generation of electron-hole pairs. More interestingly, only the excess noise of the Lorentzian voltage pulses exactly vanishes at zero temperature: the excess noise of both the sinusoidal and the square drive stays well above zero even for integer values of $q$.\\

%%%%%%%%%%%%%%%%%%%%%%%%%%%%%%%%%%%%%%%%%%%%%%%%%%%%%%%%%%%%%%%%%%%%%%%%
\subsubsection{Hong-Ou-Mandel setup}
\begin{figure}
    \centering
    \includegraphics{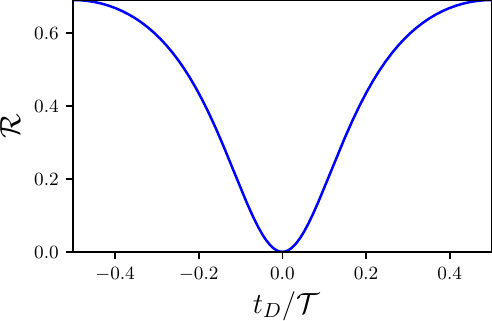}
    \caption{Normalized HOM ratio $\mathcal{R}$ at $q=1$, as a function of the time delay $t_D$ between the two sources. The width of the Lorentzian is equal to $W=0.1\mathcal{T}$. The dip at $t_D=0$ is a consequence of the Pauli principle obeyed by fermions}
    \label{fig:HOMratio}
\end{figure}

In the HOM setup
\begin{align}
V_R(t)=V_{lor}(t),\hspace{4mm}V_L(t)=V_{lor}(t+t_D),
\end{align}
where $t_D$ is the time shift between the two pulses and the parameters of the Lorentzian drive are set so that $q$ is an integer number [see Eq.~\eqref{eq:lorentzian_drive}]. The current noise computed in the HOM interferometer carries information about statistical properties of electrons. In particular, it can be used to probe the fermion anti-bunching property of electrons. For simplicity, we consider the limit of zero temperature, where the quantum interference effects dominate over the thermal fluctuations. The current noise reads
\begin{equation}
\label{eq:noise_pasn_zeroHOM}
\mathcal{S}^{HOM}=-S_{0}\sum_{l}|\tilde{p}_{l}|^2 |l|,
\end{equation}
where we defined the HOM photo-assisted coefficients
\begin{align}
\tilde{p}_{l}&=\int_{-\frac{\mathcal{T}}{2}}^{\frac{\mathcal{T}}{2}} \mathrm{d}t\, e^{i \chi_{ac} (t)}e^{-i \chi_{ac} (t+t_D)}e^{i l \Omega t}. \label{eq:p_periodHOM}
\end{align}
It is instructive to recast the expression for noise in the case of $q=1$ in terms of the single-Leviton wave-function in Eq. \eqref{eq:wave-levitond_period}. In this way, the HOM current noise becomes
\begin{equation}
\mathcal{S}^{HOM}=2\mathcal{S}^{HBT}-\mathcal{I}(t_D),
\end{equation}
where $\mathcal{S}^{HBT}$ is the zero temperature limit of the HBT noise given in Eq. \eqref{eq:noise_pasn_zero} and
\begin{equation}
\label{eq:HOM_noise_scatt_2}
\mathcal{I}(t_D)=-2S_0 \left|\int_{-\frac{\mathcal{T}}{2}}^{\frac{\mathcal{T}}{2}}\frac{\mathrm{d}t}{\mathcal{T}}\varphi_{1}(t)\varphi^{*}_{1}(t+t_D)\right|,
\end{equation}
where the wave-function $\varphi_1(t)$ is related to the one introduced in Eq.~\eqref{eq:wave-levitond_period} as $\varphi_1(t)\equiv \varphi_1(-vt)$. This last formula shows that the HOM current noise is directly related to the overlap between the two wave-functions of Levitons impinging at the QPC. In order to deal with dimensionless quantity, it is common to introduce the following ratio
\begin{equation}
\mathcal{R}(t_D)=\frac{\mathcal{S}^{HOM}}{2 \mathcal{S}^{HBT}},
\end{equation}
and study the HOM noise normalized with respect to the HBT noise. From the vanishing of excess noise $\Delta \mathcal{S}$ for Levitons at zero temperature (see the discussion about HBT setup), we know that the HBT contribution becomes~\cite{jonckheere2012}
\begin{equation}
\label{eq:HBT_charge}
\mathcal{S}^{HBT}=\mathcal{S}_{dc}=S_0 q=S_0,
\end{equation}
where in the last step we put $q=1$. By using these results the HOM ratio of two single-Leviton states colliding at the QPC becomes
\begin{equation}
\mathcal{R}(t_D)=1-\left|\int_{-\frac{\mathcal{T}}{2}}^{\frac{\mathcal{T}}{2}}\frac{\mathrm{d}t}{\mathcal{T}}\varphi_{1}(t)\varphi^{*}_{1}(t+t_D)\right|^2.\label{eq:ratio_phi1}
\end{equation}
Let us notice that for $t_D=0$, the overlap integral of the single-Leviton wave-function reduces to the norm of the wave function $\varphi_{1}(t)$, which is therefore equal to $1$. In this case, the HOM ratio vanishes, in accordance with the expected anti-bunching effect of fermions. An analytical form can be provided for this ratio at $q=1$ \cite{rech2017}
\begin{equation}
\mathcal{R}(t_D)=\frac{\sin^2\left(\pi\frac{t_D}{\mathcal{T}}\right)}{\sin^2\left(\pi\frac{t_D}{\mathcal{T}}\right)+\sinh^2\left(2\pi\frac{W}{\mathcal{T}}\right)}.\label{eq:ratio_analytical}
\end{equation}
This HOM ratio is plotted in Fig. \ref{fig:HOMratio}. Clearly, the interference effects that lead to the total suppression of noise at $t_D=0$ are reduced for greater time delays, when the distinguishability of the two Levitons is increased. 

%In this case, the ratio is close to $1$, meaning that the HOM current noise is similar to the sum of the partitioning HBT noise of two uncorrelated Levitons. Moreover, the shape of this HOM ratio is Lorentzian and it is plotted for two different values of the width $W=0.09$ (black line) and $W=0.18$ (red line): by comparing these two curves, one can argue that the HOM ratio is extremely sensitive to the shape of the overlapping wave-packets. \\ 

It is interesting to point out a peculiar property of the ratio for $q=1$ at finite temperature. When thermal effects are relevant, one has to provide a more general definition of the HOM ratio, where equilibrium thermal noise $\mathcal{S}_{th}$ has to be subtracted from signals. Therefore, for a finite temperature the HOM ratio is defined as
\begin{equation}
\mathcal{R}(t_D)=\frac{\mathcal{S}^{HOM}-\mathcal{S}_{th}}{2\mathcal{S}^{HBT}-2\mathcal{S}_{th}}.
\end{equation}
When a single Leviton is emitted by each source, this ratio has exactly the same expression as the zero temperature limit given in Eq. \eqref{eq:ratio_analytical}.
%%%%%%%%%%%%%%%%%%%%%%%%%%%%%%%%%%%%%%%%%%%%%%%%%%%%%%%%%%%%%%%%%%%%%%%%%%%%%%%%%%%%%%%%%% 

\section{Photo-Assisted Shot Noise formalism in the FQHE \label{chap_charge}}

 The fractional quantum Hall effect (FQHE)\cite{tsui1982,laughlin1983} represents a seminal example of strongly-correlated state where the interaction between electrons cannot be neglected and gives rise to a new topological phase of matter. In the Laughlin sequence of the FQHE a single chiral channel exists at the boundary of the system and the excitations are exotic quasi-particles with fractional charge and statistics called anyons~\cite{nayak2008,hashisaka2021,jonckheere2023,glidic2023}. The propagation of Levitons in these exotic states of matter is currently under investigation~\cite{ronetti2018}. Here, we present the photo-assisted shot-noise (PASN) formalism that is conveniently employed to deal with time-dependent voltage drives in the FQH regime in the presence of a QPC~\cite{crepieux2004}.

\subsection{Model and Hamiltonian \label{sec:charge_FQHE}}
 In Sec. \ref{noninteracting}, tunneling at the QPC in the IQH regime has been treated within the scattering matrix approach. This theoretical description is not valid in the FQH phase because the presence of correlation does not allow to introduce fermion scattering states as input and output states. Indeed, the excitations propagating along the FQH edge states are bosons~\cite{wen1995}, while the ones who tunnel are anyons~\cite{kane1992}, thus rendering tunneling at the QPC a highly non-linear problem. The most convenient method to compute transport properties in this configuration is a perturbative calculation based on a Hamiltonian approach. 

We consider a four-terminal FQH bar in the presence of a QPC, as shown in Fig. \ref{fig:interf_setup}. For a quantum Hall system with filling factor $\nu$ in the Laughlin sequence $\nu=1/(2n+1)$ \cite{laughlin1983}, with $n \in \mathbb N$, a single chiral mode emerges at each edge of the sample. The effective bosonic Hamiltonian for edge states reads \cite{wen1995}
\begin{equation}
	\label{eq:H_wg}
	H_{ 0} = \sum_{r=R,L} \frac{v}{4\pi} \int_{-\infty}^{+\infty} \mathrm{d}x\, \left[ \partial_x \Phi_r(x) \right]^2\, ,
\end{equation} 
where $\Phi(x)$ are the bosonic excitations.
Due to the presence of the voltage drives applied to reservoirs $1$ and $4$, the corresponding boson modes experience a shift:
\begin{equation}
\label{eq:eq_motion}
	\Phi_{R/L}(x,t) = \phi_{R/L}\left(t \mp \frac{x}{v}\right) - e \sqrt \nu \int_{-\infty}^{t \mp \frac{x}{v}} \mathrm{d}t'\, V_{R/L}(t').
\end{equation}

Tunneling between the two edges occurs through a QPC at $x=0$. The tunneling Hamiltonian describing weak backscattering at the QPC reads as: \cite{kane1994,vondelft1998}
\begin{equation}
    H_T^{(qp)}=\Lambda \Psi_R{}^{\dagger}(0)\Psi_L(0)+\text{H.c.},\label{ham_tunn_qp}
\end{equation}
where $\Lambda$ is the tunneling amplitude. Here, we introduced annihilation and creation fields for Laughlin quasi-particles carrying fractional charge $-\nu e$ which are defined through the standard procedure of bosonization \cite{wen1995}. They read 
\begin{equation}
\Psi_{R/L}(x\mp vt) = \frac{\mathcal{F}_{R/L}}{\sqrt{2\pi a}} e^{-i \sqrt \nu \Phi_{R/L}(x,t)},
\end{equation}
where $a$ is a short-distance cut-off and $\mathcal{F}_{R/L}$ are the Klein factors \cite{wen1995,vondelft1998,guyon2002,martin2005}. 

The tunneling Hamiltonian will be treated perturbatively with respect to the Hamiltonian $H_0$. Note that only quasi-particles tunneling contribution has to be considered in the weak backscattering regime, since this is the most relevant process in the renormalization group sense \cite{kane1992}.
\subsection{Transport properties in the QPC geometry}
We are interested in computing charge current and noise in EQO-like configurations for fractional filling factors in the Laughlin sequence. To this end, we introduce the charge current operator for FQH edge states and we present the perturbative approach to compute the average current and noise to lowest order in $\Lambda$.

Charge current operators for right- and left-moving modes can be defined by resorting to the continuity equation of densities $\rho_{R/L}(x,t)$, namely 
\begin{equation}
\partial_t\rho_{R/L}(x,t)+\partial_x J_{R/L}(x,t)=0.
\end{equation}
According to chirality of edge states, one finds
\begin{equation}
\label{eq:charge_current_op}
J_{R/L}(x,t)=\mp ev \rho_{R/L}(x,t),
\end{equation}
where $\rho_{R/L}(x,t)=(2\pi)^{-1}\left(\partial_x\Phi_{R/L}(x,t)\right)$ are density operators evolving with respect to the whole Hamiltonian.
Starting from the definition of the chiral current operator, we can define the operators for the charge current entering reservoir $2$ and $3$ as
\begin{equation}
J_{2/3}(t)=J_{R/L}(\pm d,t),
\label{defcurr}
\end{equation}
where the interfaces between edge states and contacts $2$ and $3$ are placed in $x=\pm d$ respectively. 

Below, we compute the average charge current and the current noise using the Kelysh formulation of non-equilibrium statistical physics  to lowest order in the tunneling at the QPC.

\subsubsection{Average charge current}

In the absence of tunneling processes ($\Lambda=0$), the zeroth-order contributions corresponds to the current generated by the two voltages along the two separated edge channels with opposite chirality. The total charge $\mathcal{C}_{R/L}$ emitted by $V_{R/L}$ into the edge states thus reads
\begin{equation}
    \label{eq:charge}
    \mathcal{C}_{R/L}=\int_{-\frac{\mathcal{T}}{2}}^{\frac{\mathcal{T}}{2}}\mathrm{d}t \langle J_{2/3}(t)\rangle^{(0)}=\frac{e^2 \nu}{\Omega } V_{dc}^{(R/L)}=-e q_{R/L},
\end{equation}
where $\langle J_{2/3}(t)\rangle^{(0)}$ is the current in the absence of the QPC,
$V_{dc}^{(R/L)}=\int_{-\frac{\mathcal{T}}{2}}^{\frac{\mathcal{T}}{2}}\frac{\mathrm{d}t}{\mathcal{T}}V_{R/L}\left(t\right)$
and $q_{R/L}=-\frac{e \nu V^{(R/L)}_{dc}}{\Omega}$ are the numbers of electronic charges injected by $V_{R/L}$.

Due to the QPC, some of the particles emitted into the edge states are backscattered. Since terms involving a different number of annihilation or creation field operators with a defined chirality have a vanishing average value, $\langle J_{2/3}^{(1)}(x,t)\rangle = 0$ and the only remaining contribution is due to  $\langle J_{2/3}^{(2)}(t)\rangle$. This term can be physically identified as the current due to the reflection of particles incoming from reservoirs $1$ and $4$, respectively. These currents are usually termed backscattering currents \cite{kane1994,kane1996} and are equal up to a sign.
Thus we can define
\begin{equation}
J_B(t)=\langle J_{3}(t)\rangle^{(2)} =-\langle J_{2}(t)\rangle^{(2)},
\end{equation}
where $\langle J_{2/3}(t)\rangle^{(2)}$ is the charge current to second order in the tunneling amplitude. 
The backscattering current can be evaluated by computing the average on the unperturbed Hamiltonian. One finds
\begin{equation}
\begin{aligned}
    J_B(t)=2i\nu e  \left|\lambda\right|^2\int_{0}^{+\infty} d\tau &\sin\left[\nu e \int_{t-\tau}^{t}\mathrm{d}t'' V_{-}(t'')\right]\\
    &\times\sum_{\epsilon=+,-}\epsilon e^{2\nu\mathcal{G}(\epsilon\tau)}\,,
\end{aligned}
\end{equation}
where we defined $\lambda\equiv \Lambda / (2\pi a)$ and the voltage difference $V_{-}(t)=V_{R}(t)-V_{L}(t)$ and 
\begin{align}
\mathcal{G}(t-t')&\equiv\left \langle\phi_{R/L}(0,t)\phi_{R/L}(0,t')-\phi^2_{R/L}(0,t)\right\rangle=\nonumber\\&= \ln \left\{\frac{\pi (t-t') \theta}{\sinh \left[\pi (t-t') \theta\right]\left[1+i \omega_c (t-t')\right]}\right\},\label{eq:W}
\end{align}
is the boson correlation function for the unperturbed edge at temperature $\theta$. The above expression is valid as long as all the energy scales are much smaller than the high-energy cut-off $\omega_c =u/a$.

Since $V_R$ and $V_L$  are time-dependent voltages with period $\mathcal{T}=\frac{2\pi}{\Omega}$, we expect the current to satisfy $J_B(t)=J_B(t+\mathcal{T})$. 
It is thus relevant to consider the average over one period of the backscattering current:
\begin{equation}\label{eq:back_curr}
\begin{aligned}
\overline{J_B(t)}=2i \nu e &\left|\lambda\right|^2 \sum_{l}\left|p_{l}\right|^2\\
&\times\int_{-\infty}^{+\infty} \mathrm{d}\tau\, \sin \left[\left(q_R-q_L+l\right)\Omega \tau\right]e^{2\nu\mathcal{G}(\tau)}.
\end{aligned}
\end{equation}
Similarly to what happens in the IQH, the average backscattered current is directly related to the properties of the voltage drive through the coefficient $\tilde{p}_l$ and to the injected charges $q_R$ and $q_L$.

The integral in Eq.~\eqref{eq:back_curr} can be computed to leading order in $1/\omega_c$ and reads\cite{rech2017}:
\begin{widetext}
\begin{equation}
\overline{J_B(t)}=\frac{2\nu e  \left|\lambda\right|^2}{\Gamma\left(2\nu\right)} \left(\frac{2\pi \theta}{\Omega}\right)^{2\nu-1}\sum_{l}\left|p_{l}\right|^2 \sinh\left[\frac{(q_R-q_L+l)\Omega}{2\theta}\right] \left|\Gamma \left[\nu-i\frac{(q_R-q_L+l)\Omega}{2\pi\theta}\right]\right|^2,
\label{Jbsfin}
\end{equation}
\end{widetext}
where $\Gamma(x)$ is the Euler Gamma function~\cite{zwillinger2007}.

\subsubsection{Current noise}

We now turn our attention to the period averaged zero-frequency current noise, which is defined as
\begin{equation}
\mathcal{S}_{\alpha \beta}=\int_{0}^{\mathcal{T}}\frac{\mathrm{d}t}{\mathcal{T}}\int_{-\infty}^{+\infty} \mathrm{d}t'\, \left \langle \delta J_{\alpha}(t) \delta J_{\beta}(t')\right \rangle,
\end{equation}
where $\alpha$ and $\beta$ refer to reservoirs $2$ or $3$. The time correlator $\langle \delta J_{\alpha}(t) \delta J_{\beta}(t')\rangle$ depends independently on two times $t$ and $t'$ and is periodic in both. 

We focus on the auto-correlator of reservoirs $2$, namely $\mathcal{S}_{22}$, and we use the shorthand notation $\mathcal{S}_C\equiv \mathcal{S}_{22}$. 
Expanding $\mathcal{S}_C$ to leading order in $\Lambda$ we find
\begin{equation}\label{eq:noise2}
\begin{aligned}
\mathcal{S}_{C}=2(\nu e)^2 \left|\lambda\right|^2&\int_{0}^{\mathcal{T}}\frac{\mathrm{d}t}{\mathcal{T}}\int_{-\infty}^{+\infty} \mathrm{d}\tau\,\\
&\times\cos \left[\nu e \int_{t+\tau}^{t} V_{-}(t'') \mathrm{d}t''\right]e^{2\nu\mathcal{G}(\tau)}.
\end{aligned}
\end{equation}
Even though the noise is generated in a double-drive configuration, it is interesting to point out that it actually depends only on the single effective drive $V_{-}(t)$. 
Therefore, the same behavior can be obtained in a single-drive configuration, where reservoir 4 is grounded ($V_L(t)=0$) and reservoir 1 is contacted to the periodic voltage $V_{-}(t)$.
%\begin{equation}
%\label{rel:double-single}
%\mathcal{S}_{C}\left(V_R,V_L\right)=\mathcal{S}_{C}\left(V_{-},0\right).
%\end{equation}
%Here, the arguments in brackets indicate the voltage applied to reservoirs $1$ and $4$, respectively. 
%One might consider Eq. \eqref{rel:double-single} as a consequence of a trivial shift of both voltages by a value corresponding to $V_L$. 
Even though one might consider this as merely a shift in voltage, such a result cannot be obtained by means of a gauge transformation. In this sense, the charge noise \textit{incidentally} acquires the same expression in these two physically distinct experimental setups. As will be clearer in the following, for the charge case this is a consequence of the presence of a single local (energy independent) QPC. Generally, we expect that the double-drive and the single-drive ($V_R(t)=V_{-}(t)$ and $V_{L}(t)=0$) configurations return different outcomes for other physical observables, such as heat noise.\cite{vannucci2018}

As was done for the average current the noise can be written in terms of the coefficient $\tilde{p}_l$ as:
\begin{equation}
    \begin{aligned}
        \mathcal{S}_{C} = 2(\nu e)^2 &\left|\lambda\right|^2 \sum_{l} \left|\tilde{p}_l\right|^2 \int_{-\infty}^{+\infty} \mathrm{d}\tau\\
        &\times\cos\left[(q_R-q_L+l)\Omega \tau \right] e^{2\nu\mathcal{G}(\tau)}.
    \end{aligned}
\end{equation}
Performing the integral yields\cite{rech2017}
\begin{widetext}
\begin{equation}
\mathcal{S}_{C}=\frac{4(\nu e)^2  \left|\lambda\right|^2}{\Gamma\left(2\nu\right)} \left(\frac{2\pi \theta}{\Omega}\right)^{2\nu-1}\sum_{l}\left|\tilde{p}_{l}\right|^2 \cosh\left[\frac{(q_R-q_L+l)\Omega}{2\theta}\right] \left|\Gamma \left[\nu-i\frac{(q_R-q_L+l)\Omega}{2\pi\theta}\right]\right|^2.\label{eq:noise_FQHE_fin}
\end{equation}
\end{widetext}

\subsubsection{DC case and zero temperature limits}

The expressions for charge current and noise that we have just derived are valid for a generic temperature $\theta\ll\omega_c$ and for all kinds of periodic voltage drives. 
Before moving on with our discussion, it is useful to discuss two limiting cases. 
The first one is the zero-temperature limit, where the temperature is assumed to be the lowest energy scale in the problem. 
In the second case, only a DC drive $V_{dc}$ is applied to reservoir $1$ and reservoir $4$ is grounded.

\paragraph{Zero temperature limit: } In this case, Eqs.~\eqref{Jbsfin} and \eqref{eq:noise_FQHE_fin} become
\begin{widetext}
\begin{align}
\overline{J_B(t)}\Big|_{\theta=0}&=\frac{\nu e}{\Omega}\left|\lambda\right|^2\frac{2\pi }{\Gamma(2\nu)}\left(\frac{\Omega}{\omega_c}\right)^{2\nu}\sum_{l}\left|p_l\right|^2 \left|q_R-q_L+l\right|^{2\nu-1}\text{sign}(q_R-q_L+l),\label{eq:curr_zerot}\\
\mathcal{S}_{C}\Big|_{\theta=0}&=\frac{(\nu e)^2}{\Omega}\left|\lambda\right|^2\frac{4\pi }{\Gamma(2\nu)}\left(\frac{\Omega}{\omega_c}\right)^{2\nu}\sum_{l}\left|p_l\right|^2 \left|q_R-q_L+l\right|^{2\nu-1}\label{eq:noise_zerot}.
\end{align}
\end{widetext}
In the fractional case, each contribution to the sum follows a power law in $q_R-q_L+l$ with exponent $2\nu-1$.
Such a (negative) power law behavior in the tunneling properties at zero-temperature is typical of Luttinger liquids\cite{miranda2003,giamarchi2003,kane1996,kane1997} (in particular of chiral Luttinger liquid in ther weak backscattering regime where the exponent is negative).

\paragraph{DC case:} Here $V_R(t)=V_{dc}$ and $V_L(t)=0$. This particular configuration entails that the photo-assisted coefficients reduce to $\tilde{p}_l=\delta_{l,0}$. Thus, the current and the noise become
\begin{widetext}
\begin{align}
J_B&=\frac{2\nu e}{\omega_c}\left|\lambda\right|^2\frac{1 }{\Gamma(2\nu)}\left(\frac{2\pi \theta}{\omega_c}\right)^{2\nu-1}\left|\Gamma\left(\nu - i \frac{\nu e V_{dc}}{2\pi \theta}\right)\right|^2\sinh\left(\frac{\nu e V_{dc}}{2 \theta}\right),\label{eq:curr_dc}\\
\mathcal{S}_{C}&=\frac{2(\nu e)^2}{\omega_c}\left|\lambda\right|^2\frac{1 }{\Gamma(2\nu)}\left(\frac{2\pi \theta}{\omega_c}\right)^{2\nu-1}\left|\Gamma\left(\nu - i \frac{\nu e V_{dc}}{2\pi \theta}\right)\right|^2\cosh\left(\frac{\nu e V_{dc}}{2 \theta}\right),\label{eq:noise_dc}
\end{align}
\end{widetext}
where $q \Omega=\nu e V_{dc}$ and $J_B$ is the average current (it is time independent as there is no AC voltage). 
These expressions are valid both at zero and finite temperature $\theta$. It is instructive to discuss the limit $\theta\rightarrow 0$ of Eqs. \eqref{eq:curr_dc} and \eqref{eq:noise_dc}, which read
\begin{align}
J_B\Big|_{\theta=0}&=\frac{\nu e}{\omega_c}\left|\lambda\right|^2\frac{2\pi }{\Gamma(2\nu)}\left|\frac{\nu e V_{dc}}{\omega_c}\right|^{2\nu-1} \text{sign}(V_{dc}),\label{eq:J_dc}\\
\mathcal{S}_{C}\Big|_{\theta=0}&=\frac{(\nu e)^2}{\omega_c}\left|\lambda\right|^2\frac{4\pi }{\Gamma(2\nu)}\left|\frac{\nu e V_{dc}}{\omega_c}\right|^{2\nu-1}.\label{eq:S_dc}
\end{align}
In particular, in the zero-temperature limit, one has \footnote{We assume $V_{dc}>0$.}
\begin{equation}
\mathcal{S}_{C}= \nu e \overline{J_B(t)},
\label{eq:Schottky}
\end{equation}
which is the well-known Schottky result for noise in the weak-backscattering regime~\cite{schottky1918}. Interestingly, noise and backscattering current are directly proportional in this regime and the constant of proportionality is exactly the charge of tunneling quasi-particles. This result was confirmed in two pioneering experiments.\cite{saminadayar1997, depicciotto1997}
To lowest order in tunneling and at zero temperature, the charge current is driven by a Poisson process.
This corresponds to the already known result in the non-interacting case at $\nu=1$, where the tunneling charge is $e$.

%%%%%%%%%%%%%%%%%%%%%%%%%%%%%%%%%%%%%%%%%%%%%%%%%%%%%%%%%%%%%%%%%%%%%%%%%%%%%%%%%%%%%%%%%%%%%%%%%%%%%%%%

\section{Levitons in the FQHE}
\label{levitons_fqhe}

In the first part of this section, we employ the results for the charge current and noise previously derived in order to investigate under which conditions Levitons are minimal excitation states of the FQHE. 
Specifically, for both integer and fractional filling factors of the Laughlin sequence, one can probe the electron-like nature of Levitons by means of an excess noise.
The latter vanishes for integer value of $q$ in the case of a Lorentizan-shaped voltage. In this way, it will be clear that Levitons are robust even to the presence of electron-electron interaction. 

In the second part of this section we focus on a theoretical proposal to reveal the interaction between two propagating Levitons in the FQH edge states. 
A lot of attention has indeed been attracted by the idea of exploiting Levitons and other types of single-electron excitations for quantum information and computation purposes, such as schemes based on the concept of electron flying qubits~\cite{mcneil2011,yamamoto2012,dasenbrook2015,dasenbrook2016,dasenbrook2016b,glattli2016b,bauerle2018,edlbauer2022}. In this context, the realization of two-qubit gates crucially relies on the non-trivial entanglement between electrons flying qubits, which can be induced by Coulomb interaction. The choice of focusing on the Laughlin sequence is motivated by the absence of decoherence induced by other propagating channels~\cite{bocquillon2013,wahl2014,marguerite2016}. 

\subsection{Levitons as minimal excitations in the FQH effect}
The charge current and noise previously evaluated for a generic drive can now be employed to test whether quantized Lorentzian pulses or other kind of driving voltages are minimal excitations even in the FQH regime. To this end, the suitable experimental configuration is the HBT setup, as for the integer case in Sec. \ref{noninteracting}. Here, a drive is applied only to reservoir $1$ and reservoir $4$ is grounded, such that $V_R(t)=V(t)$ and $V_L(t)=0$,  with $V(t)$ a generic periodic drive. Notice that, in this case,  the emitted numbers of particles are $q_R=q$ and $q_L=0$ .

In this light, one should find an extension to the concept of excess noise  introduced in the IQH case. The idea to extend the definition of excess noise to the FQH  effect is based on the Schottky result in Eq. \eqref{eq:Schottky}. In general, time-dependent drives do not satisfy that relation in contrast with a DCconstant bias. The combination of transport properties that we use to define the excess noise is given by
\begin{equation}
\Delta \mathcal{S}=\mathcal{S}_{C}- \nu e \overline{J_B(t)},\label{eq:exc_FQH}
\end{equation}
where the noise is measured with respect to a reference value given by the average value of ACcurrent. The explicit formula for the excess noise can be obtained by combining Eqs. \eqref{eq:back_curr} and \eqref{eq:noise2} according to the definition in Eq. \eqref{eq:exc_FQH}
\begin{equation}\label{eq:exc_FQH2}
\begin{aligned}
    \Delta \mathcal{S}=2(\nu e)^2&\left|\lambda\right|^2\int_{-\infty}^{+\infty} d\tau' \int_{0}^{\mathcal{T}} \frac{dt}{\mathcal{T}} \\
    &\times\exp \left[i \nu e \int_{\tau'-\tau}^{\tau'}\mathrm{d}t'\, V(t')\right]e^{2\nu \mathcal{G}(\tau)},
\end{aligned}
\end{equation}
This relation can be used to prove that Levitons are minimal excitation states for any filling factor in the Laughlin sequence\cite{rech2017}. In order to understand the importance of the excess noise at any filling factor, we will compare the fractional case to the integer one discussed in Sec.~\ref{noninteracting}. As a starting point, we recast the number of holes in Eq.~\eqref{eq:num_eh} in terms of $\mathcal{G}(\tau)$:
\begin{equation}\label{eq:nh}
\begin{aligned}
N_{h}=\frac{1}{\left(2\pi a\right)^2}\int_{-\infty}^{+\infty} &\frac{d\tau'}{2\pi} \int_{-\infty}^{+\infty} \frac{d\tau}{2\pi} \\
&\times\exp \left[i e \int_{\tau'+\tau}^{\tau'}\mathrm{d}t'\, V(t')\right]e^{2\mathcal{G}(\tau)},
\end{aligned}
\end{equation}
The latter expression can be easily generalized to the fractional case by replacing $\mathcal{G}(\tau) \rightarrow \nu \mathcal{G}(\tau)$ and $e\rightarrow e^*$, thus obtaining
\begin{equation}\label{Nhfrac}
\begin{aligned}
\mathcal{N}=\frac{1}{\left(2\pi a\right)^2}\int_{-\infty}^{+\infty}& \frac{d\tau'}{2\pi} \int_{-\infty}^{+\infty} \frac{d\tau}{2\pi} \\
&\times\exp \left[i \nu e \int_{\tau'+\tau}^{\tau'}\mathrm{d}t'\, V(t')\right]e^{2\nu\mathcal{G}(\tau)}.
\end{aligned}
\end{equation}
The expression in Eq.~\eqref{Nhfrac} is proportional to the excess noise in Eq. \eqref{eq:exc_FQH2} for a generic $\nu$ belonging to the Laughlin sequence. As a consequence, a voltage drive which emit clean pulses in the FQH regime must fulfill the condition $\Delta \mathcal{S}=0$, where the excess noise has been defined according to Eq. \eqref{eq:exc_FQH}.

At $\theta=0$, one can consider the expressions for current and noise in Eqs. \eqref{eq:curr_zerot} and \eqref{eq:noise_zerot} and write
\begin{equation}
\Delta \mathcal{S}=\frac{(\nu e)^2}{\Omega}\left|\lambda\right|^2 \frac{4\pi}{\Gamma(2\nu)}\left(\frac{\Omega}{\omega_c}\right)^{2\nu}\sum_{l<-q}\left|p_l\right|^2 \left|q+l\right|^{2\nu-1}.\label{eq:exc_S_C}
\end{equation} 
Similarly to what happens in the IQHE case, by substituting the proper form for the photo-assisted coefficient $p_l$, one can directly inspect whether a certain drive give rise to a vanishing excess noise. 
\begin{figure}
\centering
\includegraphics[width=0.9\linewidth]{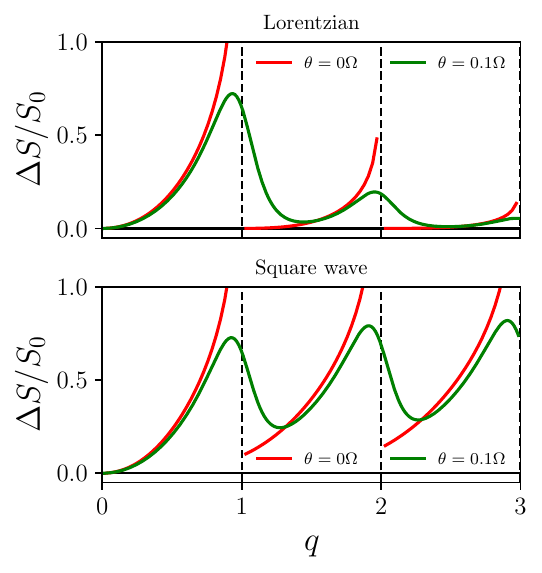}
\caption{Excess noise $\Delta \mathcal{S}_C$ in units of $\frac{e^2}{\Omega}|\lambda|^2$ at filling factor $\nu=\frac{1}{3}$ as a function of $q$. We compare the  square and Lorentzian
voltages at $\theta=0$ (solid curves) and $\theta=0.1$ (dashed curves). The width of Lorentzian pulses is $W=0.1\mathcal{T}$.}
\label{fig:exc_FQH}
\end{figure}
It can be proven mathematically that the Lorentzian pulse is the only drive showing minimal (zero) excess noise, as all contributions of the sum in Eq.~\eqref{eq:exc_S_C} are positive terms and can thus \textit{only} vanish if $|p_l|^2$ is zero for all $l$ below $-q$.

In order to confirm the validity of this analytical result, we plot the excess noise for the Lorentzian-shaped voltage of Eq. (\ref{eq:lorentzian_drive}) 
and we compare it to another relevant voltage drive, as we did in Sec. \ref{noninteracting}, such as the square drive of Eq. (\ref{eq:lorentzian_drive}).

The excess noise corresponding to these drives are presented in Fig. \ref{fig:exc_FQH} in the zero-temperature limit for filling factor $\nu=\frac{1}{3}$, which is the most stable FQH plateau of the Laughlin sequence. All the curves show local minima in correspondence of integer value of $q$. In particular, the signal for the Lorentzian drive at $\theta=0$ vanishes exactly for $q=1,2,3,...$, in accordance with the quantization condition already discussed in Sec. \ref{noninteracting} in the framework of the non-interacting edge states of the IQH  effect. Again, a square voltage drive always generates a finite excess noise, even for integer $q$. Integer Lorentzian voltage pulses still generate minimal excitation states even in the Laughlin FQH regime.  Moreover, fractional values of $q$ do not present any remarkable feature, despite the presence of quasi-particles with a fractional charge.

Interestingly, all the curves show a diverging behavior right before integer values of $q$.  This feature is connected to the orthogonality catastrophe argument discussed by Levitov\cite{levitov1996}. Non-optimal pulses generate a quantum state that is orthogonal to the unperturbed ground state, this manifests as a huge number of particle-hole pairs contributing to transport. Nevertheless, apart from the power-law divergent behavior, the excess noise is qualitatively similar to the case of a normal metal.
At finite temperature, one can modify the excess noise as follows
\begin{equation}
\Delta \mathcal{S}=\mathcal{S}_{C}-\nu e \overline{J_B(t)}\coth \left(\frac{q \Omega}{2\theta}\right).
\end{equation}
Green solid curves in Fig. \ref{fig:exc_FQH} are plotted for $\theta=0.1\omega_c$. Here, the diverging behavior is smoothed by finite temperature effects. For a non-zero temperature, the excess noise is always finite, a fact that was already observed in the Fermi liquid case. This immediately means that electron-hole pairs are always generated by thermal effects, independently of the type of drive.

Turning to electronic Hong-Ou-Mandel interferometry,\cite{hong1987} 
one can study the collision of synchronized excitations impinging onto the QPC from opposite edges of the Hall bar. 
When the delay between the two trains of Levitons is reduced, 
second order interference effects lead to
a Pauli-like dip.\cite{rech2017} which, for integer Levitons has a universal feature. It does not depend either on temperature or on the filling factor, and is therefore  identical to that displayed in Fig. \ref{fig:HOMratio}.

\subsection{Two-Leviton correlated states}

Here, we focus on the effect of the FQHE background on the Levitons and we show that an interaction is induced between them by the strong correlation of the edge states~\cite{bertinjohannet2023c}. We emphasize the detection of the effect of this interaction between Levitons rather than proposals of quantum information schemes based on the aforementioned interaction. 
In order to study the effect of the background correlations on $q$-Levitons we take into account the possibility of injecting multiple Levitons in one period separated by a delay $\Delta t$. The corresponding time-dependent potential is
\begin{equation}
	\label{eq:Levitons}
V(t)=\sum_{j=0}^{q-1}\sum\limits_{k=-\infty}^{+\infty}\frac{V_0}{\pi}\frac{W^2}{W^2+(t-k \mathcal{T}-j \Delta t)^2}\,,
\end{equation}
with period $\mathcal{T}=\frac{2\pi}{\Omega}$, amplitude $V_0$ and half width at half height $W$.  Later, we will consider also the case of an isolated pulse, which can be recovered from the above expression in the limit $W \ll \mathcal{T}$. 

As proven in Ref.~\onlinecite{ronetti2018}, a quantum transport analysis before the QPC does not carry any information about possible interaction between two propagating Levitons. Therefore, we focus on the transport properties backscattered at the QPC. Indeed, we expect that, because of the non-linear nature of tunnelling at the QPC, the backscattered charge will be affected by correlations. For a periodic time-dependent voltage, the charge which is backscattered in one period $\mathcal{T}$ is given by
\begin{align}
\label{eq:QB}
Q = \int_{-\mathcal{T}/2}^{\mathcal{T}/2} \mathrm{d}t\, \left\langle J_B(t) \right\rangle\,.
\end{align}
As described in the previous section, a pertubative calculation can  be carried out in the assumption of weak backscattering regime between the edges. To lowest order in the tunnelling amplitude $\lambda$, the backscattered charge becomes
\begin{equation}\label{eq:charge2int} 
\begin{aligned}
	Q=2ie^*\left|\lambda\right|^2&\int_{-\infty}^{+\infty}\mathrm{d}\tau\, e^{2\nu \mathcal{G}\left(\tau\right)}\\
	&\times\int_{-\mathcal{T}/2}^{\mathcal{T}/2}\mathrm{d}t\,\sin\left[\chi(t)-\chi(t-\tau)\right].
\end{aligned}
\end{equation}
Although the expressions for the backscattered charge $Q$ are valid for any voltage drive $V(t)$, our main interest is in electron-like excitations such as Levitons. Hence, we consider the time-dependent potential defined in Eq.~\eqref{eq:Levitons} for $q = 1$ and $q = 2$ in the presence of a time delay $\Delta t$ between the injections.

The calculation for a periodic signal is conveniently carried out in the PASN formalism presented in Sec,~\ref{sec:charge_FQHE}. 

The photo-assisted expressions for the backscattered charges $Q_1$ and $Q_2$ are
\begin{align}
	\label{eq:Q1}
	Q_1 &=\mathcal{Q} \sum_{m}p_m^2\left\lvert \Gamma\left(\nu+i\frac{\left(m+1\right)\Omega}{2\theta\pi}\right)\right\rvert^2\sinh\left[\frac{\left(m+1\right)\Omega}{2\theta}\right],\\ \label{eq:Q2}
	Q_2 &=\mathcal{Q} \sum_{m}\left|\tilde{p}_m\right|^2\left\lvert \Gamma\left(\nu+i\frac{\left(m+2\right)\Omega}{2\theta\pi}\right)\right\rvert^2\sinh\left[\frac{\left(m+2\right)\Omega}{2\theta}\right],
\end{align}
where $\mathcal{Q} = 4\pi e^*\left|\lambda\right|^2\left(2\pi\theta /\omega_c\right)^{2\nu-1}\frac{\Omega}{\omega_c\Gamma(2\nu)}$. Here, we introduced the photo-assisted coefficients for $q=1$
\begin{equation}
	\label{eq:pm}
p_m = \begin{cases}
		e^{-2\pi \eta m}\left(1-e^{-4\pi \eta}\right)\hspace{5mm} &m\ge 0\\
		-e^{-2\pi \eta}\hspace{5mm} &m = -1\\
		0\hspace{5mm} &m< -1.
	\end{cases}
\end{equation}
and for $q=2$
\begin{equation}
	\tilde{p}_m = \begin{cases}
		\frac{\left[1-e^{i \pi \alpha (m+1)}\right]-e^{-4\pi \eta -i \pi \alpha}\left[1-e^{i \pi \alpha (m+3)}\right]}{\left(1-e^{i \pi \alpha}\right)e^{i \pi \alpha m}}p_m &m\ge 0\\e^{i \pi \alpha}\left(e^{-i \pi \alpha}+1\right)\left(1-e^{-4\pi \eta}\right) p_{-1} & m=-1 \\ e^{i \pi \alpha} p_{-1}^2.&m=-2\\0 &m<-2
	\end{cases}.
\end{equation}
in terms of the rescaled pulse width $\eta = W / \mathcal{T}$, the reduced temperature $\theta = k_{\rm B} T /\Omega$ and the pulse separation $\alpha = 2\Delta t/\mathcal{T}$. The sums appearing in Eqs.~\eqref{eq:Q1} and~\eqref{eq:Q2} can be evaluated numerically: their convergence is ensured by the negative exponential of coefficients $p_m$ in Eq.~\eqref{eq:pm}.
\subsubsection{Correlated two-Levitons state}
In this part, we recast the expression for the backscattered charges $Q_1$ and $Q_2$ in terms of the wave-function of a single Leviton. By using Eq.~\eqref{eq:wave-levitond_period}, the charge $Q_1$ can be recast as
\begin{equation}
\begin{aligned}
	Q_1 = -2e^*\left|\lambda\right|^2 \int_{-\mathcal{T}/2}^{\mathcal{T}/2}\mathrm{d}t\, \int_{-\infty}^{\infty}\mathrm{d}\tau\,  &\Re\left[\varphi_1(t)\varphi_1^*(t-\tau)\right]\\
	\times&\sin\left(\frac{\pi \tau}{\mathcal{T}}\right)\tau e^{2\nu \mathcal{G}(\tau)}\,.
\end{aligned}
\end{equation}
We observe that the charge $Q_1$ contains a product of Leviton wave-function, thus showing that it is determined directly by the charge density of the state injected on the system's ground state. One can similarly express the charge backscattered by the two-Leviton state as
\begin{widetext}
 \begin{align}
 \label{Q2}
Q_{2}=-8i e^*\left|\lambda\right|^2 \int_{-\infty}^{\infty}dt \int_{-\infty}^{\infty}d\tau~\tau^2 \sin^2\left(\frac{\pi \tau}{\mathcal{T}}\right)\Im\left[\varphi_1(t)\varphi_1^*(t-\tau)\varphi_1(t+\Delta t)\varphi_1*(t-\tau+\Delta t) \right] e^{2\nu \mathcal{G}(\tau)}+ 2 Q_1 \, .
\end{align}
\end{widetext}
We note that, while the Leviton wave-function is strictly valid only for non-interacting systems, the expressions for the backscattered charge are equivalent to the ones obtained in a system without interactions where the Green's functions have been replaced by those of the strongly correlated fractional quantum Hall edge channels. Based on this observation, we claim that we can still use the Leviton wave-function to support the physical interpretation for our result. A general description of a Leviton, even with a fractional charge, in terms of wave-functions is possible, but requires a more elaborated formalism~\cite{sim-private}.

Moreover, we observe that the integral appearing in the above expression is zero for $\nu=1$. Indeed, in this case $e^{2\mathcal{G}(\tau)}$ is an even function of $\tau$, while the imaginary part appearing in Eq.~\eqref{Q2} is an odd function of the same variable: therefore, the integral over $\tau$ vanishes at $\nu = 1$ and one recovers the trivial result $Q_2 = 2Q_1$ at any temperature.

It is instructive to recast the excess charge $\Delta Q =Q_2-2Q_1$ in terms of the Leviton wave-function as
\begin{equation}
\begin{aligned}
	\Delta Q =e^*\left|\lambda\right|^2(i)^{2-2\nu}\int_{-\infty}^{\infty}\mathrm{d}t \big[&\varphi_1(t)g_{\nu}(t,\Delta t)\varphi_1(t+\Delta t)\\
	&\qquad\qquad -\text{H.c.}\big]\, ,
\end{aligned}
\end{equation}
where 
\begin{widetext}
\begin{equation}
	g_{\nu}(t,\Delta t) = \int_{-\infty}^{\infty}\mathrm{d}\tau\, \varphi_1^*(t-\tau)\varphi_1^*(t-\tau+\Delta t)\tau^2 \sin^2\left(\frac{\pi \tau}{\mathcal{T}}\right) e^{2\nu \mathcal{G}(\tau)}.
\end{equation}
\end{widetext}
In contrast with $Q_1$, the excess charge is related to the product of four Leviton wave-functions, thus proving that it stems from a density-density interaction between Levitons. We interpret this result by conjecturing that the strongly-correlated background mediates an effective interaction between the two Levitons. 

The function $g_{\nu}$ is different from zero for fractional filling factors because of the propagator $e^{2\nu \mathcal{G}(\tau)}$. The power law decay for fractional filling factors is slower than $\tau^2$, thus inducing long-time correlations between the two Levitons. These correlations do not affect the charge only when the isolated pulses are well separated (limit of $\Delta t \gg W,1/\theta$). Otherwise, correlations induce an effective interaction between Levitons that effectively enhance the value of the charge $Q_2$ compared to the limit of two well-isolated pulses. 

We highlight that the existence of this interaction dramatically relies on the correlations of the FQH background. We base our claim on the fact that for $\nu=1$ the interaction between Levitons is absent, as discussed below Eq.~\eqref{Q2}. Indeed, the specific type of this correlation, while influencing the form of the interaction between Levitons, is not crucial for its existence. In particular, even at finite temperature, where the power-law behaviour of the correlation functions is exponentially suppressed at times $t > \theta^{-1}$, the interaction $g_{\nu}$ is still present. Since the long-range nature of FQH correlation is not necessary for the interaction to exist, we consider in the next subsection the case of isolated pulses, which allows to derive analytical expressions.

\subsubsection{Isolated pulses}
In the case of isolated pulses the integral over $t$ in Eq.~\eqref{eq:QB} can be extended from $-\infty$ to $+\infty$ and can be solved analytically. Let us comment that this limit is well-defined only for voltage pulses that go to zero at $t=\pm \infty$, which is the case for Lorentzian-shaped pulses. The expression for the charge becomes
\begin{equation}\label{eq:charge3int} 
\begin{aligned}
	Q=2ie^*\left|\lambda\right|^2&\int_{-\infty}^{+\infty}\, \mathrm{d}\tau e^{2\nu \mathcal{G}(\tau)}\\
	&\times\int_{-\infty}^{+\infty}\mathrm{d}t\,\sin\left[\varphi(t)-\varphi(t-\tau)\right].
\end{aligned}
\end{equation}
The integral over $t$ can be solved analytically for integer values of $q$. For $q = 1$, one finds
\begin{equation}
	Q_1=16\pi ie^*\left|\lambda\right|^2W^2\int_{-\infty}^{+\infty}\mathrm{d}\tau\, e^{2\nu \mathcal{G}(\tau)}\frac{\tau}{\tau^2+4W^2}\, .
\end{equation}
Next, we consider the case $q=2$ where the isolated pulses are separated by a constant delay $\Delta t$. The integral over $t$ gives
\begin{widetext}
	\begin{equation}\label{eq:totalchargetransfer}
	Q_2=64 \pi ie^*\left|\lambda\right|^2W^2\int_{-\infty}^{+\infty}\mathrm{d}\tau\,e^{2\nu \mathcal{G}(\tau)}\frac{\tau \left[\left(4 W^2 +  \Delta t^2\right)^2 - \tau^2\left(3  \Delta t^2 + 4W^2\right) + 2 \tau^4 \right]}{(\tau^2+4W^2)\left[( \Delta t+\tau)^2+4W^2\right]\left[( \Delta t-\tau)^2+4W^2\right]}\, .
	\end{equation}
\end{widetext}
These two expressions can be solved numerically at finite temperature $\theta$. 
On the other hand, analytical expressions  for $Q_1$ and $Q_2$ can be obtained in the zero temperature limit.
In this case the bosonic Green's function reads
\begin{align}
	\mathcal{G}\left(\tau\right) = -\log \left(1+i \omega_c\tau\right)\, .
\end{align}
The residue theorem can be used to calculate the integrals over $t$ in Eq.~\eqref{eq:totalchargetransfer}.
In the case $q=1$, we obtain
\begin{equation}
 Q_1=e^*\left(\frac{\lambda}{v}\right)^2\left(2W \omega_c\right)^{2-2\nu}\,  + \mathcal{O}\left[\left(\omega_c W\right)^{2\nu-1}\right],
\end{equation}
where we kept only the leading order in $1/\left(W\omega_c\right)$. 
By performing a similar calculation for $Q_2$ in the limit of zero temperature, we find, to lowest order in $1/\left(W\omega_c\right)$,
\begin{equation}\label{eq:charge2lev}
\begin{aligned}
 Q_2&=Q_1\Bigg\{2\Re\left[\left(1+\frac{2iW}{ \Delta t}\right)^2\left(1-\frac{i \Delta t}{2W}\right)^{-2\nu}\right] \\
 &+2\left(1+\frac{4W^2}{ \Delta t^2}\right) \Bigg\} + \mathcal{O} \left[ \left(\frac{1}{W \omega_c}\right)^{2\nu-1} \right].
\end{aligned}
\end{equation}

We note that, at zero temperature, the backscattered charge for two pulses is proportional to the backscattered charge for a single pulse. For the filling factor $\nu = 1$, we recover the trivial result that $Q_2 = 2 Q_1$. 
However, for fractional filling factor, the constant of proportionality depends on $\Delta t$ and $W$ and $Q_2\ne 2Q_1$. 

Before concluding this part, it is instructive to analyze two extreme limits of the ratio $ \Delta t/W$ at zero temperature. In the limit of simultaneous pulses, which can be obtained by setting $ \Delta t/W\ll1$ in Eq.~(\ref{eq:charge2lev}), we find
\begin{equation}\label{eq:charge2levsim}
\begin{aligned}
\lim_{\Delta t/W\rightarrow 0}Q_2=2Q_1\left(2 - 3\nu + 2\nu^2\right)\, .
\end{aligned}
\end{equation}
In this limit the constant of proportionality acquires a simple expression, becoming independent of $W$ and being determined solely by the filling factor $\nu$.\\
Finally, we consider the opposite case of well-separated pulses, which can be found by taking the limit $ \Delta t/W\to\infty$ in  Eq.~\eqref{eq:charge2lev}, which yields
\begin{equation}
\lim_{\Delta t/W\rightarrow \infty}Q_2 = 2Q_1.
\end{equation}
In this case the charge backscattered for two Levitons is twice the one obtained with a single Leviton. For well-separated injection time, the system has relaxed to equilibrium when the second pulse comes in. As a result, the two Levitons behave as two independent single pulses. 
%We summarize these results in Fig.~\ref{fig:analytical}, where the black and red dotted lines represents the two limits of simultaneous and well-separated pulses at zero temperature. 

\subsubsection{Results and plots}

\begin{figure}%[h]
	\includegraphics[width=\linewidth]{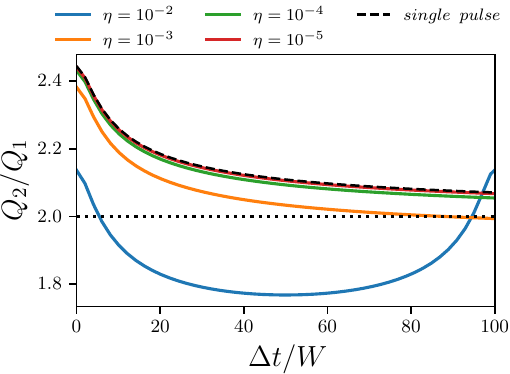}
	\caption{(Color online) Backscattered charge for a two-Leviton state $Q_2$ rescaled with respect to the same quantity for a single Leviton as a function of $\Delta t/W$ at zero temperature. The black dashed line is the limit of a single pulse. Solid lines are computed for the periodic case with $\eta = 10^{-2}, 10^{-3}, 10^{-4}, 10^{-5}$. We see that for the smallest value of $\eta = 10^{-5}$, the periodic case coincides with the analytical limit at infinite period.  The black dotted is a visual guide for $Q_2 = 2Q_1$. The ratio $Q_2 / Q_1$ always stays above this line, except for the highest value of $\eta = 10^{-2}$: in this case $Q_2$ can be smaller than twice $Q_1$.  The only other parameter is the filling factor $\nu = 1/3$. Figure taken form Ref.~\onlinecite{bertinjohannet2023c}.}
	\label{fig:fig3}
\end{figure}

We start our analysis by plotting, in Fig.~\ref{fig:fig3}, the ratio of the backscattered charges $Q_2/Q_1$ at filling factor $\nu = 1/3$ as a function of the time delay $\Delta t$ rescaled by the width of the pulses $W$. Different curves are realized at different values of the parameter $\eta = W / \mathcal{T}$. 

For potential applications in the quantum information domain, it has been conjectured that, in order to perform a single-qubit operation, one needs single-electron pulses of the order of $W\sim 10$ ps, which is at the limit of state-of-the-art technology~\cite{edlbauer2022}. 
Here, in order to provide an estimation for the experimentally realistic value of $W$, we focus on Lorentzian pulses tailored for future applications in the quantum information domain.  By considering a frequency $\Omega = 2\pi \times 5$ Ghz, typical for experiments with Levitons, the resulting value for the renormalized width is $\eta = 5 \cdot 10^{-2}$. 
For general purposes, in Fig.~\ref{fig:fig3} we consider also smaller values of $\eta$ such that a comparison can be done between the periodic and the isolated pulses cases.
One can clearly see that the case of isolated pulses (dashed line) is recovered  in the limit $W \ll \mathcal{T}$.
This is true for pulses width lower than $\eta\approx  10^{-4}, 10^{-5}$.  

The analysis of Fig.~\ref{fig:fig3} reveals some interesting features of $Q_2$ in the periodic case. Indeed, at $\eta = 10^{-4}, 10^{-5}$ the ratio $Q_2 / Q_1$ is always greater than or equal to $2$. As a result, the correlated background always enhances the value of the backscattered charge compared to the case $\nu = 1$. Remarkably, for $\eta = 10^{-2}, 10^{-3}$, this is not the case anymore as we see that the ratio $Q_2/Q_1$ can become smaller than or equal to $2$. Thus the effect of the correlated background is strongly affected by the width of Lorentzian pulses in the periodic case. In passing, we comment that this additional feature appears exactly for values of $\eta$ which are closer to realistic experimental ones. 
\begin{figure}%[h]
	\includegraphics[width=\linewidth]{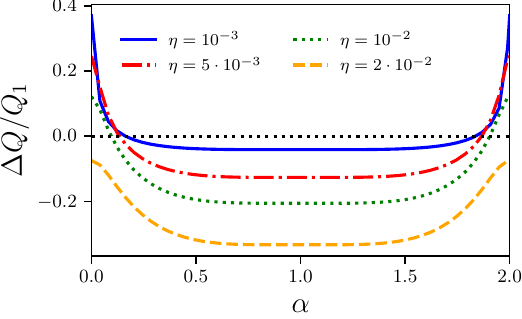}
	\caption{(Color online) Excess backscattered charge $\Delta Q$ rescaled with respect to the backscatterd charge $Q_1$ for a single Leviton as a function of $\alpha = 2\Delta  t/ \mathcal{T}$ at zero temperature. The black dotted line is a visual guide for $\Delta Q = 0$. Solid lines are computed for the periodic case with $\eta = 10^{-3}, 5\cdot 10^{-3}, 10^{-2}, 2\cdot 10^{-2}$. The excess charge $\Delta Q$ changes sign as a function of $\alpha$, except for the highest value of $\eta = 2\cdot 10^{-2}$: in this case $\Delta Q$ is always negative. The smaller the value of $\eta$ and the higher the value of $\alpha$ at which $\Delta Q = 0$. The only other parameter is the filling factor $\nu = 1/3$. Figure taken form Ref.~\onlinecite{bertinjohannet2023c}.}
	\label{fig:fig4}
\end{figure}

To continue our analysis, it is interesting to present more information about the dependence of the backscattered charge on the parameters $\alpha = 2\Delta t/\mathcal{T}$ and $\eta = W / \mathcal{T}$. To this end, let us now focus on the excess charge $\Delta Q$. It is important to remark that for $\Delta Q>0$ (resp. $\Delta Q<0$), the backscattered charge is increased (resp. reduced) with respect to the trivial case at $\nu = 1$. In Fig.~\ref{fig:fig4}, we present the excess charge as a function of $\alpha$ for different values of $\eta$. One can deduce from this plot that it exists a large range of $\eta$ for which the sign of $\Delta Q$ can be changed by tuning the parameter $\alpha$.
Interestingly, there exist some values of $\alpha$ where the excess charge $\Delta Q$ vanishes, thus showing that the effect of strong correlations on two-Leviton states can be tuned on and off by acting on the separation time $\Delta t$. Above a certain value of $\eta$, we found that the sign of $\Delta Q$ is negative for any value of $\alpha$ at zero temperature and $\nu = 1/3$. While the specific values of $\eta$ depend on temperature and filling factor, the important result is that there always exists a width of Lorentzian pulse above which the sign of $\Delta Q$ is strictly negative. 

%%%%%%%%%%%%%%%%%%%%%%%%%%%%%%%%%%%%%%%%%%%%%%%%%%%%%%%%%%%%%%%%%%%%%%%%%%%%%%%%%%%%%%%%%%
\section{Levitons and Superconducting correlations}
\label{entangled}

We now turn to a different form of electronic correlations: those present in BCS superconductors. In such a metal, the phonon-mediated interaction between electrons can be attractive.
This generates a ground state consisting of a superposition of Cooper pairs. The latter are pairs of electrons which have opposite spin and momentum in a singlet state.
This can lead to interesting tunnel processes when for example a superconductor is connected to two normal metal leads.
At the junction, the separation between the two normal metals connected to the superconductor has to be smaller than the coherence length. The bias voltage between the normal metals is much smaller than the superconducting gap. Temperatures well below the gap are also necessary. 
The two constituent electrons of a Cooper pair can either tunnel to the same lead (direct Andreev reflexion) or they can tunnel in different leads (crossed Andreev reflection)~\cite{anantram1996,martin1996a,torres1999,lesovik2001,recher2001,sauret2004,sauret2005,chevallier2011,rech2012}. If there are no magnetic impurities involved in the tunneling process, the singlet nature of the two-electrons wave function is preserved. One can then impose additional constraints on the two metallic leads -- either by filtering separately electrons (positive energies) and holes (negative energies) -- with respect to the superconductor chemical potential -- or by imposing that both leads are half metals with opposite spin polarization -- so that crossed Andreev reflection is the dominant process. 

This constitutes a Cooper pair beam splitter,\cite{lesovik2001,recher2001} which has been proposed two decades ago as a DC source of entangled electrons. Indeed in the former case, the spin structure of the electron bound state is preserved and gives rise to spin entanglement, while in the latter case, because the spin is projected on the two (opposite spin) ferromagnetic leads, this gives rise to energy entanglement. 
This Cooper pair beam splitter has generated a lot of theoretical and experimental activity since its proposal. Proposals for detecting both the spin entanglement and the energy entanglement in both setups via a Bell inequality test\cite{bell1966} analogous to that proposed earlier on for photon pairs experimentally\cite{aspect1982} have been made.\cite{chtchelkatchev2002,sauret2005,bayandin2006} Experimental detection of entanglement in mesoscopic physics remains a challenge, nevertheless positive noise crossed correlations were measured experimentally in a Cooper pair beam splitter, which constitutes a step in the right direction.\cite{das2012}

In a previous work,\cite{bertinjohannet2022} some of the authors studied a normal metal/BCS superconductor junction subjected to a time dependent periodic voltage bias consisting of a superposition of Lorentzian pulses. Ref. \onlinecite{vanevic2016a} had used circuit theory in two specific limits -- the high frequency regime ($\Delta\ll \Omega$) -- where quasi-particle transfer dominates, and the how frequency regime ($\Delta\gg \Omega$), where Andreev reflection operates. In the former regime, the injected charge required to minimize the excess noise is an integer of the electron charge, while in the latter regime, half quantized Lorentzians minimize this excess noise. This is a consequence of the fact that in the Andreev regime, the effective bias corresponds to $2eV$. The predictions of Ref. \onlinecite{vanevic2016a} were confirmed using a Keldysh-Nambu-Floquet Hamiltonian approach\cite{bertinjohannet2022} where the tunnel coupling between the normal metal and the superconductor is treated to all orders, for arbitrary frequency drive, allowing to describe the intermediate regime where $\Delta\sim \Omega$. The concept of minimal excess noise only makes sense in the small and the large gap regime: in the intermediate regime, the excess noise can grow arbitrarily, or even become negative. Deep in the Andreev regime, it was shown with realistic parameters that the normal metal/superconductor junction can inject approximately one Cooper pair per period of the drive.\cite{bertinjohannet2023a} 

\subsection{Time-dependent driven normal-superconducting junctions}

\begin{figure}
    \centering
\includegraphics[width=0.45\textwidth]{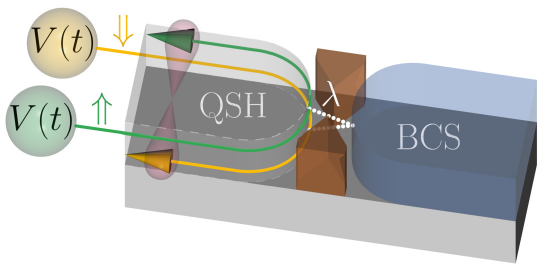}
    \caption{The setup: a superconductor (right) is tunnel coupled to a Quantum spin Hall bar with two opposite edge spin channels via an adjustable QPC. The voltages imposed on both channels correspond to trains of Lorentzian voltage pulses $V(t)$. The shaded area covering the channels (downstream from the injection point of the two electrons ejected from  the superconductor) represent the energy-entangled electrons which are generated on the normal side (left). Figure taken form Ref.~\onlinecite{bertinjohannet2023a}.}
    \label{fig:entangled-setup}
\end{figure}

In the setup consisting of a superconductor connected to two half metals, time-dependent drives can be imposed on both leads so that this Cooper pair source operates
in the reverse order: one Cooper pair per period can be ejected from the superconductor. 
In Ref. \onlinecite{bertinjohannet2023a} we generalized the formalism of Ref. \onlinecite{bertinjohannet2022} to multi-lead systems in order to propose a time dependent version of the Cooper pair beam splitter. For convenience, and so as to optimize the crossed Andreev process, the two half metals were replaced by a quantum spin Hall bar, as shown in Fig.~\ref{fig:entangled-setup}. Indeed, due to spin momentum locking, the edge channels with opposite spin propagate in opposite direction. The spin of the two constituent electrons of the Cooper pair are projected during tunneling on the two edge states and it is natural to enquire whether the two outgoing electrons have preserved some kind of entanglement. Naturally, momentum entanglement cannot be preserved as the QPC breaks translational symmetry. 

\subsubsection{Setup and model}
The leads are described in equilibrium by the Hamiltonians: 
\begin{align}
&H_{\text{N}} = \sum_{\sigma=\uparrow,\downarrow} H_{0,\sigma,\text{N}}\nonumber\\
&H_{\text{S}} =\sum_{\sigma=\uparrow,\downarrow} H_{0,\sigma,\text{S}}+\Delta\sum_i\left(c_{i,\text{S},\downarrow}^\dagger c_{i,\text{S},\uparrow}^\dagger +c_{i,\text{S},\downarrow}^{\phantom{\dagger}}c_{i,\text{S},\uparrow}^{\phantom{\dagger}}\right),
\end{align}
where $H_{0,\sigma,j}$ is the kinetic part of the Hamiltonian of lead $j$ (with spin $\sigma$), $i$ labels the various sites of these leads, $\Delta$ is the superconducting gap and the chemical potential is set to zero, $c_{i,S,\sigma}^{\phantom{\dagger}}$ is the annihilation operator for electrons in the superconducting lead.

The tunnel Hamiltonian for electrons is defined as
\begin{equation}\label{CurrentHamApp}
    H_T(t)=\sum_{\substack{j=N,S\\j'=N,S}}\sum_{\sigma}\lambda_{j,j'}e^{i\phi_{j,j'}(t)/2}c_{e_{jj'},j,\sigma}^\dagger c_{e_{j',j},j',\sigma}^{\phantom{\dagger}}+\text{H.c.}\, ,
\end{equation}
where $\lambda_{j,j'}$ is the tunneling amplitude from lead $j$ to lead $j'$, $c_{i,j,\sigma}^{\phantom{\dagger}}$ is the annihilation operator for electrons at site $i$ and with spin $\sigma$ on the lead $j$, $e_{j,j'}$ denotes the site of lead $j$ from which tunneling to lead $j'$ occurs, and $\phi_{jj'}(t) = e\int_{-\infty}^{t} \mathrm{d}t'\, V_{jj'}(t')$ is the time-dependent phase difference between the leads which accounts for the drive-induced voltage difference $V_{jj'}(t)$ between lead $j'$ and lead $j$.
One can introduce the tunnel matrix in lead space $W_{jj'}=\lambda_{jj'}\sigma_z e^{\sigma_z i\phi_{jj'}(t)/2}$ for leads carrying both spin orientation. 
For the device of Fig. \ref{fig:entangled-setup}, $\lambda_{\uparrow\downarrow}=0$, the superconducting lead only is connected to two spin-polarized chiral edge states of the QSH, so that $j$ is a spin index for the QSH. This allows us to replace the $\sigma_z$ prefactor in $W_{Sj'}$ by $(\sigma_z\pm \sigma_0)/2$ (for $j'=\uparrow,\downarrow$). From now on, we focus on $\lambda_{S \uparrow} = \lambda_{S \downarrow} = \lambda$. 
%%%%%%%%%%%%%%%%%%%%%%%%%%%

\subsection{Energy-entangled Levitons}

In order to determine what type of electron states are generated at the output, we apply a {\it single} Lorentzian voltage pulse to both of the spin-polarized channels and we expand the quantum state to second order in the tunnel coupling  at zero temperature. The two pulses are synchronised both in time and amplitude, but this assumption can be relaxed\cite{bertinjohannet2023b}.
The single pulse is a Lorentzian $e V(t)=-\frac{1}{2 \pi W} \; \frac{1}{1+t^2/W^2}$ with a width $W \gg \Delta^{-1}$ chosen well above the inverse gap in order to be in the Andreev regime.
The two-electron state created in the spin Hall bar then reads as:\cite{bertinjohannet2023a}
\begin{widetext}
\begin{equation}\label{eq:lev1}
    \left\lvert\mathcal{F}\right\rangle=i\frac{\lambda^2\sqrt{W}}{2\sqrt{2}\pi^2}\int_{-\infty}^{\infty}\mathrm{d}\varepsilon\, \varphi_T(2\varepsilon)\int_{-\varepsilon}^{\varepsilon}\mathrm{d}E c_{k(\varepsilon+E),\uparrow}^\dagger c_{k(\varepsilon -E),\downarrow}^\dagger\left\lvert F_\uparrow\right\rangle\otimes\left\lvert F_\downarrow\right\rangle\otimes\left\lvert\Psi_\text{BCS}\right\rangle\, ,
\end{equation}
\end{widetext}
where $\left\lvert F_{\uparrow}\right\rangle$ and $\left\lvert F_{\downarrow}\right\rangle$ are the Fermi seas for the spin-polarized 
channels at equilibrium with BCS ground state, $\varphi_T(\varepsilon)=\sqrt{2W}e^{-W\varepsilon}H(\varepsilon)$
is the single Leviton wave function~\cite{keeling2006,grenier2013,glattli2016a,ronetti2018}, and $H(\epsilon)$ is the Heaviside distribution.

One cannot factorize this state into a product of two states acting separately on the Fermi seas of the leads. While momentum entanglement is lost at the junction, both electron states have energies higher than the superconducting chemical potential and are entangled in energy. As is the case for normal metal devices, this state does not carry unwanted electron-hole pair excitations because of the particular choice of a Lorentzian drive. 

\subsection{Excess noise in the normal-superconducting junction}
We now want to probe whether or not the generated state is purely electronic.
Hence we consider periodic drive and compute transport observables, such as the excess noise.
Writing the Hamiltonian as $H=\sum_j H_j + H_{\text{Tun}}$, the current operator from lead $j$ is given by
\begin{equation}\label{Current}
I_j(t)=\sum_{j'}i\psi_j^\dagger(t) \sigma_z W_{jj'}(t) \psi_{j'}(t) + \text{H.c.},
\end{equation}
where we introduced the standard Nambu spinor notation $\psi_j$ for the electron operators.
The real time irreducible noise correlator between lead $j$ and $l$ is defined as
\begin{equation}
	\tilde{S}_{jl}(t,t')=\left\langle I_j \left(t+t' \right) I_l \left( t \right) \right\rangle 
	     - \left\langle I_j \left( t+t' \right) \right\rangle
	       \left\langle I_l \left( t \right) \right\rangle \, .
\end{equation}
It is computed from the Keldysh Green's function $G^{\pm\mp}_{jj'}(t,t')= -i \left\langle T_K\psi_j(t^{\pm})  \psi_{j'}^\dagger(t^{'\mp}) \right\rangle$, where $ T_K$ denotes Keldysh ordering, and $\pm$ superscripts stand for the branch on the contour. 
We look at the noise $S_{jl}(t)=\int_{\-\infty}^{+\infty} \mathrm{d}t'\, \tilde{S}_{jl}(t,t')$.
One uses Wick' theorem to cast the noise as a product of single particle Green's functions:
\begin{widetext}
\begin{equation}\label{Noise1leadfinal}
\begin{aligned}
        S_{jl}(t)=-e^2\sum_{j'l'}\int \mathrm{d}t'\,\text{Tr}_{\text{N}}\big[&\sigma_z W_{jj'}(t)G_{j'l}^{-+}(t,t') \sigma_z W_{ll'}(t')G_{l'j}^{+-}(t',t)
    +\sigma_z W_{j'j}(t) G_{jl'}^{-+}(t,t')\sigma_z W_{l'l}(t')G_{lj'}^{+-}(t',t)\\
        -&\sigma_z W_{j'j}(t)G_{jl}^{-+}(t,t') \sigma_z W_{ll'}(t')G_{l'j'}^{+-}(t',t)
        -\sigma_z W_{jj'}(t) G_{j'l'}^{-+}(t,t')\sigma_z W_{l'l}(t')G_{lj}^{+-}(t',t)
        \big]\, .
\end{aligned}
\end{equation}
\end{widetext}
As before, $V(t) = V_{\text{DC}} + V_{\text{AC}}(t)$, where $V_{\text{AC}}(t)$ averages to zero on one period $\mathcal{T}=2\pi / \Omega$ of the periodic drive. 
The injected charge per period and per spin is then given by $q =-\frac{e V_{\text{DC}}}{\Omega}$. The Floquet coefficients are defined as $\exp[-i \chi(t) ]= \sum_l p_{l} e^{-i l \Omega t}$ and the Floquet weights as $P_l = \left\lvert p_l\right\rvert^2$.

The Green's functions and Dyson's equations of the system adopt a double Fourier representation\cite{jonckheere2013,jacquet2020,bertinjohannet2022}. 
The zero-frequency period averaged noise (PAN) is defined as $\overline{\left\langle S_{jl}\right\rangle}\equiv \int_{-T/2}^{T/2}\frac{\mathrm{d}t}{T}\left\langle S_{jl} \left(t\right) + S_{lj} \left(t\right)  \right\rangle$.

We define the total excess noise as
\begin{equation}
    S_\text{exc} = \left. \overline{\left\langle S_T\right\rangle}\right|_{DC+AC}  - \left. \overline{\left\langle S_T\right\rangle}\right|_{\text{DC}} \, .
\label{eq:ExcessNoise}
\end{equation}
where $\overline{\langle S_T\rangle} \equiv \sum_{\sigma ,\sigma'=\uparrow,\downarrow} \overline{\langle S_{\sigma\sigma'}\rangle}$ is the {\it total} noise of the source. The Andreev transfer channel imposes further that: $\overline{\langle S_{\uparrow\downarrow}\rangle}=\overline{\langle S_{\downarrow\uparrow}\rangle}=\overline{\langle S_{\uparrow\uparrow}\rangle}=\overline{\langle S_{\downarrow\downarrow}\rangle}$.\cite{lesovik2001} 

\subsubsection{Optimal working point of the source}
\begin{figure}
    \centering
    \includegraphics{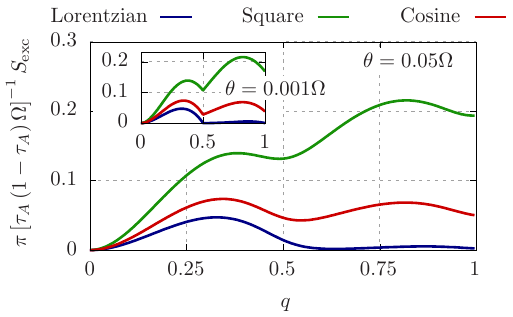}
    \caption{Excess noise as a function of the injected charge for various drive shapes and at two different temperatures.}
    \label{fig:xn-ns}
\end{figure}

In the regime $\Omega\ll\Delta$, the total noise of the source obeys the analytical expression: 
\begin{equation}
\begin{aligned}
		&\overline{\left\langle S_T \right\rangle}_{q} = \frac{e^2}{\pi}\bigg(4\tau_A^2\theta +2\tau_A(1-\tau_A)\\
		  &\qquad\times\sum_n  (2q+n)\Omega P_n (2q) \coth \left[ (2q+n)\frac{\Omega}{2\theta} \right]\bigg)\, ,
	\label{Eq_Noise_andreev}
\end{aligned}
\end{equation}
where $\tau_A=4\lambda^4/(1+\lambda^4)^2$ is the Andreev transmission coefficient. 
This excess noise is plotted as a function of $q$ in Fig.~\ref{fig:xn-ns}. 
We observe that it achieves minimal noise for half-integer and integer $q$ at  zero temperature, and for slightly higher values of $q$ when the temperature $\theta$ is finite. Periodic trains of Lorentzians always lead to minimal noise when compared to cosine, or square voltages.

%%%%%%%%%%%%%%%%%%%%%%%%%%%%%%%%%%%%%%%%%%%%%%%%%%%
%%%%%%%%%%%%%%%%%%%%%%%%%%%%%%%%%%%%%%%%%%%%%%%%%%%%%
This suggests that the optimal working point of this source of energy entangled states imposes near perfect transmission and an average charge per period 
$\langle Q \rangle = 2\pi\frac{\overline{\left\langle I \right\rangle}_{q}}{\Omega}=4 q e \tau_A$ which is a multiple of a half integer. 

What happens when we relax these constraints in order to make a realistic proposal which works at finite temperature ? The fact that $\langle Q \rangle$ corresponds to a multiple of a Cooper pair charge can also
be achieved for $\tau_A<1$ provided that $q$ is ``slightly larger'' than a half integer.  One should then ensure the minimization of excess noise so as to generate primarily minimal excitation states, which can only be achieved by a Leviton voltage drive. This average charge transferred per drive period is robust under variations of the electron temperature.

Ref.~\onlinecite{dubois2013b}, the first experimental evidence for Levitons, worked at a temperature is $\theta \approx \unit{10}{\milli\kelvin}$ with a drive frequency $f \approx \unit{5}{\giga\hertz}$ ($V_{\text{DC}} \approx\unit{10}{\micro\volt}$ for $q=0.5$). 
With a Niobum gap $\Delta_\text{Nb}\approx \unit{1.55}{\milli\electronvolt} $), this puts us well into the Andreev regime, as we have $\Delta_\text{Nb} \approx 2000\theta$ and $\Delta_\text{Nb} \approx 100\Omega$. 

The excess noise gets modified as the temperature is increased, leading to a first arch which, instead of vanishing exactly at $q=0.5$, now reaches a (nonzero) local minimum for a slightly higher value of $q$. The value $S_\text{min}$ of this local minimum is chosen as the reference point to optimize the quality of the produced entangled states. We obtain the $\lambda$ dependence of the excess noise under the constraint $\left\langle Q \right\rangle = 2 e$.
For the above temperature and drive frequency, we identify\cite{bertinjohannet2023a} an amplitude $\lambda \simeq 0.78$, ($\tau_A \simeq 0.79$) for which the excess noise is close to the minimum noise, which imposes $q^*\simeq 0.65$. When operated at this optimized transmission, the junction displays the minimal possible noise while transferring on average a charge $2e$ per period that is distributed on both leads. One might thus achieve an on demand source of energy entangled states which generate one non local electron pair per cycle, and have optimized its working conditions in a realistic finite temperature context.  

\section{Conclusion}
\label{conclusion}

In this short review, we showed how Levitons can be employed to achieve electronic quantum optics scenarios such as HBT and HOM experiments for normal-metal 
systems of condensed matter physics. For arbitrary periodic voltage drives, the HBT excess noise signal provides direct information on the presence of spurious holes in the input/injected current: only Lorentzian drives bring the excess noise to its minimal value, zero. 

Our main motivation was to examine whether the Leviton paradigm survives in the presence of electronic correlations and to take profit of the latter to design new EQO experiments. For this purpose we first considered the strong correlations which are explicit in the FQHE regime, and showed that in the weak backscattering regime -- where Laughlin quasi-particle tunnelling between edge states represent the dominant charge transfer process -- minimal excess noise is achieved for voltage pulses which carry an integer multiple of the electron charge rather than the (quasiparticle) fractional charge predicted in Ref. \onlinecite{keeling2006}. Furthermore, we computed the average charge transmitted from one edge state to the other when the voltage injects two Levitons (per period in the AC regime).
We showed that it does not correspond to twice the charge associated with a single Leviton, and that it is due to the non-linearities of the QPC associated with the scaling dimension of the tunneling operator. Moreover, we showed that the excess charge can be cast in the form of an interaction between Leviton wave packets carried by the strongly correlated medium of the FQHE.  

We note that these results can be extended to other charge transport observables such as the heat current.\cite{vannucci2017} Regarding noise, one can then consider noise observables where the correlator is the product of a charge current and a heat current (``mixed'' noise) or the product of two heat currents (``heat'' noise). It turns out (not shown) that both of these quantities are also minimized for the same condition as the charge noise.     

Localized voltage pulses such as Levitons are likely to be relevant in the different approaches aimed at making a diagnostic of the fractional statistics of anyons in the FQHE, where braiding effects between quasi-particles are explicit in the current and noise signatures.\cite{jonckheere2023} 

The case of Levitons injected in a non-chiral Luttinger liquid (a one dimensional interacting electron system such as carbon nanotubes) has been studied elsewhere by some of the authors.\cite{fukuzawa2023} Similar results apply to this situation, except that the divergences of the excess noise close  to integer values of the charge are absent even at zero temperature, and Fabry-Perot resonances are found when the charge is injected  in the bulk of one dimensional systems connected to Fermi liquid leads. 

Finally, we studied Levitons in a system where correlations have altogether a totally different nature, when the device contains as superconducting correlations. When a superconducting lead is connected to two half metals (or equivalently to two opposite spin edge channels of a quantum spin Hall bar) on which the bias voltage is imposed, this allows to realize a time dependent version of the Cooper pair beam splitter.\cite{lesovik2001,recher2001} This allowed us to propose an on demand source of energy entangled electron states whose constituent electrons propagate in a non local manner in opposite spin polarized channels, and to determine its optimal working point at finite temperatures. A detailed derivation of the state generated by a train of Levitons will appear elsewhere, together with a diagnosis of entanglement via noise crossed correlation measurements.\cite{ronetti2024} 
The entangled electron states thus generated could allow the implementation of  quantum protocols\cite{bennett1993,long2002,ekert1991,bennett1992,pan2020,bennett1992b} in a condensed matter physics setting.

%%%%%%%%%%%%%%%%%%%%%%%%%%%%%%%%%%%%%%%%%%%%%%%%%%%%%%%%%%%%%%%%%%%%%%%%%%%%%%%%%%%%%%%%%%%
\begin{acknowledgments}
One of us (TM) is indebted to Professor D.K. Campbell for hiring him as a graduate student early on in Los Alamos National Laboratory during two summer internships in 1986 and 1987 where we worked on conducting polymers\cite{martin1987} and high $T_c$ superconductivity\cite{loh1988} under his guidance. More importantly, D. K. Campbell showed to this author his passion for quantum field theoretical approaches of condensed matter physics (as P.W. Anderson used to say: ``More is different...''\cite{anderson1972}) which conditioned TM's choice of field of physics and advisor during his PhD at UCLA, together with his future career plans. This review article is the result of an ongoing collaboration between TM and three of his former PhD students, as well as several tenured members of the nanophysics team of CPT. 

This work received support from the French government under the France 2030 investment plan, as part of the Initiative d'Excellence d'Aix-Marseille Universit\'e - A*MIDEX, through the institutes IPhU (AMX-19-IET-008) and AMUtech (AMX-19-IET-01X). Funding is also acknowledged from the French Agence Nationale de Recherche (ANR), Programmes et équipements prioritaire de recherche/PEPR Technologies Quantiques, Project E-QUBIT-FLY, No. ANR-22-PETQ-0012.
\end{acknowledgments}
%\nocite{*}
\bibliography{main_bib}% Produces the bibliography via BibTeX.

%merlin.mbs apsrev4-1.bst 2010-07-25 4.21a (PWD, AO, DPC) hacked
%Control: key (0)
%Control: author (72) initials jnrlst
%Control: editor formatted (1) identically to author
%Control: production of article title (-1) disabled
%Control: page (0) single
%Control: year (1) truncated
%Control: production of eprint (0) enabled
\providecommand{\noopsort}[1]{}\providecommand{\singleletter}[1]{#1}%
\begin{thebibliography}{97}%
\makeatletter
\providecommand \@ifxundefined [1]{%
 \@ifx{#1\undefined}
}%
\providecommand \@ifnum [1]{%
 \ifnum #1\expandafter \@firstoftwo
 \else \expandafter \@secondoftwo
 \fi
}%
\providecommand \@ifx [1]{%
 \ifx #1\expandafter \@firstoftwo
 \else \expandafter \@secondoftwo
 \fi
}%
\providecommand \natexlab [1]{#1}%
\providecommand \enquote  [1]{``#1''}%
\providecommand \bibnamefont  [1]{#1}%
\providecommand \bibfnamefont [1]{#1}%
\providecommand \citenamefont [1]{#1}%
\providecommand \href@noop [0]{\@secondoftwo}%
\providecommand \href [0]{\begingroup \@sanitize@url \@href}%
\providecommand \@href[1]{\@@startlink{#1}\@@href}%
\providecommand \@@href[1]{\endgroup#1\@@endlink}%
\providecommand \@sanitize@url [0]{\catcode `\\12\catcode `\$12\catcode
  `\&12\catcode `\#12\catcode `\^12\catcode `\_12\catcode `\%12\relax}%
\providecommand \@@startlink[1]{}%
\providecommand \@@endlink[0]{}%
\providecommand \url  [0]{\begingroup\@sanitize@url \@url }%
\providecommand \@url [1]{\endgroup\@href {#1}{\urlprefix }}%
\providecommand \urlprefix  [0]{URL }%
\providecommand \Eprint [0]{\href }%
\providecommand \doibase [0]{http://dx.doi.org/}%
\providecommand \selectlanguage [0]{\@gobble}%
\providecommand \bibinfo  [0]{\@secondoftwo}%
\providecommand \bibfield  [0]{\@secondoftwo}%
\providecommand \translation [1]{[#1]}%
\providecommand \BibitemOpen [0]{}%
\providecommand \bibitemStop [0]{}%
\providecommand \bibitemNoStop [0]{.\EOS\space}%
\providecommand \EOS [0]{\spacefactor3000\relax}%
\providecommand \BibitemShut  [1]{\csname bibitem#1\endcsname}%
\let\auto@bib@innerbib\@empty
%</preamble>
\bibitem [{\citenamefont {Lee}\ and\ \citenamefont {Levitov}(1993)}]{lee1993}%
  \BibitemOpen
  \bibfield  {author} {\bibinfo {author} {\bibfnamefont {H.}~\bibnamefont
  {Lee}}\ and\ \bibinfo {author} {\bibfnamefont {L.~S.}\ \bibnamefont
  {Levitov}},\ }\href@noop {} {\enquote {\bibinfo {title} {Orthogonality
  catastrophe in a mesoscopic conductor due to a time-dependent flux},}\ }
  (\bibinfo {year} {1993}),\ \bibinfo {note}
  {arXiv:cond-mat/9312013}\BibitemShut {NoStop}%
\bibitem [{\citenamefont {Levitov}\ \emph {et~al.}(1996)\citenamefont
  {Levitov}, \citenamefont {Lee},\ and\ \citenamefont {Lesovik}}]{levitov1996}%
  \BibitemOpen
  \bibfield  {author} {\bibinfo {author} {\bibfnamefont {L.~S.}\ \bibnamefont
  {Levitov}}, \bibinfo {author} {\bibfnamefont {H.}~\bibnamefont {Lee}}, \ and\
  \bibinfo {author} {\bibfnamefont {G.~B.}\ \bibnamefont {Lesovik}},\ }\href
  {\doibase 10.1063/1.531672} {\bibfield  {journal} {\bibinfo  {journal}
  {Journal of Mathematical Physics}\ }\textbf {\bibinfo {volume} {37}},\
  \bibinfo {pages} {4845} (\bibinfo {year} {1996})}\BibitemShut {NoStop}%
\bibitem [{\citenamefont {Anderson}(1967)}]{anderson1967}%
  \BibitemOpen
  \bibfield  {author} {\bibinfo {author} {\bibfnamefont {P.~W.}\ \bibnamefont
  {Anderson}},\ }\href {\doibase 10.1103/PhysRevLett.18.1049} {\bibfield
  {journal} {\bibinfo  {journal} {Phys. Rev. Lett.}\ }\textbf {\bibinfo
  {volume} {18}},\ \bibinfo {pages} {1049} (\bibinfo {year}
  {1967})}\BibitemShut {NoStop}%
\bibitem [{\citenamefont {Brown}\ and\ \citenamefont
  {Twiss}(1956)}]{brown1956}%
  \BibitemOpen
  \bibfield  {author} {\bibinfo {author} {\bibfnamefont {R.~H.}\ \bibnamefont
  {Brown}}\ and\ \bibinfo {author} {\bibfnamefont {R.~Q.}\ \bibnamefont
  {Twiss}},\ }\href@noop {} {\bibfield  {journal} {\bibinfo  {journal}
  {Nature}\ }\textbf {\bibinfo {volume} {177}},\ \bibinfo {pages} {27}
  (\bibinfo {year} {1956})}\BibitemShut {NoStop}%
\bibitem [{\citenamefont {Hong}\ \emph {et~al.}(1987)\citenamefont {Hong},
  \citenamefont {Ou},\ and\ \citenamefont {Mandel}}]{hong1987}%
  \BibitemOpen
  \bibfield  {author} {\bibinfo {author} {\bibfnamefont {C.~K.}\ \bibnamefont
  {Hong}}, \bibinfo {author} {\bibfnamefont {Z.~Y.}\ \bibnamefont {Ou}}, \ and\
  \bibinfo {author} {\bibfnamefont {L.}~\bibnamefont {Mandel}},\ }\href
  {\doibase 10.1103/PhysRevLett.59.2044} {\bibfield  {journal} {\bibinfo
  {journal} {Phys. Rev. Lett.}\ }\textbf {\bibinfo {volume} {59}},\ \bibinfo
  {pages} {2044} (\bibinfo {year} {1987})}\BibitemShut {NoStop}%
\bibitem [{\citenamefont {Klitzing}\ \emph {et~al.}(1980)\citenamefont
  {Klitzing}, \citenamefont {Dorda},\ and\ \citenamefont
  {Pepper}}]{klitzing1980}%
  \BibitemOpen
  \bibfield  {author} {\bibinfo {author} {\bibfnamefont {K.~v.}\ \bibnamefont
  {Klitzing}}, \bibinfo {author} {\bibfnamefont {G.}~\bibnamefont {Dorda}}, \
  and\ \bibinfo {author} {\bibfnamefont {M.}~\bibnamefont {Pepper}},\ }\href
  {\doibase 10.1103/PhysRevLett.45.494} {\bibfield  {journal} {\bibinfo
  {journal} {Phys. Rev. Lett.}\ }\textbf {\bibinfo {volume} {45}},\ \bibinfo
  {pages} {494} (\bibinfo {year} {1980})}\BibitemShut {NoStop}%
\bibitem [{\citenamefont {Dubois}\ \emph
  {et~al.}(2013{\natexlab{a}})\citenamefont {Dubois}, \citenamefont {Jullien},
  \citenamefont {Portier}, \citenamefont {Roche}, \citenamefont {Cavanna},
  \citenamefont {Jin}, \citenamefont {Wegscheider}, \citenamefont {Roulleau},\
  and\ \citenamefont {Glattli}}]{dubois2013b}%
  \BibitemOpen
  \bibfield  {author} {\bibinfo {author} {\bibfnamefont {J.}~\bibnamefont
  {Dubois}}, \bibinfo {author} {\bibfnamefont {T.}~\bibnamefont {Jullien}},
  \bibinfo {author} {\bibfnamefont {F.}~\bibnamefont {Portier}}, \bibinfo
  {author} {\bibfnamefont {P.}~\bibnamefont {Roche}}, \bibinfo {author}
  {\bibfnamefont {A.}~\bibnamefont {Cavanna}}, \bibinfo {author} {\bibfnamefont
  {Y.}~\bibnamefont {Jin}}, \bibinfo {author} {\bibfnamefont {W.}~\bibnamefont
  {Wegscheider}}, \bibinfo {author} {\bibfnamefont {P.}~\bibnamefont
  {Roulleau}}, \ and\ \bibinfo {author} {\bibfnamefont {D.~C.}\ \bibnamefont
  {Glattli}},\ }\href {\doibase 10.1038/nature12713} {\bibfield  {journal}
  {\bibinfo  {journal} {{Nature}}\ }\textbf {\bibinfo {volume} {502}},\
  \bibinfo {pages} {L659 } (\bibinfo {year} {2013}{\natexlab{a}})}\BibitemShut
  {NoStop}%
\bibitem [{\citenamefont {Grenier}\ \emph {et~al.}(2013)\citenamefont
  {Grenier}, \citenamefont {Dubois}, \citenamefont {Jullien}, \citenamefont
  {Roulleau}, \citenamefont {Glattli},\ and\ \citenamefont
  {Degiovanni}}]{grenier2013}%
  \BibitemOpen
  \bibfield  {author} {\bibinfo {author} {\bibfnamefont {C.}~\bibnamefont
  {Grenier}}, \bibinfo {author} {\bibfnamefont {J.}~\bibnamefont {Dubois}},
  \bibinfo {author} {\bibfnamefont {T.}~\bibnamefont {Jullien}}, \bibinfo
  {author} {\bibfnamefont {P.}~\bibnamefont {Roulleau}}, \bibinfo {author}
  {\bibfnamefont {D.~C.}\ \bibnamefont {Glattli}}, \ and\ \bibinfo {author}
  {\bibfnamefont {P.}~\bibnamefont {Degiovanni}},\ }\href {\doibase
  10.1103/PhysRevB.88.085302} {\bibfield  {journal} {\bibinfo  {journal} {Phys.
  Rev. B}\ }\textbf {\bibinfo {volume} {88}},\ \bibinfo {pages} {085302}
  (\bibinfo {year} {2013})}\BibitemShut {NoStop}%
\bibitem [{\citenamefont {Misiorny}\ \emph {et~al.}(2018)\citenamefont
  {Misiorny}, \citenamefont {F{\`e}ve},\ and\ \citenamefont
  {Splettstoesser}}]{misiorny2018}%
  \BibitemOpen
  \bibfield  {author} {\bibinfo {author} {\bibfnamefont {M.}~\bibnamefont
  {Misiorny}}, \bibinfo {author} {\bibfnamefont {G.}~\bibnamefont {F{\`e}ve}},
  \ and\ \bibinfo {author} {\bibfnamefont {J.}~\bibnamefont {Splettstoesser}},\
  }\href@noop {} {\bibfield  {journal} {\bibinfo  {journal} {Physical Review
  B}\ }\textbf {\bibinfo {volume} {97}},\ \bibinfo {pages} {075426} (\bibinfo
  {year} {2018})}\BibitemShut {NoStop}%
\bibitem [{\citenamefont {Glattli}\ and\ \citenamefont
  {Roulleau}(2017)}]{glattli2016b}%
  \BibitemOpen
  \bibfield  {author} {\bibinfo {author} {\bibfnamefont {D.~C.}\ \bibnamefont
  {Glattli}}\ and\ \bibinfo {author} {\bibfnamefont {P.~S.}\ \bibnamefont
  {Roulleau}},\ }\href {\doibase https://doi.org/10.1002/pssb.201600650}
  {\bibfield  {journal} {\bibinfo  {journal} {Physica Status Solidi (b)}\
  }\textbf {\bibinfo {volume} {254}},\ \bibinfo {pages} {1600650} (\bibinfo
  {year} {2017})}\BibitemShut {NoStop}%
\bibitem [{\citenamefont {B\"auerle}\ \emph {et~al.}(2018)\citenamefont
  {B\"auerle}, \citenamefont {Glattli}, \citenamefont {Meunier}, \citenamefont
  {Portier}, \citenamefont {Roche}, \citenamefont {Roulleau}, \citenamefont
  {Takada},\ and\ \citenamefont {Waintal}}]{bauerle2018}%
  \BibitemOpen
  \bibfield  {author} {\bibinfo {author} {\bibfnamefont {C.}~\bibnamefont
  {B\"auerle}}, \bibinfo {author} {\bibfnamefont {D.~C.}\ \bibnamefont
  {Glattli}}, \bibinfo {author} {\bibfnamefont {T.}~\bibnamefont {Meunier}},
  \bibinfo {author} {\bibfnamefont {F.}~\bibnamefont {Portier}}, \bibinfo
  {author} {\bibfnamefont {P.}~\bibnamefont {Roche}}, \bibinfo {author}
  {\bibfnamefont {P.}~\bibnamefont {Roulleau}}, \bibinfo {author}
  {\bibfnamefont {S.}~\bibnamefont {Takada}}, \ and\ \bibinfo {author}
  {\bibfnamefont {X.}~\bibnamefont {Waintal}},\ }\href {\doibase
  10.1088/1361-6633/aaa98a} {\bibfield  {journal} {\bibinfo  {journal} {Reports
  on Progress in Physics}\ }\textbf {\bibinfo {volume} {81}},\ \bibinfo {pages}
  {056503} (\bibinfo {year} {2018})}\BibitemShut {NoStop}%
\bibitem [{\citenamefont {Glattli}\ and\ \citenamefont
  {Roulleau}(2016)}]{glattli2016a}%
  \BibitemOpen
  \bibfield  {author} {\bibinfo {author} {\bibfnamefont {D.}~\bibnamefont
  {Glattli}}\ and\ \bibinfo {author} {\bibfnamefont {P.}~\bibnamefont
  {Roulleau}},\ }\href {\doibase https://doi.org/10.1016/j.physe.2015.10.034}
  {\bibfield  {journal} {\bibinfo  {journal} {Physica E: Low-dimensional
  Systems and Nanostructures}\ }\textbf {\bibinfo {volume} {76}},\ \bibinfo
  {pages} {216} (\bibinfo {year} {2016})}\BibitemShut {NoStop}%
\bibitem [{\citenamefont {Ferraro}\ \emph {et~al.}(2018)\citenamefont
  {Ferraro}, \citenamefont {Ronetti}, \citenamefont {Vannucci}, \citenamefont
  {Acciai}, \citenamefont {Rech}, \citenamefont {Jockheere}, \citenamefont
  {Martin},\ and\ \citenamefont {Sassetti}}]{ferraro2018}%
  \BibitemOpen
  \bibfield  {author} {\bibinfo {author} {\bibfnamefont {D.}~\bibnamefont
  {Ferraro}}, \bibinfo {author} {\bibfnamefont {F.}~\bibnamefont {Ronetti}},
  \bibinfo {author} {\bibfnamefont {L.}~\bibnamefont {Vannucci}}, \bibinfo
  {author} {\bibfnamefont {M.}~\bibnamefont {Acciai}}, \bibinfo {author}
  {\bibfnamefont {J.}~\bibnamefont {Rech}}, \bibinfo {author} {\bibfnamefont
  {T.}~\bibnamefont {Jockheere}}, \bibinfo {author} {\bibfnamefont
  {T.}~\bibnamefont {Martin}}, \ and\ \bibinfo {author} {\bibfnamefont
  {M.}~\bibnamefont {Sassetti}},\ }\href {\doibase
  10.1140/epjst/e2018-800074-1} {\bibfield  {journal} {\bibinfo  {journal} {The
  European Physical Journal Special Topics}\ }\textbf {\bibinfo {volume}
  {227}},\ \bibinfo {pages} {1345} (\bibinfo {year} {2018})}\BibitemShut
  {NoStop}%
\bibitem [{\citenamefont {Moskalets}(2016)}]{moskalets2016}%
  \BibitemOpen
  \bibfield  {author} {\bibinfo {author} {\bibfnamefont {M.}~\bibnamefont
  {Moskalets}},\ }\href {\doibase 10.1103/PhysRevLett.117.046801} {\bibfield
  {journal} {\bibinfo  {journal} {Phys. Rev. Lett.}\ }\textbf {\bibinfo
  {volume} {117}},\ \bibinfo {pages} {046801} (\bibinfo {year}
  {2016})}\BibitemShut {NoStop}%
\bibitem [{\citenamefont {Ivanov}\ \emph {et~al.}(1997)\citenamefont {Ivanov},
  \citenamefont {Lee},\ and\ \citenamefont {Levitov}}]{ivanov1997}%
  \BibitemOpen
  \bibfield  {author} {\bibinfo {author} {\bibfnamefont {D.~A.}\ \bibnamefont
  {Ivanov}}, \bibinfo {author} {\bibfnamefont {H.~W.}\ \bibnamefont {Lee}}, \
  and\ \bibinfo {author} {\bibfnamefont {L.~S.}\ \bibnamefont {Levitov}},\
  }\href {\doibase 10.1103/PhysRevB.56.6839} {\bibfield  {journal} {\bibinfo
  {journal} {Phys. Rev. B}\ }\textbf {\bibinfo {volume} {56}},\ \bibinfo
  {pages} {6839} (\bibinfo {year} {1997})}\BibitemShut {NoStop}%
\bibitem [{\citenamefont {Keeling}\ \emph {et~al.}(2006)\citenamefont
  {Keeling}, \citenamefont {Klich},\ and\ \citenamefont
  {Levitov}}]{keeling2006}%
  \BibitemOpen
  \bibfield  {author} {\bibinfo {author} {\bibfnamefont {J.}~\bibnamefont
  {Keeling}}, \bibinfo {author} {\bibfnamefont {I.}~\bibnamefont {Klich}}, \
  and\ \bibinfo {author} {\bibfnamefont {L.~S.}\ \bibnamefont {Levitov}},\
  }\href {\doibase 10.1103/PhysRevLett.97.116403} {\bibfield  {journal}
  {\bibinfo  {journal} {Phys. Rev. Lett.}\ }\textbf {\bibinfo {volume} {97}},\
  \bibinfo {pages} {116403} (\bibinfo {year} {2006})}\BibitemShut {NoStop}%
\bibitem [{\citenamefont {Moskalets}(2015)}]{moskalets2015}%
  \BibitemOpen
  \bibfield  {author} {\bibinfo {author} {\bibfnamefont {M.}~\bibnamefont
  {Moskalets}},\ }\href {\doibase 10.1103/PhysRevB.91.195431} {\bibfield
  {journal} {\bibinfo  {journal} {Phys. Rev. B}\ }\textbf {\bibinfo {volume}
  {91}},\ \bibinfo {pages} {195431} (\bibinfo {year} {2015})}\BibitemShut
  {NoStop}%
\bibitem [{\citenamefont {Glattli}\ and\ \citenamefont
  {Roulleau}(2018)}]{glattli2018}%
  \BibitemOpen
  \bibfield  {author} {\bibinfo {author} {\bibfnamefont {D.~C.}\ \bibnamefont
  {Glattli}}\ and\ \bibinfo {author} {\bibfnamefont {P.}~\bibnamefont
  {Roulleau}},\ }\href {\doibase 10.1103/PhysRevB.97.125407} {\bibfield
  {journal} {\bibinfo  {journal} {Phys. Rev. B}\ }\textbf {\bibinfo {volume}
  {97}},\ \bibinfo {pages} {125407} (\bibinfo {year} {2018})}\BibitemShut
  {NoStop}%
\bibitem [{\citenamefont {Dasenbrook}\ and\ \citenamefont
  {Flindt}(2015)}]{dasenbrook2015}%
  \BibitemOpen
  \bibfield  {author} {\bibinfo {author} {\bibfnamefont {D.}~\bibnamefont
  {Dasenbrook}}\ and\ \bibinfo {author} {\bibfnamefont {C.}~\bibnamefont
  {Flindt}},\ }\href {\doibase 10.1103/PhysRevB.92.161412} {\bibfield
  {journal} {\bibinfo  {journal} {Phys. Rev. B}\ }\textbf {\bibinfo {volume}
  {92}},\ \bibinfo {pages} {161412} (\bibinfo {year} {2015})}\BibitemShut
  {NoStop}%
\bibitem [{\citenamefont {Dasenbrook}\ \emph {et~al.}(2016)\citenamefont
  {Dasenbrook}, \citenamefont {Bowles}, \citenamefont {Brask}, \citenamefont
  {Hofer}, \citenamefont {Flindt},\ and\ \citenamefont
  {Brunner}}]{dasenbrook2016b}%
  \BibitemOpen
  \bibfield  {author} {\bibinfo {author} {\bibfnamefont {D.}~\bibnamefont
  {Dasenbrook}}, \bibinfo {author} {\bibfnamefont {J.}~\bibnamefont {Bowles}},
  \bibinfo {author} {\bibfnamefont {J.~B.}\ \bibnamefont {Brask}}, \bibinfo
  {author} {\bibfnamefont {P.~P.}\ \bibnamefont {Hofer}}, \bibinfo {author}
  {\bibfnamefont {C.}~\bibnamefont {Flindt}}, \ and\ \bibinfo {author}
  {\bibfnamefont {N.}~\bibnamefont {Brunner}},\ }\href {\doibase
  10.1088/1367-2630/18/4/043036} {\bibfield  {journal} {\bibinfo  {journal}
  {New Journal of Physics}\ }\textbf {\bibinfo {volume} {18}},\ \bibinfo
  {pages} {043036} (\bibinfo {year} {2016})}\BibitemShut {NoStop}%
\bibitem [{\citenamefont {Dasenbrook}\ and\ \citenamefont
  {Flindt}(2016)}]{dasenbrook2016}%
  \BibitemOpen
  \bibfield  {author} {\bibinfo {author} {\bibfnamefont {D.}~\bibnamefont
  {Dasenbrook}}\ and\ \bibinfo {author} {\bibfnamefont {C.}~\bibnamefont
  {Flindt}},\ }\href {\doibase 10.1103/PhysRevLett.117.146801} {\bibfield
  {journal} {\bibinfo  {journal} {Phys. Rev. Lett.}\ }\textbf {\bibinfo
  {volume} {117}},\ \bibinfo {pages} {146801} (\bibinfo {year}
  {2016})}\BibitemShut {NoStop}%
\bibitem [{\citenamefont {Grenier}\ \emph {et~al.}(2011)\citenamefont
  {Grenier}, \citenamefont {Herv{\'e}}, \citenamefont {Bocquillon},
  \citenamefont {Parmentier}, \citenamefont {Pla\c{c}ais}, \citenamefont
  {Berroir}, \citenamefont {F{\`e}ve},\ and\ \citenamefont
  {Degiovanni}}]{grenier2011}%
  \BibitemOpen
  \bibfield  {author} {\bibinfo {author} {\bibfnamefont {C.}~\bibnamefont
  {Grenier}}, \bibinfo {author} {\bibfnamefont {R.}~\bibnamefont {Herv{\'e}}},
  \bibinfo {author} {\bibfnamefont {E.}~\bibnamefont {Bocquillon}}, \bibinfo
  {author} {\bibfnamefont {F.~D.}\ \bibnamefont {Parmentier}}, \bibinfo
  {author} {\bibfnamefont {B.}~\bibnamefont {Pla\c{c}ais}}, \bibinfo {author}
  {\bibfnamefont {J.-M.}\ \bibnamefont {Berroir}}, \bibinfo {author}
  {\bibfnamefont {G.}~\bibnamefont {F{\`e}ve}}, \ and\ \bibinfo {author}
  {\bibfnamefont {P.}~\bibnamefont {Degiovanni}},\ }\href@noop {} {\bibfield
  {journal} {\bibinfo  {journal} {New J. of Phys.}\ }\textbf {\bibinfo {volume}
  {13}},\ \bibinfo {pages} {093007} (\bibinfo {year} {2011})}\BibitemShut
  {NoStop}%
\bibitem [{\citenamefont {Ferraro}\ \emph {et~al.}(2013)\citenamefont
  {Ferraro}, \citenamefont {Feller}, \citenamefont {Ghibaudo}, \citenamefont
  {Thibierge}, \citenamefont {Bocquillon}, \citenamefont {F{\`e}ve},
  \citenamefont {Grenier},\ and\ \citenamefont {Degiovanni}}]{ferraro2013}%
  \BibitemOpen
  \bibfield  {author} {\bibinfo {author} {\bibfnamefont {D.}~\bibnamefont
  {Ferraro}}, \bibinfo {author} {\bibfnamefont {A.}~\bibnamefont {Feller}},
  \bibinfo {author} {\bibfnamefont {A.}~\bibnamefont {Ghibaudo}}, \bibinfo
  {author} {\bibfnamefont {E.}~\bibnamefont {Thibierge}}, \bibinfo {author}
  {\bibfnamefont {E.}~\bibnamefont {Bocquillon}}, \bibinfo {author}
  {\bibfnamefont {G.}~\bibnamefont {F{\`e}ve}}, \bibinfo {author}
  {\bibfnamefont {C.}~\bibnamefont {Grenier}}, \ and\ \bibinfo {author}
  {\bibfnamefont {P.}~\bibnamefont {Degiovanni}},\ }\href@noop {} {\bibfield
  {journal} {\bibinfo  {journal} {Physical Review B}\ }\textbf {\bibinfo
  {volume} {88}},\ \bibinfo {pages} {205303} (\bibinfo {year}
  {2013})}\BibitemShut {NoStop}%
\bibitem [{\citenamefont {Ferraro}\ \emph {et~al.}(2014)\citenamefont
  {Ferraro}, \citenamefont {Roussel}, \citenamefont {Cabart}, \citenamefont
  {Thibierge}, \citenamefont {F\`eve}, \citenamefont {Grenier},\ and\
  \citenamefont {Degiovanni}}]{ferraro2014}%
  \BibitemOpen
  \bibfield  {author} {\bibinfo {author} {\bibfnamefont {D.}~\bibnamefont
  {Ferraro}}, \bibinfo {author} {\bibfnamefont {B.}~\bibnamefont {Roussel}},
  \bibinfo {author} {\bibfnamefont {C.}~\bibnamefont {Cabart}}, \bibinfo
  {author} {\bibfnamefont {E.}~\bibnamefont {Thibierge}}, \bibinfo {author}
  {\bibfnamefont {G.}~\bibnamefont {F\`eve}}, \bibinfo {author} {\bibfnamefont
  {C.}~\bibnamefont {Grenier}}, \ and\ \bibinfo {author} {\bibfnamefont
  {P.}~\bibnamefont {Degiovanni}},\ }\href {\doibase
  10.1103/PhysRevLett.113.166403} {\bibfield  {journal} {\bibinfo  {journal}
  {Phys. Rev. Lett.}\ }\textbf {\bibinfo {volume} {113}},\ \bibinfo {pages}
  {166403} (\bibinfo {year} {2014})}\BibitemShut {NoStop}%
\bibitem [{\citenamefont {Jullien}\ \emph {et~al.}(2014)\citenamefont
  {Jullien}, \citenamefont {Roulleau}, \citenamefont {Roche}, \citenamefont
  {Cavanna}, \citenamefont {Jin},\ and\ \citenamefont {Glattli}}]{jullien2014}%
  \BibitemOpen
  \bibfield  {author} {\bibinfo {author} {\bibfnamefont {T.}~\bibnamefont
  {Jullien}}, \bibinfo {author} {\bibfnamefont {P.}~\bibnamefont {Roulleau}},
  \bibinfo {author} {\bibfnamefont {B.}~\bibnamefont {Roche}}, \bibinfo
  {author} {\bibfnamefont {A.}~\bibnamefont {Cavanna}}, \bibinfo {author}
  {\bibfnamefont {Y.}~\bibnamefont {Jin}}, \ and\ \bibinfo {author}
  {\bibfnamefont {D.~C.}\ \bibnamefont {Glattli}},\ }\href {\doibase
  10.1038/nature13821} {\bibfield  {journal} {\bibinfo  {journal} {Nature}\
  }\textbf {\bibinfo {volume} {514}},\ \bibinfo {pages} {603} (\bibinfo {year}
  {2014})}\BibitemShut {NoStop}%
\bibitem [{\citenamefont {Bocquillon}\ \emph {et~al.}(2012)\citenamefont
  {Bocquillon}, \citenamefont {Parmentier}, \citenamefont {Grenier},
  \citenamefont {Berroir}, \citenamefont {Degiovanni}, \citenamefont {Glattli},
  \citenamefont {Pla\ifmmode~\mbox{\c{c}}\else \c{c}\fi{}ais}, \citenamefont
  {Cavanna}, \citenamefont {Jin},\ and\ \citenamefont
  {F\`eve}}]{bocquillon2012}%
  \BibitemOpen
  \bibfield  {author} {\bibinfo {author} {\bibfnamefont {E.}~\bibnamefont
  {Bocquillon}}, \bibinfo {author} {\bibfnamefont {F.~D.}\ \bibnamefont
  {Parmentier}}, \bibinfo {author} {\bibfnamefont {C.}~\bibnamefont {Grenier}},
  \bibinfo {author} {\bibfnamefont {J.-M.}\ \bibnamefont {Berroir}}, \bibinfo
  {author} {\bibfnamefont {P.}~\bibnamefont {Degiovanni}}, \bibinfo {author}
  {\bibfnamefont {D.~C.}\ \bibnamefont {Glattli}}, \bibinfo {author}
  {\bibfnamefont {B.}~\bibnamefont {Pla\ifmmode~\mbox{\c{c}}\else
  \c{c}\fi{}ais}}, \bibinfo {author} {\bibfnamefont {A.}~\bibnamefont
  {Cavanna}}, \bibinfo {author} {\bibfnamefont {Y.}~\bibnamefont {Jin}}, \ and\
  \bibinfo {author} {\bibfnamefont {G.}~\bibnamefont {F\`eve}},\ }\href
  {\doibase 10.1103/PhysRevLett.108.196803} {\bibfield  {journal} {\bibinfo
  {journal} {Phys. Rev. Lett.}\ }\textbf {\bibinfo {volume} {108}},\ \bibinfo
  {pages} {196803} (\bibinfo {year} {2012})}\BibitemShut {NoStop}%
\bibitem [{\citenamefont {Bocquillon}\ \emph {et~al.}(2013)\citenamefont
  {Bocquillon}, \citenamefont {Freulon}, \citenamefont {Berroir}, \citenamefont
  {Degiovanni}, \citenamefont {Pla\c{c}ais}, \citenamefont {Cavanna},
  \citenamefont {Jin},\ and\ \citenamefont {F\`eve}}]{bocquillon2013}%
  \BibitemOpen
  \bibfield  {author} {\bibinfo {author} {\bibfnamefont {E.}~\bibnamefont
  {Bocquillon}}, \bibinfo {author} {\bibfnamefont {V.}~\bibnamefont {Freulon}},
  \bibinfo {author} {\bibfnamefont {J.-M.}\ \bibnamefont {Berroir}}, \bibinfo
  {author} {\bibfnamefont {P.}~\bibnamefont {Degiovanni}}, \bibinfo {author}
  {\bibfnamefont {B.}~\bibnamefont {Pla\c{c}ais}}, \bibinfo {author}
  {\bibfnamefont {A.}~\bibnamefont {Cavanna}}, \bibinfo {author} {\bibfnamefont
  {Y.}~\bibnamefont {Jin}}, \ and\ \bibinfo {author} {\bibfnamefont
  {G.}~\bibnamefont {F\`eve}},\ }\href {\doibase 10.1126/science.1232572}
  {\bibfield  {journal} {\bibinfo  {journal} {Science}\ }\textbf {\bibinfo
  {volume} {339}},\ \bibinfo {pages} {1054} (\bibinfo {year}
  {2013})}\BibitemShut {NoStop}%
\bibitem [{\citenamefont {Jonckheere}\ \emph {et~al.}(2012)\citenamefont
  {Jonckheere}, \citenamefont {Rech}, \citenamefont {Wahl},\ and\ \citenamefont
  {Martin}}]{jonckheere2012}%
  \BibitemOpen
  \bibfield  {author} {\bibinfo {author} {\bibfnamefont {T.}~\bibnamefont
  {Jonckheere}}, \bibinfo {author} {\bibfnamefont {J.}~\bibnamefont {Rech}},
  \bibinfo {author} {\bibfnamefont {C.}~\bibnamefont {Wahl}}, \ and\ \bibinfo
  {author} {\bibfnamefont {T.}~\bibnamefont {Martin}},\ }\href@noop {}
  {\bibfield  {journal} {\bibinfo  {journal} {Phys. Rev. B}\ }\textbf {\bibinfo
  {volume} {86}},\ \bibinfo {pages} {125425} (\bibinfo {year}
  {2012})}\BibitemShut {NoStop}%
\bibitem [{\citenamefont {Ferraro}\ \emph {et~al.}(2015)\citenamefont
  {Ferraro}, \citenamefont {Rech}, \citenamefont {Jonckheere},\ and\
  \citenamefont {Martin}}]{ferraro2015}%
  \BibitemOpen
  \bibfield  {author} {\bibinfo {author} {\bibfnamefont {D.}~\bibnamefont
  {Ferraro}}, \bibinfo {author} {\bibfnamefont {J.}~\bibnamefont {Rech}},
  \bibinfo {author} {\bibfnamefont {T.}~\bibnamefont {Jonckheere}}, \ and\
  \bibinfo {author} {\bibfnamefont {T.}~\bibnamefont {Martin}},\ }\href
  {\doibase 10.1103/PhysRevB.91.205409} {\bibfield  {journal} {\bibinfo
  {journal} {Phys. Rev. B}\ }\textbf {\bibinfo {volume} {91}},\ \bibinfo
  {pages} {205409} (\bibinfo {year} {2015})}\BibitemShut {NoStop}%
\bibitem [{\citenamefont {Blanter}\ and\ \citenamefont
  {B{\"u}ttiker}(2000)}]{blanter2000}%
  \BibitemOpen
  \bibfield  {author} {\bibinfo {author} {\bibfnamefont {Y.~M.}\ \bibnamefont
  {Blanter}}\ and\ \bibinfo {author} {\bibfnamefont {M.}~\bibnamefont
  {B{\"u}ttiker}},\ }\href@noop {} {\bibfield  {journal} {\bibinfo  {journal}
  {Physics reports}\ }\textbf {\bibinfo {volume} {336}},\ \bibinfo {pages} {1}
  (\bibinfo {year} {2000})}\BibitemShut {NoStop}%
\bibitem [{\citenamefont {Martin}(2005)}]{martin2005}%
  \BibitemOpen
  \bibfield  {author} {\bibinfo {author} {\bibfnamefont {T.}~\bibnamefont
  {Martin}},\ }\href {\doibase 10.1016/S0924-8099(05)80047-2} {\bibfield
  {journal} {\bibinfo  {journal} {Les Houches Summer School Proceedings}\
  }\textbf {\bibinfo {volume} {81}},\ \bibinfo {pages} {283} (\bibinfo {year}
  {2005})}\BibitemShut {NoStop}%
\bibitem [{\citenamefont {Nazarov}\ and\ \citenamefont
  {Blanter}(2009)}]{nazarov2009}%
  \BibitemOpen
  \bibfield  {author} {\bibinfo {author} {\bibfnamefont {Y.~V.}\ \bibnamefont
  {Nazarov}}\ and\ \bibinfo {author} {\bibfnamefont {Y.~M.}\ \bibnamefont
  {Blanter}},\ }\href@noop {} {\emph {\bibinfo {title} {Quantum transport:
  introduction to nanoscience}}}\ (\bibinfo  {publisher} {Cambridge university
  press},\ \bibinfo {year} {2009})\BibitemShut {NoStop}%
\bibitem [{\citenamefont {Lesovik}\ and\ \citenamefont
  {Sadovskyy}(2011)}]{lesovik2011}%
  \BibitemOpen
  \bibfield  {author} {\bibinfo {author} {\bibfnamefont {G.~B.}\ \bibnamefont
  {Lesovik}}\ and\ \bibinfo {author} {\bibfnamefont {I.~A.}\ \bibnamefont
  {Sadovskyy}},\ }\href@noop {} {\bibfield  {journal} {\bibinfo  {journal}
  {Physics-Uspekhi}\ }\textbf {\bibinfo {volume} {54}},\ \bibinfo {pages}
  {1007} (\bibinfo {year} {2011})}\BibitemShut {NoStop}%
\bibitem [{\citenamefont {Dubois}\ \emph
  {et~al.}(2013{\natexlab{b}})\citenamefont {Dubois}, \citenamefont {Jullien},
  \citenamefont {Portier}, \citenamefont {Roche}, \citenamefont {Cavanna},
  \citenamefont {Jin}, \citenamefont {Wegscheider}, \citenamefont {Roulleau},\
  and\ \citenamefont {Glattli}}]{dubois2013}%
  \BibitemOpen
  \bibfield  {author} {\bibinfo {author} {\bibfnamefont {J.}~\bibnamefont
  {Dubois}}, \bibinfo {author} {\bibfnamefont {T.}~\bibnamefont {Jullien}},
  \bibinfo {author} {\bibfnamefont {F.}~\bibnamefont {Portier}}, \bibinfo
  {author} {\bibfnamefont {P.}~\bibnamefont {Roche}}, \bibinfo {author}
  {\bibfnamefont {A.}~\bibnamefont {Cavanna}}, \bibinfo {author} {\bibfnamefont
  {Y.}~\bibnamefont {Jin}}, \bibinfo {author} {\bibfnamefont {W.}~\bibnamefont
  {Wegscheider}}, \bibinfo {author} {\bibfnamefont {P.}~\bibnamefont
  {Roulleau}}, \ and\ \bibinfo {author} {\bibfnamefont {D.}~\bibnamefont
  {Glattli}},\ }\href@noop {} {\bibfield  {journal} {\bibinfo  {journal}
  {Nature}\ }\textbf {\bibinfo {volume} {502}},\ \bibinfo {pages} {659}
  (\bibinfo {year} {2013}{\natexlab{b}})}\BibitemShut {NoStop}%
\bibitem [{\citenamefont {Rech}\ \emph {et~al.}(2017)\citenamefont {Rech},
  \citenamefont {Ferraro}, \citenamefont {Jonckheere}, \citenamefont
  {Vannucci}, \citenamefont {Sassetti},\ and\ \citenamefont
  {Martin}}]{rech2017}%
  \BibitemOpen
  \bibfield  {author} {\bibinfo {author} {\bibfnamefont {J.}~\bibnamefont
  {Rech}}, \bibinfo {author} {\bibfnamefont {D.}~\bibnamefont {Ferraro}},
  \bibinfo {author} {\bibfnamefont {T.}~\bibnamefont {Jonckheere}}, \bibinfo
  {author} {\bibfnamefont {L.}~\bibnamefont {Vannucci}}, \bibinfo {author}
  {\bibfnamefont {M.}~\bibnamefont {Sassetti}}, \ and\ \bibinfo {author}
  {\bibfnamefont {T.}~\bibnamefont {Martin}},\ }\href {\doibase
  10.1103/PhysRevLett.118.076801} {\bibfield  {journal} {\bibinfo  {journal}
  {Phys. Rev. Lett.}\ }\textbf {\bibinfo {volume} {118}},\ \bibinfo {pages}
  {076801} (\bibinfo {year} {2017})}\BibitemShut {NoStop}%
\bibitem [{\citenamefont {Tsui}\ \emph {et~al.}(1982)\citenamefont {Tsui},
  \citenamefont {Stormer},\ and\ \citenamefont {Gossard}}]{tsui1982}%
  \BibitemOpen
  \bibfield  {author} {\bibinfo {author} {\bibfnamefont {D.~C.}\ \bibnamefont
  {Tsui}}, \bibinfo {author} {\bibfnamefont {H.~L.}\ \bibnamefont {Stormer}}, \
  and\ \bibinfo {author} {\bibfnamefont {A.~C.}\ \bibnamefont {Gossard}},\
  }\href {\doibase 10.1103/PhysRevLett.48.1559} {\bibfield  {journal} {\bibinfo
   {journal} {Phys. Rev. Lett.}\ }\textbf {\bibinfo {volume} {48}},\ \bibinfo
  {pages} {1559} (\bibinfo {year} {1982})}\BibitemShut {NoStop}%
\bibitem [{\citenamefont {Laughlin}(1983)}]{laughlin1983}%
  \BibitemOpen
  \bibfield  {author} {\bibinfo {author} {\bibfnamefont {R.~B.}\ \bibnamefont
  {Laughlin}},\ }\href {\doibase 10.1103/PhysRevLett.50.1395} {\bibfield
  {journal} {\bibinfo  {journal} {Phys. Rev. Lett.}\ }\textbf {\bibinfo
  {volume} {50}},\ \bibinfo {pages} {1395} (\bibinfo {year}
  {1983})}\BibitemShut {NoStop}%
\bibitem [{\citenamefont {Nayak}\ \emph {et~al.}(2008)\citenamefont {Nayak},
  \citenamefont {Simon}, \citenamefont {Stern}, \citenamefont {Freedman},\ and\
  \citenamefont {Das~Sarma}}]{nayak2008}%
  \BibitemOpen
  \bibfield  {author} {\bibinfo {author} {\bibfnamefont {C.}~\bibnamefont
  {Nayak}}, \bibinfo {author} {\bibfnamefont {S.~H.}\ \bibnamefont {Simon}},
  \bibinfo {author} {\bibfnamefont {A.}~\bibnamefont {Stern}}, \bibinfo
  {author} {\bibfnamefont {M.}~\bibnamefont {Freedman}}, \ and\ \bibinfo
  {author} {\bibfnamefont {S.}~\bibnamefont {Das~Sarma}},\ }\href {\doibase
  10.1103/RevModPhys.80.1083} {\bibfield  {journal} {\bibinfo  {journal} {Rev.
  Mod. Phys.}\ }\textbf {\bibinfo {volume} {80}},\ \bibinfo {pages} {1083}
  (\bibinfo {year} {2008})}\BibitemShut {NoStop}%
\bibitem [{\citenamefont {Hashisaka}\ \emph {et~al.}(2021)\citenamefont
  {Hashisaka}, \citenamefont {Jonckheere}, \citenamefont {Akiho}, \citenamefont
  {Sasaki}, \citenamefont {Rech}, \citenamefont {Martin},\ and\ \citenamefont
  {Muraki}}]{hashisaka2021}%
  \BibitemOpen
  \bibfield  {author} {\bibinfo {author} {\bibfnamefont {M.}~\bibnamefont
  {Hashisaka}}, \bibinfo {author} {\bibfnamefont {T.}~\bibnamefont
  {Jonckheere}}, \bibinfo {author} {\bibfnamefont {T.}~\bibnamefont {Akiho}},
  \bibinfo {author} {\bibfnamefont {S.}~\bibnamefont {Sasaki}}, \bibinfo
  {author} {\bibfnamefont {J.}~\bibnamefont {Rech}}, \bibinfo {author}
  {\bibfnamefont {T.}~\bibnamefont {Martin}}, \ and\ \bibinfo {author}
  {\bibfnamefont {K.}~\bibnamefont {Muraki}},\ }\href {\doibase
  10.1038/s41467-021-23160-6} {\bibfield  {journal} {\bibinfo  {journal}
  {Nature Communications}\ }\textbf {\bibinfo {volume} {12}},\ \bibinfo {pages}
  {2794} (\bibinfo {year} {2021})}\BibitemShut {NoStop}%
\bibitem [{\citenamefont {Jonckheere}\ \emph {et~al.}(2023)\citenamefont
  {Jonckheere}, \citenamefont {Rech}, \citenamefont {Gr\'emaud},\ and\
  \citenamefont {Martin}}]{jonckheere2023}%
  \BibitemOpen
  \bibfield  {author} {\bibinfo {author} {\bibfnamefont {T.}~\bibnamefont
  {Jonckheere}}, \bibinfo {author} {\bibfnamefont {J.}~\bibnamefont {Rech}},
  \bibinfo {author} {\bibfnamefont {B.}~\bibnamefont {Gr\'emaud}}, \ and\
  \bibinfo {author} {\bibfnamefont {T.}~\bibnamefont {Martin}},\ }\href
  {\doibase 10.1103/PhysRevLett.130.186203} {\bibfield  {journal} {\bibinfo
  {journal} {Phys. Rev. Lett.}\ }\textbf {\bibinfo {volume} {130}},\ \bibinfo
  {pages} {186203} (\bibinfo {year} {2023})}\BibitemShut {NoStop}%
\bibitem [{\citenamefont {Glidic}\ \emph {et~al.}(2023)\citenamefont {Glidic},
  \citenamefont {Maillet}, \citenamefont {Piquard}, \citenamefont {Aassime},
  \citenamefont {Cavanna}, \citenamefont {Jin}, \citenamefont {Gennser},
  \citenamefont {Anthore},\ and\ \citenamefont {Pierre}}]{glidic2023}%
  \BibitemOpen
  \bibfield  {author} {\bibinfo {author} {\bibfnamefont {P.}~\bibnamefont
  {Glidic}}, \bibinfo {author} {\bibfnamefont {O.}~\bibnamefont {Maillet}},
  \bibinfo {author} {\bibfnamefont {C.}~\bibnamefont {Piquard}}, \bibinfo
  {author} {\bibfnamefont {A.}~\bibnamefont {Aassime}}, \bibinfo {author}
  {\bibfnamefont {A.}~\bibnamefont {Cavanna}}, \bibinfo {author} {\bibfnamefont
  {Y.}~\bibnamefont {Jin}}, \bibinfo {author} {\bibfnamefont {U.}~\bibnamefont
  {Gennser}}, \bibinfo {author} {\bibfnamefont {A.}~\bibnamefont {Anthore}}, \
  and\ \bibinfo {author} {\bibfnamefont {F.}~\bibnamefont {Pierre}},\ }\href
  {\doibase 10.1038/s41467-023-36080-4} {\bibfield  {journal} {\bibinfo
  {journal} {Nature Communications}\ }\textbf {\bibinfo {volume} {14}},\
  \bibinfo {pages} {514} (\bibinfo {year} {2023})}\BibitemShut {NoStop}%
\bibitem [{\citenamefont {Ronetti}\ \emph {et~al.}(2018)\citenamefont
  {Ronetti}, \citenamefont {Vannucci}, \citenamefont {Ferraro}, \citenamefont
  {Jonckheere}, \citenamefont {Rech}, \citenamefont {Martin},\ and\
  \citenamefont {Sassetti}}]{ronetti2018}%
  \BibitemOpen
  \bibfield  {author} {\bibinfo {author} {\bibfnamefont {F.}~\bibnamefont
  {Ronetti}}, \bibinfo {author} {\bibfnamefont {L.}~\bibnamefont {Vannucci}},
  \bibinfo {author} {\bibfnamefont {D.}~\bibnamefont {Ferraro}}, \bibinfo
  {author} {\bibfnamefont {T.}~\bibnamefont {Jonckheere}}, \bibinfo {author}
  {\bibfnamefont {J.}~\bibnamefont {Rech}}, \bibinfo {author} {\bibfnamefont
  {T.}~\bibnamefont {Martin}}, \ and\ \bibinfo {author} {\bibfnamefont
  {M.}~\bibnamefont {Sassetti}},\ }\href {\doibase 10.1103/PhysRevB.98.075401}
  {\bibfield  {journal} {\bibinfo  {journal} {Phys. Rev. B}\ }\textbf {\bibinfo
  {volume} {98}},\ \bibinfo {pages} {075401} (\bibinfo {year}
  {2018})}\BibitemShut {NoStop}%
\bibitem [{\citenamefont {Cr{\'e}pieux}\ \emph {et~al.}(2004)\citenamefont
  {Cr{\'e}pieux}, \citenamefont {Devillard},\ and\ \citenamefont
  {Martin}}]{crepieux2004}%
  \BibitemOpen
  \bibfield  {author} {\bibinfo {author} {\bibfnamefont {A.}~\bibnamefont
  {Cr{\'e}pieux}}, \bibinfo {author} {\bibfnamefont {P.}~\bibnamefont
  {Devillard}}, \ and\ \bibinfo {author} {\bibfnamefont {T.}~\bibnamefont
  {Martin}},\ }\href@noop {} {\bibfield  {journal} {\bibinfo  {journal} {Phys.
  Rev. B}\ }\textbf {\bibinfo {volume} {69}},\ \bibinfo {pages} {205302}
  (\bibinfo {year} {2004})}\BibitemShut {NoStop}%
\bibitem [{\citenamefont {Wen}(1995)}]{wen1995}%
  \BibitemOpen
  \bibfield  {author} {\bibinfo {author} {\bibfnamefont {X.-G.}\ \bibnamefont
  {Wen}},\ }\href {\doibase 10.1080/00018739500101566} {\bibfield  {journal}
  {\bibinfo  {journal} {Advances in Physics}\ }\textbf {\bibinfo {volume}
  {44}},\ \bibinfo {pages} {405} (\bibinfo {year} {1995})}\BibitemShut
  {NoStop}%
\bibitem [{\citenamefont {Kane}\ and\ \citenamefont {Fisher}(1992)}]{kane1992}%
  \BibitemOpen
  \bibfield  {author} {\bibinfo {author} {\bibfnamefont {C.~L.}\ \bibnamefont
  {Kane}}\ and\ \bibinfo {author} {\bibfnamefont {M.~P.~A.}\ \bibnamefont
  {Fisher}},\ }\href {\doibase 10.1103/PhysRevB.46.15233} {\bibfield  {journal}
  {\bibinfo  {journal} {Phys. Rev. B}\ }\textbf {\bibinfo {volume} {46}},\
  \bibinfo {pages} {15233} (\bibinfo {year} {1992})}\BibitemShut {NoStop}%
\bibitem [{\citenamefont {Kane}\ and\ \citenamefont {Fisher}(1994)}]{kane1994}%
  \BibitemOpen
  \bibfield  {author} {\bibinfo {author} {\bibfnamefont {C.~L.}\ \bibnamefont
  {Kane}}\ and\ \bibinfo {author} {\bibfnamefont {M.~P.~A.}\ \bibnamefont
  {Fisher}},\ }\href {\doibase 10.1103/PhysRevLett.72.724} {\bibfield
  {journal} {\bibinfo  {journal} {Phys. Rev. Lett.}\ }\textbf {\bibinfo
  {volume} {72}},\ \bibinfo {pages} {724} (\bibinfo {year} {1994})}\BibitemShut
  {NoStop}%
\bibitem [{\citenamefont {von Delft}\ and\ \citenamefont
  {Schoeller}(1998)}]{vondelft1998}%
  \BibitemOpen
  \bibfield  {author} {\bibinfo {author} {\bibfnamefont {J.}~\bibnamefont {von
  Delft}}\ and\ \bibinfo {author} {\bibfnamefont {H.}~\bibnamefont
  {Schoeller}},\ }\href {\doibase https://doi.org/10.1002/andp.19985100401}
  {\bibfield  {journal} {\bibinfo  {journal} {Annalen der Physik}\ }\textbf
  {\bibinfo {volume} {510}},\ \bibinfo {pages} {225} (\bibinfo {year}
  {1998})}\BibitemShut {NoStop}%
\bibitem [{\citenamefont {Guyon}\ \emph {et~al.}(2002)\citenamefont {Guyon},
  \citenamefont {Devillard}, \citenamefont {Martin},\ and\ \citenamefont
  {Safi}}]{guyon2002}%
  \BibitemOpen
  \bibfield  {author} {\bibinfo {author} {\bibfnamefont {R.}~\bibnamefont
  {Guyon}}, \bibinfo {author} {\bibfnamefont {P.}~\bibnamefont {Devillard}},
  \bibinfo {author} {\bibfnamefont {T.}~\bibnamefont {Martin}}, \ and\ \bibinfo
  {author} {\bibfnamefont {I.}~\bibnamefont {Safi}},\ }\href {\doibase
  10.1103/PhysRevB.65.153304} {\bibfield  {journal} {\bibinfo  {journal} {Phys.
  Rev. B}\ }\textbf {\bibinfo {volume} {65}},\ \bibinfo {pages} {153304}
  (\bibinfo {year} {2002})}\BibitemShut {NoStop}%
\bibitem [{\citenamefont {Kane}\ and\ \citenamefont {Fisher}(1996)}]{kane1996}%
  \BibitemOpen
  \bibfield  {author} {\bibinfo {author} {\bibfnamefont {C.}~\bibnamefont
  {Kane}}\ and\ \bibinfo {author} {\bibfnamefont {M.~P.}\ \bibnamefont
  {Fisher}},\ }\href@noop {} {\bibfield  {journal} {\bibinfo  {journal}
  {Physical review letters}\ }\textbf {\bibinfo {volume} {76}},\ \bibinfo
  {pages} {3192} (\bibinfo {year} {1996})}\BibitemShut {NoStop}%
\bibitem [{\citenamefont {Zwillinger}\ and\ \citenamefont
  {Jeffrey}(2007)}]{zwillinger2007}%
  \BibitemOpen
  \bibfield  {author} {\bibinfo {author} {\bibfnamefont {D.}~\bibnamefont
  {Zwillinger}}\ and\ \bibinfo {author} {\bibfnamefont {A.}~\bibnamefont
  {Jeffrey}},\ }\href@noop {} {\emph {\bibinfo {title} {Table of Integrals,
  Series, and Products}}}\ (\bibinfo  {publisher} {Elsevier Science},\ \bibinfo
  {year} {2007})\BibitemShut {NoStop}%
\bibitem [{\citenamefont {Vannucci}\ \emph {et~al.}(2018)\citenamefont
  {Vannucci}, \citenamefont {Ronetti}, \citenamefont {Ferraro}, \citenamefont
  {Rech}, \citenamefont {Jonckheere}, \citenamefont {Martin},\ and\
  \citenamefont {Sassetti}}]{vannucci2018}%
  \BibitemOpen
  \bibfield  {author} {\bibinfo {author} {\bibfnamefont {L.}~\bibnamefont
  {Vannucci}}, \bibinfo {author} {\bibfnamefont {F.}~\bibnamefont {Ronetti}},
  \bibinfo {author} {\bibfnamefont {D.}~\bibnamefont {Ferraro}}, \bibinfo
  {author} {\bibfnamefont {J.}~\bibnamefont {Rech}}, \bibinfo {author}
  {\bibfnamefont {T.}~\bibnamefont {Jonckheere}}, \bibinfo {author}
  {\bibfnamefont {T.}~\bibnamefont {Martin}}, \ and\ \bibinfo {author}
  {\bibfnamefont {M.}~\bibnamefont {Sassetti}},\ }\href {\doibase
  10.1088/1742-6596/969/1/012143} {\bibfield  {journal} {\bibinfo  {journal}
  {Journal of Physics: Conference Series}\ }\textbf {\bibinfo {volume} {969}},\
  \bibinfo {pages} {012143} (\bibinfo {year} {2018})}\BibitemShut {NoStop}%
\bibitem [{\citenamefont {Miranda}(2003)}]{miranda2003}%
  \BibitemOpen
  \bibfield  {author} {\bibinfo {author} {\bibfnamefont {E.}~\bibnamefont
  {Miranda}},\ }\href@noop {} {\bibfield  {journal} {\bibinfo  {journal}
  {Brazilian Journal of Physics}\ }\textbf {\bibinfo {volume} {33}},\ \bibinfo
  {pages} {3} (\bibinfo {year} {2003})}\BibitemShut {NoStop}%
\bibitem [{\citenamefont {Giamarchi}(2003)}]{giamarchi2003}%
  \BibitemOpen
  \bibfield  {author} {\bibinfo {author} {\bibfnamefont {T.}~\bibnamefont
  {Giamarchi}},\ }\href {https://books.google.co.uk/books?id=GVeuKZLGMZ0C}
  {\emph {\bibinfo {title} {Quantum Physics in One Dimension}}},\ International
  Series of Monographs on Physics\ (\bibinfo  {publisher} {Clarendon Press},\
  \bibinfo {year} {2003})\BibitemShut {NoStop}%
\bibitem [{\citenamefont {Kane}\ and\ \citenamefont {Fisher}(1997)}]{kane1997}%
  \BibitemOpen
  \bibfield  {author} {\bibinfo {author} {\bibfnamefont {C.~L.}\ \bibnamefont
  {Kane}}\ and\ \bibinfo {author} {\bibfnamefont {M.~P.}\ \bibnamefont
  {Fisher}},\ }\href@noop {} {\bibfield  {journal} {\bibinfo  {journal}
  {Nature}\ }\textbf {\bibinfo {volume} {389}},\ \bibinfo {pages} {119}
  (\bibinfo {year} {1997})}\BibitemShut {NoStop}%
\bibitem [{Note1()}]{Note1}%
  \BibitemOpen
  \bibinfo {note} {We assume $V_{dc}>0$.}\BibitemShut {Stop}%
\bibitem [{\citenamefont {Schottky}(1918)}]{schottky1918}%
  \BibitemOpen
  \bibfield  {author} {\bibinfo {author} {\bibfnamefont {W.}~\bibnamefont
  {Schottky}},\ }\href@noop {} {\bibfield  {journal} {\bibinfo  {journal}
  {Annalen der physik}\ }\textbf {\bibinfo {volume} {362}},\ \bibinfo {pages}
  {541} (\bibinfo {year} {1918})}\BibitemShut {NoStop}%
\bibitem [{\citenamefont {Saminadayar}\ \emph {et~al.}(1997)\citenamefont
  {Saminadayar}, \citenamefont {Glattli}, \citenamefont {Jin},\ and\
  \citenamefont {Etienne}}]{saminadayar1997}%
  \BibitemOpen
  \bibfield  {author} {\bibinfo {author} {\bibfnamefont {L.}~\bibnamefont
  {Saminadayar}}, \bibinfo {author} {\bibfnamefont {D.~C.}\ \bibnamefont
  {Glattli}}, \bibinfo {author} {\bibfnamefont {Y.}~\bibnamefont {Jin}}, \ and\
  \bibinfo {author} {\bibfnamefont {B.}~\bibnamefont {Etienne}},\ }\href
  {\doibase 10.1103/PhysRevLett.79.2526} {\bibfield  {journal} {\bibinfo
  {journal} {Phys. Rev. Lett.}\ }\textbf {\bibinfo {volume} {79}},\ \bibinfo
  {pages} {2526} (\bibinfo {year} {1997})}\BibitemShut {NoStop}%
\bibitem [{\citenamefont {{de-Picciotto}}\ \emph {et~al.}(1997)\citenamefont
  {{de-Picciotto}}, \citenamefont {{Reznikov}}, \citenamefont {{Heiblum}},
  \citenamefont {{Umansky}}, \citenamefont {{Bunin}},\ and\ \citenamefont
  {{Mahalu}}}]{depicciotto1997}%
  \BibitemOpen
  \bibfield  {author} {\bibinfo {author} {\bibfnamefont {R.}~\bibnamefont
  {{de-Picciotto}}}, \bibinfo {author} {\bibfnamefont {M.}~\bibnamefont
  {{Reznikov}}}, \bibinfo {author} {\bibfnamefont {M.}~\bibnamefont
  {{Heiblum}}}, \bibinfo {author} {\bibfnamefont {V.}~\bibnamefont
  {{Umansky}}}, \bibinfo {author} {\bibfnamefont {G.}~\bibnamefont {{Bunin}}},
  \ and\ \bibinfo {author} {\bibfnamefont {D.}~\bibnamefont {{Mahalu}}},\
  }\href {\doibase 10.1038/38241} {\bibfield  {journal} {\bibinfo  {journal}
  {Nature (London)}\ }\textbf {\bibinfo {volume} {389}},\ \bibinfo {pages}
  {162} (\bibinfo {year} {1997})}\BibitemShut {NoStop}%
\bibitem [{\citenamefont {McNeil}\ \emph {et~al.}(2011)\citenamefont {McNeil},
  \citenamefont {Kataoka}, \citenamefont {Ford}, \citenamefont {Barnes},
  \citenamefont {Anderson}, \citenamefont {Jones}, \citenamefont {Farrer},\
  and\ \citenamefont {Ritchie}}]{mcneil2011}%
  \BibitemOpen
  \bibfield  {author} {\bibinfo {author} {\bibfnamefont {R.~P.~G.}\
  \bibnamefont {McNeil}}, \bibinfo {author} {\bibfnamefont {M.}~\bibnamefont
  {Kataoka}}, \bibinfo {author} {\bibfnamefont {C.~J.~B.}\ \bibnamefont
  {Ford}}, \bibinfo {author} {\bibfnamefont {C.~H.~W.}\ \bibnamefont {Barnes}},
  \bibinfo {author} {\bibfnamefont {D.}~\bibnamefont {Anderson}}, \bibinfo
  {author} {\bibfnamefont {G.~A.~C.}\ \bibnamefont {Jones}}, \bibinfo {author}
  {\bibfnamefont {I.}~\bibnamefont {Farrer}}, \ and\ \bibinfo {author}
  {\bibfnamefont {D.~A.}\ \bibnamefont {Ritchie}},\ }\href {\doibase
  10.1038/nature10444} {\bibfield  {journal} {\bibinfo  {journal} {Nature}\
  }\textbf {\bibinfo {volume} {477}},\ \bibinfo {pages} {439} (\bibinfo {year}
  {2011})}\BibitemShut {NoStop}%
\bibitem [{\citenamefont {Yamamoto}\ \emph {et~al.}(2012)\citenamefont
  {Yamamoto}, \citenamefont {Takada}, \citenamefont {B{\"a}uerle},
  \citenamefont {Watanabe}, \citenamefont {Wieck},\ and\ \citenamefont
  {Tarucha}}]{yamamoto2012}%
  \BibitemOpen
  \bibfield  {author} {\bibinfo {author} {\bibfnamefont {M.}~\bibnamefont
  {Yamamoto}}, \bibinfo {author} {\bibfnamefont {S.}~\bibnamefont {Takada}},
  \bibinfo {author} {\bibfnamefont {C.}~\bibnamefont {B{\"a}uerle}}, \bibinfo
  {author} {\bibfnamefont {K.}~\bibnamefont {Watanabe}}, \bibinfo {author}
  {\bibfnamefont {A.~D.}\ \bibnamefont {Wieck}}, \ and\ \bibinfo {author}
  {\bibfnamefont {S.}~\bibnamefont {Tarucha}},\ }\href {\doibase
  10.1038/nnano.2012.28} {\bibfield  {journal} {\bibinfo  {journal} {Nature
  Nanotechnology}\ }\textbf {\bibinfo {volume} {7}},\ \bibinfo {pages} {247}
  (\bibinfo {year} {2012})}\BibitemShut {NoStop}%
\bibitem [{\citenamefont {Edlbauer}\ \emph {et~al.}(2022)\citenamefont
  {Edlbauer}, \citenamefont {Wang}, \citenamefont {Crozes}, \citenamefont
  {Perrier}, \citenamefont {Ouacel}, \citenamefont {Geffroy}, \citenamefont
  {Georgiou}, \citenamefont {Chatzikyriakou}, \citenamefont {Lacerda-Santos},
  \citenamefont {Waintal}, \citenamefont {Glattli}, \citenamefont {Roulleau},
  \citenamefont {Nath}, \citenamefont {Kataoka}, \citenamefont
  {Splettstoesser}, \citenamefont {Acciai}, \citenamefont {da~Silva~Figueira},
  \citenamefont {{\"O}ztas}, \citenamefont {Trellakis}, \citenamefont {Grange},
  \citenamefont {Yevtushenko}, \citenamefont {Birner},\ and\ \citenamefont
  {B{\"a}uerle}}]{edlbauer2022}%
  \BibitemOpen
  \bibfield  {author} {\bibinfo {author} {\bibfnamefont {H.}~\bibnamefont
  {Edlbauer}}, \bibinfo {author} {\bibfnamefont {J.}~\bibnamefont {Wang}},
  \bibinfo {author} {\bibfnamefont {T.}~\bibnamefont {Crozes}}, \bibinfo
  {author} {\bibfnamefont {P.}~\bibnamefont {Perrier}}, \bibinfo {author}
  {\bibfnamefont {S.}~\bibnamefont {Ouacel}}, \bibinfo {author} {\bibfnamefont
  {C.}~\bibnamefont {Geffroy}}, \bibinfo {author} {\bibfnamefont
  {G.}~\bibnamefont {Georgiou}}, \bibinfo {author} {\bibfnamefont
  {E.}~\bibnamefont {Chatzikyriakou}}, \bibinfo {author} {\bibfnamefont
  {A.}~\bibnamefont {Lacerda-Santos}}, \bibinfo {author} {\bibfnamefont
  {X.}~\bibnamefont {Waintal}}, \bibinfo {author} {\bibfnamefont {D.~C.}\
  \bibnamefont {Glattli}}, \bibinfo {author} {\bibfnamefont {P.}~\bibnamefont
  {Roulleau}}, \bibinfo {author} {\bibfnamefont {J.}~\bibnamefont {Nath}},
  \bibinfo {author} {\bibfnamefont {M.}~\bibnamefont {Kataoka}}, \bibinfo
  {author} {\bibfnamefont {J.}~\bibnamefont {Splettstoesser}}, \bibinfo
  {author} {\bibfnamefont {M.}~\bibnamefont {Acciai}}, \bibinfo {author}
  {\bibfnamefont {M.~C.}\ \bibnamefont {da~Silva~Figueira}}, \bibinfo {author}
  {\bibfnamefont {K.}~\bibnamefont {{\"O}ztas}}, \bibinfo {author}
  {\bibfnamefont {A.}~\bibnamefont {Trellakis}}, \bibinfo {author}
  {\bibfnamefont {T.}~\bibnamefont {Grange}}, \bibinfo {author} {\bibfnamefont
  {O.~M.}\ \bibnamefont {Yevtushenko}}, \bibinfo {author} {\bibfnamefont
  {S.}~\bibnamefont {Birner}}, \ and\ \bibinfo {author} {\bibfnamefont
  {C.}~\bibnamefont {B{\"a}uerle}},\ }\href {\doibase
  10.1140/epjqt/s40507-022-00139-w} {\bibfield  {journal} {\bibinfo  {journal}
  {EPJ Quantum Technology}\ }\textbf {\bibinfo {volume} {9}},\ \bibinfo {pages}
  {21} (\bibinfo {year} {2022})}\BibitemShut {NoStop}%
\bibitem [{\citenamefont {Wahl}\ \emph {et~al.}(2014)\citenamefont {Wahl},
  \citenamefont {Rech}, \citenamefont {Jonckheere},\ and\ \citenamefont
  {Martin}}]{wahl2014}%
  \BibitemOpen
  \bibfield  {author} {\bibinfo {author} {\bibfnamefont {C.}~\bibnamefont
  {Wahl}}, \bibinfo {author} {\bibfnamefont {J.}~\bibnamefont {Rech}}, \bibinfo
  {author} {\bibfnamefont {T.}~\bibnamefont {Jonckheere}}, \ and\ \bibinfo
  {author} {\bibfnamefont {T.}~\bibnamefont {Martin}},\ }\href {\doibase
  10.1103/PhysRevLett.112.046802} {\bibfield  {journal} {\bibinfo  {journal}
  {Phys. Rev. Lett.}\ }\textbf {\bibinfo {volume} {112}},\ \bibinfo {pages}
  {046802} (\bibinfo {year} {2014})}\BibitemShut {NoStop}%
\bibitem [{\citenamefont {Marguerite}\ \emph {et~al.}(2016)\citenamefont
  {Marguerite}, \citenamefont {Cabart}, \citenamefont {Wahl}, \citenamefont
  {Roussel}, \citenamefont {Freulon}, \citenamefont {Ferraro}, \citenamefont
  {Grenier}, \citenamefont {Berroir}, \citenamefont
  {Pla\ifmmode~\mbox{\c{c}}\else \c{c}\fi{}ais}, \citenamefont {Jonckheere},
  \citenamefont {Rech}, \citenamefont {Martin}, \citenamefont {Degiovanni},
  \citenamefont {Cavanna}, \citenamefont {Jin},\ and\ \citenamefont
  {F\`eve}}]{marguerite2016}%
  \BibitemOpen
  \bibfield  {author} {\bibinfo {author} {\bibfnamefont {A.}~\bibnamefont
  {Marguerite}}, \bibinfo {author} {\bibfnamefont {C.}~\bibnamefont {Cabart}},
  \bibinfo {author} {\bibfnamefont {C.}~\bibnamefont {Wahl}}, \bibinfo {author}
  {\bibfnamefont {B.}~\bibnamefont {Roussel}}, \bibinfo {author} {\bibfnamefont
  {V.}~\bibnamefont {Freulon}}, \bibinfo {author} {\bibfnamefont
  {D.}~\bibnamefont {Ferraro}}, \bibinfo {author} {\bibfnamefont
  {C.}~\bibnamefont {Grenier}}, \bibinfo {author} {\bibfnamefont {J.-M.}\
  \bibnamefont {Berroir}}, \bibinfo {author} {\bibfnamefont {B.}~\bibnamefont
  {Pla\ifmmode~\mbox{\c{c}}\else \c{c}\fi{}ais}}, \bibinfo {author}
  {\bibfnamefont {T.}~\bibnamefont {Jonckheere}}, \bibinfo {author}
  {\bibfnamefont {J.}~\bibnamefont {Rech}}, \bibinfo {author} {\bibfnamefont
  {T.}~\bibnamefont {Martin}}, \bibinfo {author} {\bibfnamefont
  {P.}~\bibnamefont {Degiovanni}}, \bibinfo {author} {\bibfnamefont
  {A.}~\bibnamefont {Cavanna}}, \bibinfo {author} {\bibfnamefont
  {Y.}~\bibnamefont {Jin}}, \ and\ \bibinfo {author} {\bibfnamefont
  {G.}~\bibnamefont {F\`eve}},\ }\href {\doibase 10.1103/PhysRevB.94.115311}
  {\bibfield  {journal} {\bibinfo  {journal} {Phys. Rev. B}\ }\textbf {\bibinfo
  {volume} {94}},\ \bibinfo {pages} {115311} (\bibinfo {year}
  {2016})}\BibitemShut {NoStop}%
\bibitem [{\citenamefont {Bertin-Johannet}\ \emph {et~al.}(2024)\citenamefont
  {Bertin-Johannet}, \citenamefont {Popoff}, \citenamefont {Ronetti},
  \citenamefont {Rech}, \citenamefont {Jonckheere}, \citenamefont {Raymond},
  \citenamefont {Gr\'emaud},\ and\ \citenamefont
  {Martin}}]{bertinjohannet2023c}%
  \BibitemOpen
  \bibfield  {author} {\bibinfo {author} {\bibfnamefont {B.}~\bibnamefont
  {Bertin-Johannet}}, \bibinfo {author} {\bibfnamefont {A.}~\bibnamefont
  {Popoff}}, \bibinfo {author} {\bibfnamefont {F.}~\bibnamefont {Ronetti}},
  \bibinfo {author} {\bibfnamefont {J.}~\bibnamefont {Rech}}, \bibinfo {author}
  {\bibfnamefont {T.}~\bibnamefont {Jonckheere}}, \bibinfo {author}
  {\bibfnamefont {L.}~\bibnamefont {Raymond}}, \bibinfo {author} {\bibfnamefont
  {B.}~\bibnamefont {Gr\'emaud}}, \ and\ \bibinfo {author} {\bibfnamefont
  {T.}~\bibnamefont {Martin}},\ }\href {\doibase 10.1103/PhysRevB.109.035436}
  {\bibfield  {journal} {\bibinfo  {journal} {Phys. Rev. B}\ }\textbf {\bibinfo
  {volume} {109}},\ \bibinfo {pages} {035436} (\bibinfo {year}
  {2024})}\BibitemShut {NoStop}%
\bibitem [{\citenamefont {Sim}(2023)}]{sim-private}%
  \BibitemOpen
  \bibfield  {author} {\bibinfo {author} {\bibfnamefont {H.~S.}\ \bibnamefont
  {Sim}},\ }\href@noop {} {}\bibinfo {howpublished} {{Private Communication}}
  (\bibinfo {year} {2023})\BibitemShut {NoStop}%
\bibitem [{\citenamefont {Anantram}\ and\ \citenamefont
  {Datta}(1996)}]{anantram1996}%
  \BibitemOpen
  \bibfield  {author} {\bibinfo {author} {\bibfnamefont {M.~P.}\ \bibnamefont
  {Anantram}}\ and\ \bibinfo {author} {\bibfnamefont {S.}~\bibnamefont
  {Datta}},\ }\href {\doibase 10.1103/PhysRevB.53.16390} {\bibfield  {journal}
  {\bibinfo  {journal} {Phys. Rev. B}\ }\textbf {\bibinfo {volume} {53}},\
  \bibinfo {pages} {16390} (\bibinfo {year} {1996})}\BibitemShut {NoStop}%
\bibitem [{\citenamefont {Martin}(1996)}]{martin1996a}%
  \BibitemOpen
  \bibfield  {author} {\bibinfo {author} {\bibfnamefont {T.}~\bibnamefont
  {Martin}},\ }\href@noop {} {\bibfield  {journal} {\bibinfo  {journal}
  {Physics Letters A}\ }\textbf {\bibinfo {volume} {220}},\ \bibinfo {pages}
  {137} (\bibinfo {year} {1996})}\BibitemShut {NoStop}%
\bibitem [{\citenamefont {Torres}\ and\ \citenamefont
  {Martin}(1999)}]{torres1999}%
  \BibitemOpen
  \bibfield  {author} {\bibinfo {author} {\bibfnamefont {J.}~\bibnamefont
  {Torres}}\ and\ \bibinfo {author} {\bibfnamefont {T.}~\bibnamefont
  {Martin}},\ }\href {\doibase https://doi.org/10.1007/s100510051010}
  {\bibfield  {journal} {\bibinfo  {journal} {Eur. Phys. J. B}\ }\textbf
  {\bibinfo {volume} {12}},\ \bibinfo {pages} {319} (\bibinfo {year}
  {1999})}\BibitemShut {NoStop}%
\bibitem [{\citenamefont {Lesovik}\ \emph {et~al.}(2001)\citenamefont
  {Lesovik}, \citenamefont {Martin},\ and\ \citenamefont
  {Blatter}}]{lesovik2001}%
  \BibitemOpen
  \bibfield  {author} {\bibinfo {author} {\bibfnamefont {G.}~\bibnamefont
  {Lesovik}}, \bibinfo {author} {\bibfnamefont {T.}~\bibnamefont {Martin}}, \
  and\ \bibinfo {author} {\bibfnamefont {G.}~\bibnamefont {Blatter}},\ }\href
  {\doibase https://doi.org/10.1007/s10051-001-8675-4} {\bibfield  {journal}
  {\bibinfo  {journal} {Eur. Phys. J. B}\ }\textbf {\bibinfo {volume} {24}},\
  \bibinfo {pages} {287} (\bibinfo {year} {2001})}\BibitemShut {NoStop}%
\bibitem [{\citenamefont {Recher}\ \emph {et~al.}(2001)\citenamefont {Recher},
  \citenamefont {Sukhorukov},\ and\ \citenamefont {Loss}}]{recher2001}%
  \BibitemOpen
  \bibfield  {author} {\bibinfo {author} {\bibfnamefont {P.}~\bibnamefont
  {Recher}}, \bibinfo {author} {\bibfnamefont {E.~V.}\ \bibnamefont
  {Sukhorukov}}, \ and\ \bibinfo {author} {\bibfnamefont {D.}~\bibnamefont
  {Loss}},\ }\href {\doibase 10.1103/PhysRevB.63.165314} {\bibfield  {journal}
  {\bibinfo  {journal} {Phys. Rev. B}\ }\textbf {\bibinfo {volume} {63}},\
  \bibinfo {pages} {165314} (\bibinfo {year} {2001})}\BibitemShut {NoStop}%
\bibitem [{\citenamefont {Sauret}\ and\ \citenamefont
  {Feinberg}(2004)}]{sauret2004}%
  \BibitemOpen
  \bibfield  {author} {\bibinfo {author} {\bibfnamefont {O.}~\bibnamefont
  {Sauret}}\ and\ \bibinfo {author} {\bibfnamefont {D.}~\bibnamefont
  {Feinberg}},\ }\href {\doibase 10.1103/PhysRevLett.92.106601} {\bibfield
  {journal} {\bibinfo  {journal} {Phys. Rev. Lett.}\ }\textbf {\bibinfo
  {volume} {92}},\ \bibinfo {pages} {106601} (\bibinfo {year}
  {2004})}\BibitemShut {NoStop}%
\bibitem [{\citenamefont {Sauret}\ \emph {et~al.}(2005)\citenamefont {Sauret},
  \citenamefont {Martin},\ and\ \citenamefont {Feinberg}}]{sauret2005}%
  \BibitemOpen
  \bibfield  {author} {\bibinfo {author} {\bibfnamefont {O.}~\bibnamefont
  {Sauret}}, \bibinfo {author} {\bibfnamefont {T.}~\bibnamefont {Martin}}, \
  and\ \bibinfo {author} {\bibfnamefont {D.}~\bibnamefont {Feinberg}},\ }\href
  {\doibase 10.1103/PhysRevB.72.024544} {\bibfield  {journal} {\bibinfo
  {journal} {Phys. Rev. B}\ }\textbf {\bibinfo {volume} {72}},\ \bibinfo
  {pages} {024544} (\bibinfo {year} {2005})}\BibitemShut {NoStop}%
\bibitem [{\citenamefont {Chevallier}\ \emph {et~al.}(2011)\citenamefont
  {Chevallier}, \citenamefont {Rech}, \citenamefont {Jonckheere},\ and\
  \citenamefont {Martin}}]{chevallier2011}%
  \BibitemOpen
  \bibfield  {author} {\bibinfo {author} {\bibfnamefont {D.}~\bibnamefont
  {Chevallier}}, \bibinfo {author} {\bibfnamefont {J.}~\bibnamefont {Rech}},
  \bibinfo {author} {\bibfnamefont {T.}~\bibnamefont {Jonckheere}}, \ and\
  \bibinfo {author} {\bibfnamefont {T.}~\bibnamefont {Martin}},\ }\href
  {\doibase 10.1103/PhysRevB.83.125421} {\bibfield  {journal} {\bibinfo
  {journal} {Phys. Rev. B}\ }\textbf {\bibinfo {volume} {83}},\ \bibinfo
  {pages} {125421} (\bibinfo {year} {2011})}\BibitemShut {NoStop}%
\bibitem [{\citenamefont {Rech}\ \emph {et~al.}(2012)\citenamefont {Rech},
  \citenamefont {Chevallier}, \citenamefont {Jonckheere},\ and\ \citenamefont
  {Martin}}]{rech2012}%
  \BibitemOpen
  \bibfield  {author} {\bibinfo {author} {\bibfnamefont {J.}~\bibnamefont
  {Rech}}, \bibinfo {author} {\bibfnamefont {D.}~\bibnamefont {Chevallier}},
  \bibinfo {author} {\bibfnamefont {T.}~\bibnamefont {Jonckheere}}, \ and\
  \bibinfo {author} {\bibfnamefont {T.}~\bibnamefont {Martin}},\ }\href
  {\doibase 10.1103/PhysRevB.85.035419} {\bibfield  {journal} {\bibinfo
  {journal} {Phys. Rev. B}\ }\textbf {\bibinfo {volume} {85}},\ \bibinfo
  {pages} {035419} (\bibinfo {year} {2012})}\BibitemShut {NoStop}%
\bibitem [{\citenamefont {Bell}(1966)}]{bell1966}%
  \BibitemOpen
  \bibfield  {author} {\bibinfo {author} {\bibfnamefont {J.~S.}\ \bibnamefont
  {Bell}},\ }\href@noop {} {\bibfield  {journal} {\bibinfo  {journal} {Reviews
  of Modern physics}\ }\textbf {\bibinfo {volume} {38}},\ \bibinfo {pages}
  {447} (\bibinfo {year} {1966})}\BibitemShut {NoStop}%
\bibitem [{\citenamefont {Aspect}\ \emph {et~al.}(1982)\citenamefont {Aspect},
  \citenamefont {Dalibard},\ and\ \citenamefont {Roger}}]{aspect1982}%
  \BibitemOpen
  \bibfield  {author} {\bibinfo {author} {\bibfnamefont {A.}~\bibnamefont
  {Aspect}}, \bibinfo {author} {\bibfnamefont {J.}~\bibnamefont {Dalibard}}, \
  and\ \bibinfo {author} {\bibfnamefont {G.}~\bibnamefont {Roger}},\ }\href
  {\doibase 10.1103/PhysRevLett.49.1804} {\bibfield  {journal} {\bibinfo
  {journal} {Phys. Rev. Lett.}\ }\textbf {\bibinfo {volume} {49}},\ \bibinfo
  {pages} {1804} (\bibinfo {year} {1982})}\BibitemShut {NoStop}%
\bibitem [{\citenamefont {Chtchelkatchev}\ \emph {et~al.}(2002)\citenamefont
  {Chtchelkatchev}, \citenamefont {Blatter}, \citenamefont {Lesovik},\ and\
  \citenamefont {Martin}}]{chtchelkatchev2002}%
  \BibitemOpen
  \bibfield  {author} {\bibinfo {author} {\bibfnamefont {N.~M.}\ \bibnamefont
  {Chtchelkatchev}}, \bibinfo {author} {\bibfnamefont {G.}~\bibnamefont
  {Blatter}}, \bibinfo {author} {\bibfnamefont {G.~B.}\ \bibnamefont
  {Lesovik}}, \ and\ \bibinfo {author} {\bibfnamefont {T.}~\bibnamefont
  {Martin}},\ }\href {\doibase 10.1103/PhysRevB.66.161320} {\bibfield
  {journal} {\bibinfo  {journal} {Phys. Rev. B}\ }\textbf {\bibinfo {volume}
  {66}},\ \bibinfo {pages} {161320} (\bibinfo {year} {2002})}\BibitemShut
  {NoStop}%
\bibitem [{\citenamefont {Bayandin}\ \emph {et~al.}(2006)\citenamefont
  {Bayandin}, \citenamefont {Lesovik},\ and\ \citenamefont
  {Martin}}]{bayandin2006}%
  \BibitemOpen
  \bibfield  {author} {\bibinfo {author} {\bibfnamefont {K.~V.}\ \bibnamefont
  {Bayandin}}, \bibinfo {author} {\bibfnamefont {G.~B.}\ \bibnamefont
  {Lesovik}}, \ and\ \bibinfo {author} {\bibfnamefont {T.}~\bibnamefont
  {Martin}},\ }\href {\doibase 10.1103/PhysRevB.74.085326} {\bibfield
  {journal} {\bibinfo  {journal} {Phys. Rev. B}\ }\textbf {\bibinfo {volume}
  {74}},\ \bibinfo {pages} {085326} (\bibinfo {year} {2006})}\BibitemShut
  {NoStop}%
\bibitem [{\citenamefont {Das}\ \emph {et~al.}(2012)\citenamefont {Das},
  \citenamefont {Ronen}, \citenamefont {Heiblum}, \citenamefont {Mahalu},
  \citenamefont {Kretinin},\ and\ \citenamefont {Shtrikman}}]{das2012}%
  \BibitemOpen
  \bibfield  {author} {\bibinfo {author} {\bibfnamefont {A.}~\bibnamefont
  {Das}}, \bibinfo {author} {\bibfnamefont {Y.}~\bibnamefont {Ronen}}, \bibinfo
  {author} {\bibfnamefont {M.}~\bibnamefont {Heiblum}}, \bibinfo {author}
  {\bibfnamefont {D.}~\bibnamefont {Mahalu}}, \bibinfo {author} {\bibfnamefont
  {A.~V.}\ \bibnamefont {Kretinin}}, \ and\ \bibinfo {author} {\bibfnamefont
  {H.}~\bibnamefont {Shtrikman}},\ }\href@noop {} {\bibfield  {journal}
  {\bibinfo  {journal} {Nature communications}\ }\textbf {\bibinfo {volume}
  {3}},\ \bibinfo {pages} {1} (\bibinfo {year} {2012})}\BibitemShut {NoStop}%
\bibitem [{\citenamefont {Bertin-Johannet}\ \emph {et~al.}(2022)\citenamefont
  {Bertin-Johannet}, \citenamefont {Rech}, \citenamefont {Jonckheere},
  \citenamefont {Gr\'emaud}, \citenamefont {Raymond},\ and\ \citenamefont
  {Martin}}]{bertinjohannet2022}%
  \BibitemOpen
  \bibfield  {author} {\bibinfo {author} {\bibfnamefont {B.}~\bibnamefont
  {Bertin-Johannet}}, \bibinfo {author} {\bibfnamefont {J.}~\bibnamefont
  {Rech}}, \bibinfo {author} {\bibfnamefont {T.}~\bibnamefont {Jonckheere}},
  \bibinfo {author} {\bibfnamefont {B.}~\bibnamefont {Gr\'emaud}}, \bibinfo
  {author} {\bibfnamefont {L.}~\bibnamefont {Raymond}}, \ and\ \bibinfo
  {author} {\bibfnamefont {T.}~\bibnamefont {Martin}},\ }\href {\doibase
  10.1103/PhysRevB.105.115112} {\bibfield  {journal} {\bibinfo  {journal}
  {Phys. Rev. B}\ }\textbf {\bibinfo {volume} {105}},\ \bibinfo {pages}
  {115112} (\bibinfo {year} {2022})}\BibitemShut {NoStop}%
\bibitem [{\citenamefont {Vanevi{\'c}}\ \emph {et~al.}(2016)\citenamefont
  {Vanevi{\'c}}, \citenamefont {Gabelli}, \citenamefont {Belzig},\ and\
  \citenamefont {Reulet}}]{vanevic2016a}%
  \BibitemOpen
  \bibfield  {author} {\bibinfo {author} {\bibfnamefont {M.}~\bibnamefont
  {Vanevi{\'c}}}, \bibinfo {author} {\bibfnamefont {J.}~\bibnamefont
  {Gabelli}}, \bibinfo {author} {\bibfnamefont {W.}~\bibnamefont {Belzig}}, \
  and\ \bibinfo {author} {\bibfnamefont {B.}~\bibnamefont {Reulet}},\
  }\href@noop {} {\bibfield  {journal} {\bibinfo  {journal} {Physical Review
  B}\ }\textbf {\bibinfo {volume} {93}},\ \bibinfo {pages} {041416} (\bibinfo
  {year} {2016})}\BibitemShut {NoStop}%
\bibitem [{\citenamefont {Bertin-Johannet}\ \emph
  {et~al.}(2023{\natexlab{a}})\citenamefont {Bertin-Johannet}, \citenamefont
  {Raymond}, \citenamefont {Ronetti}, \citenamefont {Rech}, \citenamefont
  {Jonckheere}, \citenamefont {Gr{\'e}maud},\ and\ \citenamefont
  {Martin}}]{bertinjohannet2023a}%
  \BibitemOpen
  \bibfield  {author} {\bibinfo {author} {\bibfnamefont {B.}~\bibnamefont
  {Bertin-Johannet}}, \bibinfo {author} {\bibfnamefont {L.}~\bibnamefont
  {Raymond}}, \bibinfo {author} {\bibfnamefont {F.}~\bibnamefont {Ronetti}},
  \bibinfo {author} {\bibfnamefont {J.}~\bibnamefont {Rech}}, \bibinfo {author}
  {\bibfnamefont {T.}~\bibnamefont {Jonckheere}}, \bibinfo {author}
  {\bibfnamefont {B.}~\bibnamefont {Gr{\'e}maud}}, \ and\ \bibinfo {author}
  {\bibfnamefont {T.}~\bibnamefont {Martin}},\ }\href@noop {} {\bibfield
  {journal} {\bibinfo  {journal} {Applied Physics Letters}\ }\textbf {\bibinfo
  {volume} {122}} (\bibinfo {year} {2023}{\natexlab{a}})}\BibitemShut {NoStop}%
\bibitem [{\citenamefont {Bertin-Johannet}\ \emph
  {et~al.}(2023{\natexlab{b}})\citenamefont {Bertin-Johannet}, \citenamefont
  {Gr{\'e}maud}, \citenamefont {Ronneti}, \citenamefont {Raymond},
  \citenamefont {Rech}, \citenamefont {Jonckheere},\ and\ \citenamefont
  {Martin}}]{bertinjohannet2023b}%
  \BibitemOpen
  \bibfield  {author} {\bibinfo {author} {\bibfnamefont {B.}~\bibnamefont
  {Bertin-Johannet}}, \bibinfo {author} {\bibfnamefont {B.}~\bibnamefont
  {Gr{\'e}maud}}, \bibinfo {author} {\bibfnamefont {F.}~\bibnamefont
  {Ronneti}}, \bibinfo {author} {\bibfnamefont {L.}~\bibnamefont {Raymond}},
  \bibinfo {author} {\bibfnamefont {J.}~\bibnamefont {Rech}}, \bibinfo {author}
  {\bibfnamefont {T.}~\bibnamefont {Jonckheere}}, \ and\ \bibinfo {author}
  {\bibfnamefont {T.}~\bibnamefont {Martin}},\ }\href@noop {} {\bibfield
  {journal} {\bibinfo  {journal} {arXiv:2311.15684}\ } (\bibinfo {year}
  {2023}{\natexlab{b}})}\BibitemShut {NoStop}%
\bibitem [{\citenamefont {Jonckheere}\ \emph {et~al.}(2013)\citenamefont
  {Jonckheere}, \citenamefont {Rech}, \citenamefont {Martin}, \citenamefont
  {Doucot}, \citenamefont {Feinberg},\ and\ \citenamefont
  {Melin}}]{jonckheere2013}%
  \BibitemOpen
  \bibfield  {author} {\bibinfo {author} {\bibfnamefont {T.}~\bibnamefont
  {Jonckheere}}, \bibinfo {author} {\bibfnamefont {J.}~\bibnamefont {Rech}},
  \bibinfo {author} {\bibfnamefont {T.}~\bibnamefont {Martin}}, \bibinfo
  {author} {\bibfnamefont {B.}~\bibnamefont {Doucot}}, \bibinfo {author}
  {\bibfnamefont {D.}~\bibnamefont {Feinberg}}, \ and\ \bibinfo {author}
  {\bibfnamefont {R.}~\bibnamefont {Melin}},\ }\href {\doibase
  10.1103/PhysRevB.87.214501} {\bibfield  {journal} {\bibinfo  {journal}
  {Physical Review B - Condensed Matter and Materials Physics}\ }\textbf
  {\bibinfo {volume} {87}} (\bibinfo {year} {2013}),\
  10.1103/PhysRevB.87.214501}\BibitemShut {NoStop}%
\bibitem [{\citenamefont {Jacquet}\ \emph {et~al.}(2020)\citenamefont
  {Jacquet}, \citenamefont {Popoff}, \citenamefont {Imura}, \citenamefont
  {Rech}, \citenamefont {Jonckheere}, \citenamefont {Raymond}, \citenamefont
  {Zazunov},\ and\ \citenamefont {Martin}}]{jacquet2020}%
  \BibitemOpen
  \bibfield  {author} {\bibinfo {author} {\bibfnamefont {R.}~\bibnamefont
  {Jacquet}}, \bibinfo {author} {\bibfnamefont {A.}~\bibnamefont {Popoff}},
  \bibinfo {author} {\bibfnamefont {K.-I.}\ \bibnamefont {Imura}}, \bibinfo
  {author} {\bibfnamefont {J.}~\bibnamefont {Rech}}, \bibinfo {author}
  {\bibfnamefont {T.}~\bibnamefont {Jonckheere}}, \bibinfo {author}
  {\bibfnamefont {L.}~\bibnamefont {Raymond}}, \bibinfo {author} {\bibfnamefont
  {A.}~\bibnamefont {Zazunov}}, \ and\ \bibinfo {author} {\bibfnamefont
  {T.}~\bibnamefont {Martin}},\ }\href {\doibase 10.1103/PhysRevB.102.064510}
  {\bibfield  {journal} {\bibinfo  {journal} {Phys. Rev. B}\ }\textbf {\bibinfo
  {volume} {102}},\ \bibinfo {pages} {064510} (\bibinfo {year}
  {2020})}\BibitemShut {NoStop}%
\bibitem [{\citenamefont {Vannucci}\ \emph {et~al.}(2017)\citenamefont
  {Vannucci}, \citenamefont {Ronetti}, \citenamefont {Rech}, \citenamefont
  {Ferraro}, \citenamefont {Jonckheere}, \citenamefont {Martin},\ and\
  \citenamefont {Sassetti}}]{vannucci2017}%
  \BibitemOpen
  \bibfield  {author} {\bibinfo {author} {\bibfnamefont {L.}~\bibnamefont
  {Vannucci}}, \bibinfo {author} {\bibfnamefont {F.}~\bibnamefont {Ronetti}},
  \bibinfo {author} {\bibfnamefont {J.}~\bibnamefont {Rech}}, \bibinfo {author}
  {\bibfnamefont {D.}~\bibnamefont {Ferraro}}, \bibinfo {author} {\bibfnamefont
  {T.}~\bibnamefont {Jonckheere}}, \bibinfo {author} {\bibfnamefont
  {T.}~\bibnamefont {Martin}}, \ and\ \bibinfo {author} {\bibfnamefont
  {M.}~\bibnamefont {Sassetti}},\ }\href {\doibase 10.1103/PhysRevB.95.245415}
  {\bibfield  {journal} {\bibinfo  {journal} {Phys. Rev. B}\ }\textbf {\bibinfo
  {volume} {95}},\ \bibinfo {pages} {245415} (\bibinfo {year}
  {2017})}\BibitemShut {NoStop}%
\bibitem [{\citenamefont {Fukuzawa}\ \emph {et~al.}(2023)\citenamefont
  {Fukuzawa}, \citenamefont {Kato}, \citenamefont {Jonckheere}, \citenamefont
  {Rech},\ and\ \citenamefont {Martin}}]{fukuzawa2023}%
  \BibitemOpen
  \bibfield  {author} {\bibinfo {author} {\bibfnamefont {K.}~\bibnamefont
  {Fukuzawa}}, \bibinfo {author} {\bibfnamefont {T.}~\bibnamefont {Kato}},
  \bibinfo {author} {\bibfnamefont {T.}~\bibnamefont {Jonckheere}}, \bibinfo
  {author} {\bibfnamefont {J.}~\bibnamefont {Rech}}, \ and\ \bibinfo {author}
  {\bibfnamefont {T.}~\bibnamefont {Martin}},\ }\href {\doibase
  10.1103/PhysRevB.108.125307} {\bibfield  {journal} {\bibinfo  {journal}
  {Phys. Rev. B}\ }\textbf {\bibinfo {volume} {108}},\ \bibinfo {pages}
  {125307} (\bibinfo {year} {2023})}\BibitemShut {NoStop}%
\bibitem [{\citenamefont {Ronetti}(2024)}]{ronetti2024}%
  \BibitemOpen
  \bibfield  {author} {\bibinfo {author} {\bibfnamefont {F.}~\bibnamefont
  {Ronetti}},\ }\href@noop {} {}\bibinfo {howpublished} {{Private
  Communication}} (\bibinfo {year} {2024})\BibitemShut {NoStop}%
\bibitem [{\citenamefont {Bennett}\ \emph {et~al.}(1993)\citenamefont
  {Bennett}, \citenamefont {Brassard}, \citenamefont {Cr\'epeau}, \citenamefont
  {Jozsa}, \citenamefont {Peres},\ and\ \citenamefont
  {Wootters}}]{bennett1993}%
  \BibitemOpen
  \bibfield  {author} {\bibinfo {author} {\bibfnamefont {C.~H.}\ \bibnamefont
  {Bennett}}, \bibinfo {author} {\bibfnamefont {G.}~\bibnamefont {Brassard}},
  \bibinfo {author} {\bibfnamefont {C.}~\bibnamefont {Cr\'epeau}}, \bibinfo
  {author} {\bibfnamefont {R.}~\bibnamefont {Jozsa}}, \bibinfo {author}
  {\bibfnamefont {A.}~\bibnamefont {Peres}}, \ and\ \bibinfo {author}
  {\bibfnamefont {W.~K.}\ \bibnamefont {Wootters}},\ }\href {\doibase
  10.1103/PhysRevLett.70.1895} {\bibfield  {journal} {\bibinfo  {journal}
  {Phys. Rev. Lett.}\ }\textbf {\bibinfo {volume} {70}},\ \bibinfo {pages}
  {1895} (\bibinfo {year} {1993})}\BibitemShut {NoStop}%
\bibitem [{\citenamefont {Long}\ and\ \citenamefont {Liu}(2002)}]{long2002}%
  \BibitemOpen
  \bibfield  {author} {\bibinfo {author} {\bibfnamefont {G.-L.}\ \bibnamefont
  {Long}}\ and\ \bibinfo {author} {\bibfnamefont {X.-S.}\ \bibnamefont {Liu}},\
  }\href@noop {} {\bibfield  {journal} {\bibinfo  {journal} {Physical Review
  A}\ }\textbf {\bibinfo {volume} {65}},\ \bibinfo {pages} {032302} (\bibinfo
  {year} {2002})}\BibitemShut {NoStop}%
\bibitem [{\citenamefont {Ekert}(1991)}]{ekert1991}%
  \BibitemOpen
  \bibfield  {author} {\bibinfo {author} {\bibfnamefont {A.~K.}\ \bibnamefont
  {Ekert}},\ }\href@noop {} {\bibfield  {journal} {\bibinfo  {journal}
  {Physical review letters}\ }\textbf {\bibinfo {volume} {67}},\ \bibinfo
  {pages} {661} (\bibinfo {year} {1991})}\BibitemShut {NoStop}%
\bibitem [{\citenamefont {Bennett}\ \emph {et~al.}(1992)\citenamefont
  {Bennett}, \citenamefont {Brassard},\ and\ \citenamefont
  {Mermin}}]{bennett1992}%
  \BibitemOpen
  \bibfield  {author} {\bibinfo {author} {\bibfnamefont {C.~H.}\ \bibnamefont
  {Bennett}}, \bibinfo {author} {\bibfnamefont {G.}~\bibnamefont {Brassard}}, \
  and\ \bibinfo {author} {\bibfnamefont {N.~D.}\ \bibnamefont {Mermin}},\
  }\href@noop {} {\bibfield  {journal} {\bibinfo  {journal} {Physical review
  letters}\ }\textbf {\bibinfo {volume} {68}},\ \bibinfo {pages} {557}
  (\bibinfo {year} {1992})}\BibitemShut {NoStop}%
\bibitem [{\citenamefont {Pan}\ \emph {et~al.}(2020)\citenamefont {Pan},
  \citenamefont {Li}, \citenamefont {Ruan}, \citenamefont {Ng},\ and\
  \citenamefont {Hanzo}}]{pan2020}%
  \BibitemOpen
  \bibfield  {author} {\bibinfo {author} {\bibfnamefont {D.}~\bibnamefont
  {Pan}}, \bibinfo {author} {\bibfnamefont {K.}~\bibnamefont {Li}}, \bibinfo
  {author} {\bibfnamefont {D.}~\bibnamefont {Ruan}}, \bibinfo {author}
  {\bibfnamefont {S.~X.}\ \bibnamefont {Ng}}, \ and\ \bibinfo {author}
  {\bibfnamefont {L.}~\bibnamefont {Hanzo}},\ }\href {\doibase
  10.1109/ACCESS.2020.3006136} {\bibfield  {journal} {\bibinfo  {journal} {IEEE
  Access}\ }\textbf {\bibinfo {volume} {8}},\ \bibinfo {pages} {121146}
  (\bibinfo {year} {2020})}\BibitemShut {NoStop}%
\bibitem [{\citenamefont {Bennett}\ and\ \citenamefont
  {Wiesner}(1992)}]{bennett1992b}%
  \BibitemOpen
  \bibfield  {author} {\bibinfo {author} {\bibfnamefont {C.~H.}\ \bibnamefont
  {Bennett}}\ and\ \bibinfo {author} {\bibfnamefont {S.~J.}\ \bibnamefont
  {Wiesner}},\ }\href {\doibase 10.1103/PhysRevLett.69.2881} {\bibfield
  {journal} {\bibinfo  {journal} {Phys. Rev. Lett.}\ }\textbf {\bibinfo
  {volume} {69}},\ \bibinfo {pages} {2881} (\bibinfo {year}
  {1992})}\BibitemShut {NoStop}%
\bibitem [{\citenamefont {Martin}\ and\ \citenamefont
  {Campbell}(1987)}]{martin1987}%
  \BibitemOpen
  \bibfield  {author} {\bibinfo {author} {\bibfnamefont {T.}~\bibnamefont
  {Martin}}\ and\ \bibinfo {author} {\bibfnamefont {D.~K.}\ \bibnamefont
  {Campbell}},\ }\href {\doibase 10.1103/PhysRevB.35.7732} {\bibfield
  {journal} {\bibinfo  {journal} {Phys. Rev. B}\ }\textbf {\bibinfo {volume}
  {35}},\ \bibinfo {pages} {7732} (\bibinfo {year} {1987})}\BibitemShut
  {NoStop}%
\bibitem [{\citenamefont {Loh}\ \emph {et~al.}(1988)\citenamefont {Loh},
  \citenamefont {Martin}, \citenamefont {Prelovsek},\ and\ \citenamefont
  {Campbell}}]{loh1988}%
  \BibitemOpen
  \bibfield  {author} {\bibinfo {author} {\bibfnamefont {E.~Y.}\ \bibnamefont
  {Loh}}, \bibinfo {author} {\bibfnamefont {T.}~\bibnamefont {Martin}},
  \bibinfo {author} {\bibfnamefont {P.}~\bibnamefont {Prelovsek}}, \ and\
  \bibinfo {author} {\bibfnamefont {D.~K.}\ \bibnamefont {Campbell}},\ }\href
  {\doibase 10.1103/PhysRevB.38.2494} {\bibfield  {journal} {\bibinfo
  {journal} {Phys. Rev. B}\ }\textbf {\bibinfo {volume} {38}},\ \bibinfo
  {pages} {2494} (\bibinfo {year} {1988})}\BibitemShut {NoStop}%
\bibitem [{\citenamefont {Anderson}(1972)}]{anderson1972}%
  \BibitemOpen
  \bibfield  {author} {\bibinfo {author} {\bibfnamefont {P.~W.}\ \bibnamefont
  {Anderson}},\ }\href@noop {} {\bibfield  {journal} {\bibinfo  {journal}
  {Science}\ }\textbf {\bibinfo {volume} {177}},\ \bibinfo {pages} {393}
  (\bibinfo {year} {1972})}\BibitemShut {NoStop}%
\end{thebibliography}%

\end{document}